\documentclass{aastex631}

\usepackage{bm}
\usepackage{amsmath}

\accepted{June 30, 2022}

\shorttitle{Semi-analytical NEO propagation}
\shortauthors{Fuentes-Muñoz, Meyer, Scheeres}
\graphicspath{{./}{FinalFigures/}}
\begin{document}

% \title{A semi-analytical near-Earth asteroid propagation method}
\title{Semi-analytical near-Earth objects propagation: \\the orbit history of (35107) 1991 VH and (175706) 1996 FG3}

\author[0000-0001-5875-1083]{Oscar Fuentes-Muñoz}
\affiliation{Research Assistant \\ 
Ann and H. J. Smead Department of Aerospace Engineering Sciences, University of Colorado Boulder, 3775 Discovery Dr., \\ Boulder, CO 80303, USA}

\author[0000-0001-8437-1076]{Alex J. Meyer}
\affiliation{Research Assistant \\ 
Ann and H. J. Smead Department of Aerospace Engineering Sciences, University of Colorado Boulder, 3775 Discovery Dr., \\ Boulder, CO 80303, USA}

\author[0000-0003-0558-3842]{Daniel J. Scheeres}
\affiliation{University of Colorado Distinguished Professor \\ 
Ann and H. J. Smead Department of Aerospace Engineering Sciences, University of Colorado Boulder, 3775 Discovery Dr., \\ Boulder, CO 80303, USA}

\begin{abstract}
The propagation of small bodies in the Solar system is driven by the combination of planetary encounters that cause abrupt changes in their orbits and secular long-term perturbations. We propose a propagation strategy that combines both of these effects into a single framework for long-term, rapid propagation of small bodies in the inner Solar System. The analytical secular perturbation of Jupiter is interrupted to numerically solve planetary encounters, which last a small fraction of the simulation time. The proposed propagation method is compared to numerical integrations in the Solar system, effectively capturing properties of the numerical solutions in a fraction of the computational time. We study the orbital history of the Janus mission targets: (35107) 1991 VH and (175706) 1996 FG3, obtaining a stochastic representation of their long-term dynamics and frequencies of very close encounters. Over the last million years the probability of a strongly perturbing flyby is found to be small.
\end{abstract}

% \begin{abstract}
% The long-term propagation of small bodies in the Solar system is challenged by a combination of planetary encounters that cause abrupt changes in their orbits combined with secular drifts of the orbits due to planetary perturbations. The proposed propagation strategy combines both of these effects into a single framework for long-term, rapid propagation of small bodies in the inner Solar System. The analytical secular perturbation of Jupiter is interrupted to numerically solve planetary encounters, which last a small fraction of the simulation time. The proposed propagation method is compared to numerical integrations in the Solar system, effectively capturing properties of the numerical solutions in a fraction of the computational time.
% \end{abstract}

%\keywords{Asteroids --- Near-Earth objects --- Astrodynamics --- Asteroid dynamics --- Close encounters --- Opik theory}

\section{Introduction} \label{sec:intro}
As remnants of primordial planetary formation, Near Earth Objects (NEOs) are relevant targets for scientific exploration. The number of discovered NEOs is expected to continue increasing with future surveys, offering new opportunities to the scientific community \citep{Jones2018}. The interest and availability of specific objects is assessed with long term predictions of their orbits. Trajectories of Near Earth Objects are dominated by close encounters with the inner Solar System planets and secular perturbations \citep{Michel1996}.  Planetary encounters cause a high dependence on the initial conditions as the flybys cause neighboring trajectories to diverge \citep{Tancredi1998}. This divergence of the dynamics also causes the uncertainty to grow very rapidly. Thus, an analytical model that captures the main dynamical effects can avoid the computational cost of using high-fidelity models while capturing the overall statistical evolution of the orbits accurately. The semi-analytical propagation of asteroids allows the rapid propagation of NEO orbits, in the interest of the analysis of large databases of NEOs.

The long-term study of asteroid orbits has been achieved in the past using a wide variety of analytical, semi-analytical and numerical methods. Analytical methods are based on the study of the gravity potential to obtain secular and resonant perturbations \citep{Milani1990}. Semi-analytical methods are used to map orbital elements to {the locations of linear secular resonances, which are resonances involving one planetary and one asteroid frequency} \citep{Michel1997,Michel1997-2}. Both types of solutions represent the dynamics of asteroids in the absence of planetary encounters by averaging the perturbing potential.

On the other hand, previous studies focus on the accumulation of planetary encounters in contrast to numerical integration \citep{Dones1999}. The effect of close encounters on the orbit of asteroids can be computed using analytical \citep{opik1976interplanetary}, semi-analytical or numerical methods. Semi-analytical solutions \citep{Alessi2015} allow the computation of flybys treating the planet as a perturbing force in the Lagrange Planetary Equations. Specific numerical integrators are convenient to propagate orbits of asteroids in the long-term, in which symplecticity is desired along with the capacity to accurately solve close encounters \citep{Wisdom1991,Chambers1999}. Under multiple resonances asteroids start to encounter planets while their eccentricity increases. This increase often causes the asteroids to eventually collide with the Sun, planets or to be ejected from the Solar System on a hyperbolic orbit \citep{Farinella1994,Gladman1997,Milani1989,Dones1999,Michel2005}.

In this paper we aim to provide a simulation framework for the propagation of particles in the Solar System. Our approach consists in the analytical propagation of the particle until a close encounter is found. The propagation is stopped when the trajectory is close to a planet, then the close encounter is evaluated numerically. The evaluation of the encounters is based on a quadrature of the Lagrange Planetary Equations (LPE) around the closest approach date. After the encounter the analytical propagation of the orbit is resumed. The propagation under secular perturbations provides a realistic prediction of when the next encounter can occur as the orbit of the asteroid drifts between different regions of the inner Solar System. This approach reduces substantially the computational time of solutions obtained entirely by numerical integration while providing deeper insight into the dynamics.

The use of the analytical secular model allows the prediction of long-term properties of the asteroid dynamics. Eccentricities, inclinations and angles of asteroid and planets drift secularly. Thus, we can propagate the minimum orbit intersection distance (MOID). The MOID constrains the minimum closest approach distance between the asteroid and the planets and defines if asteroids are potentially hazardous (PHAs). The long-term dynamics of the orbits of NEOs and the MOID are studied by sampling a large number of virtual asteroids from their uncertainty distributions. We use the semi-analytical propagation of these asteroids to show the stochastic nature of the orbital evolution of NEOs.

The semi-analytical propagation allows us to track the encounters experienced by asteroids in the inner Solar System, which can perturb the physical properties of asteroids. The orbits of binary asteroids can be disrupted by a very close encounter \citep{MEYER2021114554}. In this paper we study the orbital history of the targets of the exploration mission Janus \citep{Scheeres2020Janus}: the two binaries (35107) 1991 VH and (175706) 1996 FG3. The  stochastic long-term dynamics in the last million years are modelled by sampling a large number of particles from their current orbit uncertainties. We model the evolution of these statistical distributions by a random walk in semi-major axis, eccentricity and inclination and a uniform distribution in longitude of perihelion. Then, we compute the probability that (35107) 1991 VH and (175706) 1996 FG3 could have been potentially disrupted by a close encounter in this period of a million years. 

This section presents the scope of the paper and is followed by the background of this work in section \ref{s:2background}. Next, section \ref{s:3Method} describes the methodology including a detailed study of flybys evaluation and the derivation of an analytical N-body secular problem solution. Section \ref{s:4longterm-prop} shows examples of the long-term propagation of asteroids and how the long-term dynamics can be characterized stochastically. Section \ref{s:5disc} discusses the limitations of the semi-analytical propagation tool. Section \ref{s:janustargs} studies the orbital history of the Janus targets and the frequency of close encounters. Last, section \ref{s:6conclusions} concludes by evaluating the aspects in which this methodology proves beneficial, questions that remained unanswered, and future work.

%========================================================
% BACKGROUND
%========================================================
%\newpage
\section{Background}\label{s:2background}

The long-term dynamics of NEOs are governed by their gravitational interactions with the other bodies of the Solar System. The effects of the most massive and external planets have timescales of millennia. However, planetary close encounters can abruptly change an orbit over a timescale of days. The accumulation of such planetary encounters cause the orbits of NEOs to be chaotic \citep{Tancredi1998}. This section describes this phenomenon in more detail. The evaluation of close encounters is necessary for the propagation of NEOs, hence the variety of possible flybys is demonstrated later for the validation of the method.

Many asteroids experience long periods of time without flybys. The dominant dynamics in those periods of time are the secular perturbations from massive planets in the Solar System. Likewise, the orbits of the planets evolve secularly over similar timescales. The Laplace-Lagrange secular theory qualitatively describes the evolution of the elements of the planets at any distant time in the future or past. As for the asteroid, the secular solution from external perturbers represents the orbital dynamics of asteroids between encounters.

\newpage
The presence of repeated encounters is one of the main characteristics of the long-term propagation of asteroids in the inner Solar System. Repeated close encounters cause a random walk in the elements of the asteroids. Very close encounters occur less frequently but change substantially the orbits of NEOs, modifying predictions on the long-term evolution of their orbits. Thus, we propose an informed analytical propagation of the orbits while characterizing planetary close encounters. The proposed methodology is born from the combination of these two dynamical regimes: the long-term effects of secular dynamics and the frequent changes in elements experienced in planetary encounters. Considering the secular drift of the asteroid we model the seasonal variation of the possible encounters with planets.

%\newpage
\subsection{Chaotic dynamics in the inner Solar System}

An accurate description of the evolution of orbits of near-Earth asteroids beyond a few centuries is challenging. This is because the succession of planetary encounters disperse neighboring trajectories to become chaotic \citep{Tancredi1998}. Small deviations in the orbital period change the timing of the flybys, spreading the uncertainty along the Line of Variation \citep{Milani2005}. After successive flybys the resulting imaginary stream of particles is spread in highly non-linear distributions. For this reason the study of long-term dynamics is often left to a statistical analysis requiring a large number of particles and computational efforts. In this context we propose the use of this semi-analytical tool to obtain long-term simulations in short computational times.

\begin{figure}[!h]
	\centering
		\includegraphics[width=5.8in]{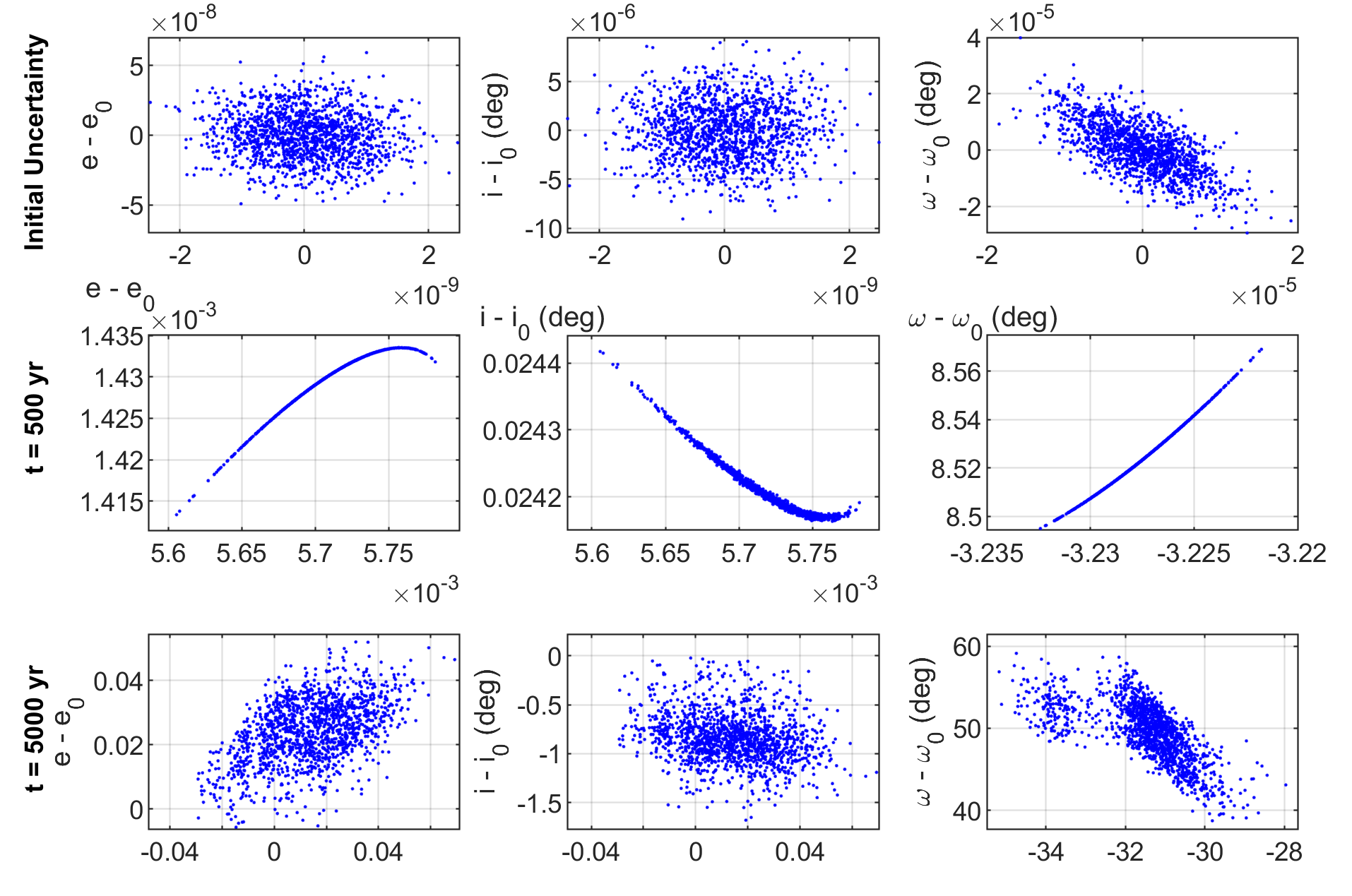}
		\includegraphics[width=5.8in]{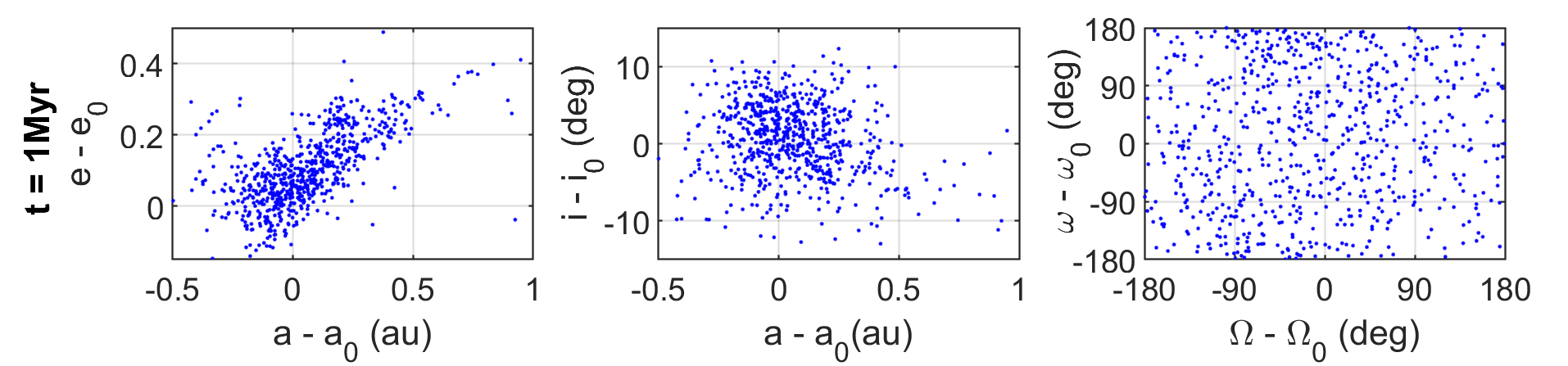}
	\caption{Chaotic dynamics of (35107) 1991 VH as obtained from numerical integration. Each axis represents the variation from the initial value of elements pairs: (left) semi-major axis-eccentricity, (center) semi-major axis-inclination, (right) argument of the node-argument of perihelion in degrees. The orbital evolution is shown at four instants of time: initial (first row), after 500 years (second row), after 5000 years (third row) and after 1 million years (bottom).}
	\label{fig:2.chaos5k}
\end{figure}

We exemplify the sensitivity to initial conditions in a numerical integration of asteroid (35107) 1991 VH, which is one of the two targets of mission Janus \citep{Scheeres2020Janus}, a NASA SIMPLEx mission. Figure \ref{fig:2.chaos5k} shows 1440 particles generated from the uncertainty in the orbit solution of (35107) 1991 VH, which is included in appendix \ref{app:uncerts}. These particles are propagated in the N-body integrator IAS15 \citep{Rein2014} including the Solar System planets from Venus to Neptune. The particles are propagated for a million years, although in this section we study in more detail the distributions after shorter periods of time.

%\newpage
After 500 years the initial normal distribution already becomes a stream of particles. While the variation in the elements from the nominal is similar for all the particles, there is a dispersion orders of magnitude smaller that represents the stream of particles. After 5000 years, the distribution becomes completely different: the presence of planetary encounters disperses the particles around the initial orbit. The variation in eccentricities and inclinations has a secular component. However, the variation on the argument of the node and argument of perihelion is dominantly secular after a few millennia. After a million years, the particles are spread along a large region of near-Earth space. In argument of perihelion and ascending node we observe that the distribution becomes almost uniform in the whole 2D angular space. 

The secular drift in the arguments defines the possibility of encounters over time. For this reason, it is important to characterize this drift and the secular cycles under the perturbation of the large bodies of the Solar System. When encounters are possible with the inner Solar System planets, these need to be accounted as perturbers of the orbit evolution. 

{The stochastic nature of the long-term dynamics of NEOs under close encounters implies that the precise determination of their position after hundreds of thousands of years is unachievable. However, we can still collect statistics that give us insight on their orbital history. Another implication is that the inclusion of higher order dynamics is shadowed under the stochastic dispersion caused by the main gravitational perturbations. For example, the magnitude of the Yarkovsky effect is typically $~10^{-4}$ au/Myr  \citep{Vokrouhlicky2000,Nesvorny2004}, which is still two orders of magnitude smaller than a typical dispersion after 10,000yr under repeated close encounters, as observed in the example of Figure \ref{fig:2.chaos5k}. In section \ref{s:stochs} we show that (35107) 1991 VH is not under a particularly high frequency of close encounters compared to other NEOs.}

{Similarly, relativistic effects can have a non-negligible effect in the secular rates of the argument of perihelion. These are usually measured in arcseconds per year or century, and typical values are 1-2 orders of magnitude smaller than the typical secular periods of the order of ~100,000 years \citep{Benitez2008}. Even if the secular rate has an error, the presence of encounters already causes the distributions to become uniform in argument of the node and perihelion after a few secular periods.}

\newpage
\subsection{NEO close encounters in the inner Solar System}
Flybys can occur with multiple planets over short periods of time. Even if the encounters are with the same planet, the closest approach distance and relative velocities can change depending on the timing of the flyby. The geometry of the flyby is constrained by the heliocentric elements of the asteroid. If shallow encounters are considered, the position in the asteroid orbit in which the planet is encountered can significantly change the relative velocity. These variations are not well captured by analytical theories, but the proposed propagation tool aims to accurately model these variations. These are different regimes of flybys in which the evaluation tool needs to be accurate.

\begin{figure}[!htb]
	\centering
		\includegraphics[width=5.5in]{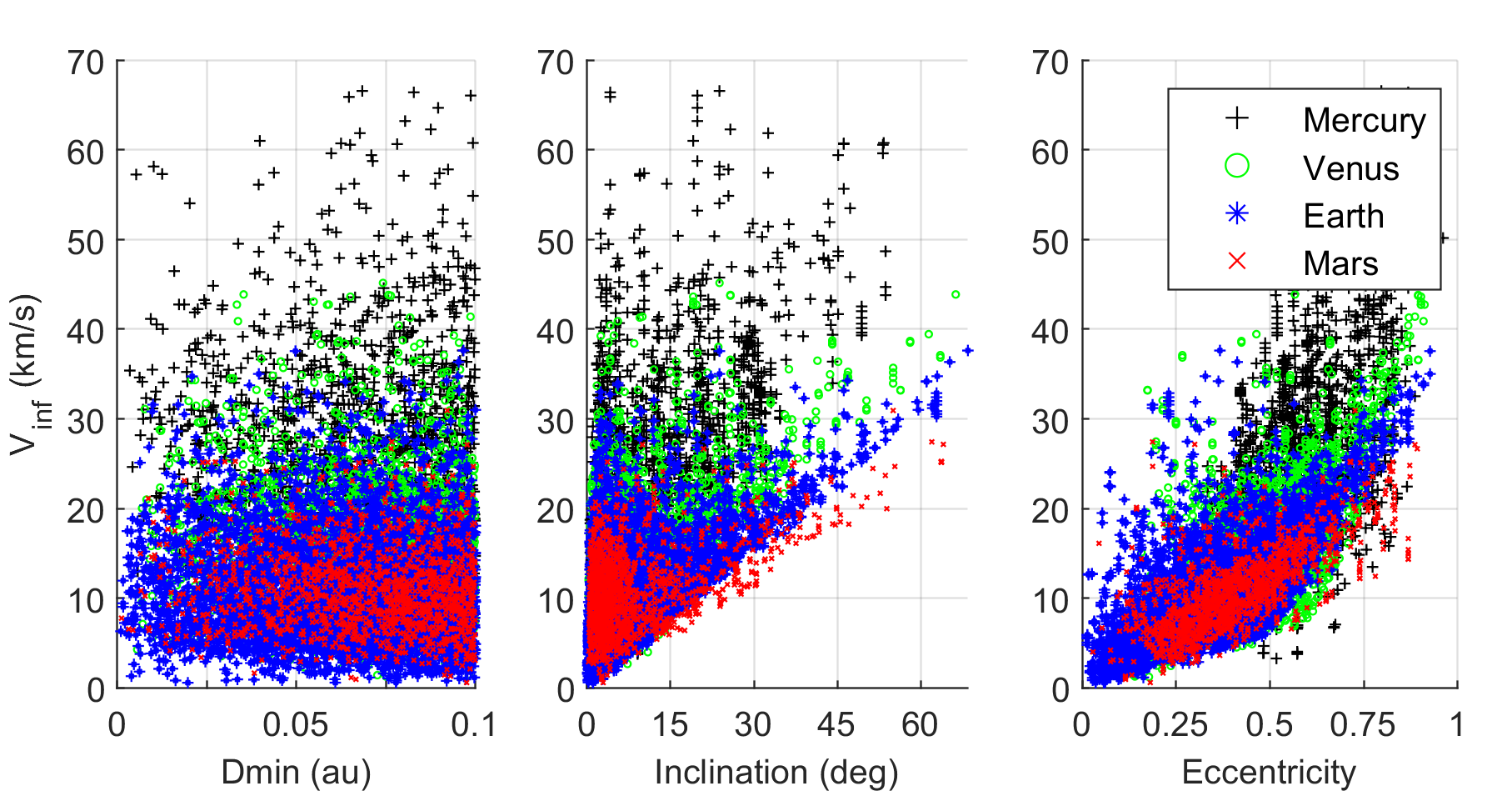}
	\caption{Relative velocity at closest approach of flybys generated from the propagation of NEOs with semi-major axis smaller than 2 au for 50 years. The symbol indicates the planet that the asteroid is encountering and the horizontal axis shows respectively closest approach distance (au), inclination (deg) and eccentricity.}
	\label{fig:3.CAD-dca-vinf}
\end{figure}

In order to broadly show the diversity in flybys that different NEOs experience, we generate a list of flybys that will be used to validate the evaluation of close encounters. From the database of NEOs we select the ones with semi-major axis smaller than 2 au \citep{JPLSolarSystemDynamics}. Then, we propagate their positions using the secular model for 50 years. For such a brief period of time the change in the elements is insignificant for our purposes. Figure \ref{fig:3.CAD-dca-vinf} shows more than 30,000 flybys generated with the described method. {Shallow encounters are much more frequent than the very close encounters that cause large variations in orbit elements. Thus, we want to consider them even if their individual contribution is not as significant.}

%\newpage
The range of possible relative velocities in figure \ref{fig:3.CAD-dca-vinf} depends on the planet in question, with increasing maximum relative velocity for the planet closest to the Sun. The relative velocity is defined by the heliocentric orbit of the asteroid, with an increasing range of possible values depending on the inclination and eccentricity of the orbit. Overall, after millions of years asteroids experience a variety of encounters that can be computed with different methods. With this purpose the list of generated flybys is used to decide the method to compute the post-encounter elements of flybys. In section \ref{s:3Method} we compute the error of different close encounter evaluation methods referenced to numerical integration of the trajectories.

%========================================================
% METHODOLOGY
%========================================================
\newpage
\section{Methodology} \label{s:3Method}

This semi-analytical propagation tool consists in the following process. First, the orbit of the asteroid is propagated by an analytical secular solution. This perturbed motion is interrupted when an encounter is found with a nearby planet. Then, the trajectory during the planetary encounter is modeled using a numerical method. Next, the secular propagation is continued until the subsequent encounter.

The simplest way to find encounters is to track the distance between the asteroid and planets at all times. While the regions in which encounters are possible are determined by the geometry of the asteroid around the central body, searching at all times is the most generic approach. The state of the planets is obtained from the secular solution of the 8 main planets interacting with each other. The state of the small body is corrected given the secular dynamics model. Once we determine the initial conditions of the encounter, the change in orbit elements is computed through the proposed numerical procedure.

There are many methods to compute planetary encounters available in the literature. Analytical solutions for Keplerian elements before and after close encounters in \"{O}pik's Theory \citep{opik1976interplanetary} were extended for multiple applications by \cite{Valsecchi2003,Valsecchi2015}. However, these analytical expressions are constrained to encounters that are very close and small bodies that are not co-moving with the planet. {Asteroid and planet are co-moving when they} have a small inclination and at least one of the node crossings close to the planet orbit. 
% OF-02/02: Mention explicitly Valsecchi I thought was fair, but maybe it's not the right style

We name shallow encounters those with large close approach distance but non-negligible effects. Shallow encounters are very frequent and influence the long-term evolution of small bodies in the Solar System. In order to account for shallow encounters, semi-analytical methodologies can be used to map before and after encounter conditions \citep{Alessi2015}. These methods are based on the quadrature of Lagrange Planetary Equations around the encounter. In this work we derive a quadrature of Lagrange Planetary Equations in Delaunay elements {which is solved using a numerical integration scheme. In the case of extremely close or slow encounters we solve the encounters numerically}.

%\newpage
Once the solution of most encounters is obtained satisfactorily we focus efforts in the computation of the perturbed motion of the asteroid in absence of encounters. We evaluate the solution of N-bodies interacting secularly to generate the orbits of the planets. Then, we obtain the perturbed motion of the asteroid including only the planets relevant to its secular influence. Taking into account the influence of only Jupiter is a valid generic approach to estimate the secular dynamics of NEOs \citep{Vokrouhlicky2012,Pokorny2013,Fuentes-Munoz}. In this work we use the Laplace-Lagrange secular model. The secular rates as obtained by the analytical theory are compared to numerical integration to validate the range of validity of the solution. {This defines a range of applicability of the tool, as we discuss later.}

{In this section we compare the individual pieces of the semi-analytical propagation tool to numerical methods. Last, we compare the combined semi-analytical propagation tool with trajectories obtained through numerical integration and evaluate the computational efficiency of the method.}

\subsection{Analytical secular dynamics of multibody systems}
\label{s:3c-propagation}

%{\color{red} Mention that works in absence of resonances}
The dynamical landscape of the Solar System is complex with gravitational interactions between all planets. This landscape leads to resonances and secular motion in asteroids in the system. Well inside the inner Solar System, the dynamics are dominantly secular. The secular solution of a planetary system formed by N-planets can be obtained analytically to first order in inclinations and eccentricities and in the absence of resonances. This section derives an implementation of the solution following the procedure in Chapter 7 of \cite{Murray2000}. The perturbing potential is written for the $N$ bodies considered. Then Lagrange Planetary Equations are used to compute the equations of motion of the elements of each particle, leading to a system of differential equations solved together.

%\subsection{3.b.i. Solar System secular model derivation}

The secular model is obtained as follows: (1) The perturbing potential is split in a direct part and an indirect part based on the dependency on fast angles, (2) then the perturbing potential is expanded in Keplerian Elements. (3) The important terms of the expansion are selected based on the averaging principle. (4) The terms are rewritten in semi-equinoctial elements to ease the solution of the global system of equations. (5) Take the necessary partials to solve the set of Lagrange Planetary Equations. The perturbing potential experienced by a mass $j$ by a second mass $k$ is:

\begin{equation}\label{eq:Pot_direct}
R_{jk} = \frac{G m_k}{a_{k}} \left( {R_{jk}}_D + {R_{jk}}_{I} \right)
\end{equation}

%\newpage
\noindent
Where $a_{k}$ is the semi-major axis of body the external body. The perturbing potential is separated in the direct  ${R_{jk}}_D$ and indirect ${R_{jk}}_{I}$ parts:

\begin{equation}
\begin{aligned}[c]
{R_{jk}}_D = \frac{a_k}{\left| {r}_j-{r}_k \right|}
\end{aligned}
\qquad\qquad
\begin{aligned}[c]
{R_{jk}}_I = - \frac{a_k^2}{a_j}  \frac{{r}_j\cdot {r}_k}{\left| {r}_k \right|^3}
\end{aligned}
\end{equation}

\noindent
The separation is convenient to expand in the ratio of semi-major axes $\alpha_{jk}$ as well as sines and cosines of $\left\lbrace \varpi_j,\Omega_j,\lambda_j,\varpi_k, \Omega_k,\lambda_k,\right\rbrace$. The ratio of semi-major axes is $\alpha_{jk} = a_j/a_k$ if the perturber is external, or $\alpha_{jk} = a_j/a_k$ if the perturber is internal. All the terms that depend on the longitudes $\left\lbrace \lambda,\lambda_j\right\rbrace$ are of short-period, so it can be argued that they do not contribute to the averaged potential $R_j$. The secular potential lowest order in eccentricities and inclinations is:

\begin{equation}
R_j = R_{0,j} + R_{1,j} = \sum^{N}_{k=1,k\neq j} G m_k \frac{1}{2a_k} b^{(0)}_{1/2}\left( \alpha_{jk} \right) + R_{1,j}
\end{equation}

\begin{equation} \label{eq:pot-exp}
R_{1,j} = n_ja^2_j \left [ \frac{1}{2}A_{jj}e^2_j + \frac{1}{2}B_{jj}I^2_j + \sum^{N}_{\substack{k=1 \\ k\neq j}}  A_{jk} e_je_k\cos{ \left ( \varpi_j-\varpi_k\right )}  + B_{jk} I_jI_k\cos{ \left ( \Omega_j-\Omega_k\right )} \right ]
\end{equation}

%\noindent
Where $\bar{\alpha}_{jk} = a_j/a_k$ if the perturber is external, or $\bar{\alpha}_{jk} = 1$ if the perturber is internal. The coefficients $A_{jj}, A_{jk}, B_{jj}, B_{jk}$ are:

\begin{equation}\label{eq:Ajk}
A_{jk} = - n_j \frac{1}{4} \frac{m_k}{m_c+m_j} \alpha_{jk} \bar{\alpha}_{jk} b^{(2)}_{3/2}\left( \alpha_{jk} \right)
\end{equation}
\begin{equation}
B_{jk} = + n_j \frac{1}{4} \frac{m_k}{m_c+m_j} \alpha_{jk} \bar{\alpha}_{jk} b^{(1)}_{3/2}\left( \alpha_{jk} \right)
\end{equation}
\begin{equation}
A_{jj} = + n_j \frac{1}{4} \sum^{N}_{k=1,k\neq j} \frac{m_k}{m_c+m_j} \alpha_{jk} \bar{\alpha}_{jk} b^{(1)}_{3/2}\left( \alpha_{jk} \right) = \sum^{N}_{k=1,k\neq j} B_{jk}
\end{equation}
\begin{equation}\label{eq:Bjj}
B_{jj} = - n_j \frac{1}{4} \sum^{N}_{k=1,k\neq j} \frac{m_k}{m_c+m_j} \alpha_{jk} \bar{\alpha}_{jk} b^{(1)}_{3/2}\left( \alpha_{jk} \right) = - \sum^{N}_{k=1,k\neq j} B_{jk}
\end{equation}

\noindent
where the coefficients $b^{(k)}_{s}$ are Laplace Coefficients. More details on their computation can be found in Appendix \ref{App:LaplaceCoeffs}. 
The coefficients $A_{jj}, A_{jk}, B_{jj}, B_{jk}$ form the matrices ${A}$ and ${B}$. We can rewrite the potential in semi-equinoctial elements,

\begin{equation}\label{eq:equinoc_def}
\begin{aligned}[c]
h_j&=e_j \sin \varpi_j\\
k_j&=e_j \cos \varpi_j\\
\end{aligned}
\qquad\qquad
\begin{aligned}[c]
p_j&=I_j \sin \Omega_j\\
q_j&=I_j \sin \Omega_j\\
\end{aligned}
\end{equation}

\noindent
the potential becomes:

\begin{equation}
R_{1,j} = n_ja^2_j \left [ \frac{1}{2} A_{jj}(h_j^2+k_j^2) + \frac{1}{2} B_{jj}(p_j^2+q_j^2) + \sum^{N}_{k=1,k\neq j}  A_{jk} (h_jh_k + k_jk_k)  + B_{jk} (p_jp_k + q_jq_k) \right ]
\end{equation}

%\subsection{3.b.ii. Solution of the semi-equinoctial elements}

Our complete set of states includes the mean anomaly at epoch $\sigma_j$ and $L_j=\sqrt{GMa_j}$. The equations of motion become:

\begin{equation}
\begin{aligned}[c]
\dot{p}_j&=\frac{1}{n_ja^2_j} \frac{\partial R_j}{\partial q_j}\\
\dot{q}_j&=-\frac{1}{n_ja^2_j} \frac{\partial R_j}{\partial p_j}\\
\end{aligned}
\qquad\qquad
\begin{aligned}[c]
\dot{h}_j&=\frac{1}{n_ja^2_j} \frac{\partial R_j}{\partial k_j}\\
\dot{k}_j&=-\frac{1}{n_ja^2_j} \frac{\partial R_j}{\partial h_j}\\
\end{aligned}
\qquad\qquad
\begin{aligned}[c]
\dot{L}_j&= \frac{\partial R_j}{\partial \sigma_j}\\
\dot{\sigma}_j&=- \frac{\partial R_j}{\partial L_j}\\
\end{aligned}
\end{equation}

%\subsubsection{Solution of N-Body systems}

The solution of $h_j,k_j,p_j,q_j$ only depends {on $R_{1,j}$. For this reason the perturbing potential} is often only expressed with those components. However, if we want the solution of the mean anomaly at epoch $\sigma_j$ it is necessary to take into account $R_{0,j}$. In the process of averaging the terms that would effect the semi-major axis are removed, meaning that under this assumption that element remains constant. The solution of $h_j,k_j,p_j,q_j$ is:

\begin{equation}\label{eq:sec-sol}
\begin{aligned}[c]
h_j(t)&=\sum^{N}_{i=1} e_{ji} \sin \left( g_it+\beta_i \right)\\
k_j(t)&=\sum^{N}_{i=1} e_{ji} \cos \left( g_it+\beta_i \right)\\
\end{aligned}
\qquad\qquad
\begin{aligned}[c]
p_j(t)&=\sum^{N}_{i=1} I_{ji} \sin \left( f_it+\gamma_i \right)\\
q_j(t)&=\sum^{N}_{i=1} I_{ji} \cos \left( f_it+\gamma_i \right)\\
\end{aligned}
\end{equation}

\noindent
where two sets of eigenvalue problems are solved for $e_{ji}, I_{ji}, f_i, g_i$.
The frequencies $g_i$ are the eigenvalues of ${A}$, and the frequencies $f_i$ are the eigenvalues of ${B}$. $e_{ji}$ and $I_{ji}$ are related to the eigenvectors of ${A}$ and ${B}$, but need to be solved with $\beta_i,\gamma_i$ given a set of initial conditions. In order to solve for $e_{ji}, I_{ji}, \beta_i,\gamma_i$ we proceed as follows. From the matrices of normalized eigenvectors $\bar{e}_{ji}, \bar{I}_{ji}$ and the initial conditions ${h},{k},{p},{q}$ we form:

\begin{equation}\label{eq:Nbodysec}
\begin{aligned}[c]
{h} &= \bar{e}_{ji} \left[ S_i \sin \beta_i \right] \\
{k} &= \bar{e}_{ji} \left[ S_i \cos \beta_i \right] \\
\end{aligned}
\qquad\qquad
\begin{aligned}[c]
{p} &= \bar{I}_{ji} \left[ T_i \sin \gamma_i \right] \\
{q} &= \bar{I}_{ji} \left[ T_i \cos \gamma_i \right] \\
\end{aligned}
\end{equation}

These are four linear systems of equations, where $S_i,T_i$ are the scaling factors of each eigenvector. Solving for the combined factors $\left[ S_i \sin \beta_i \right],\left[ S_i \cos \beta_i \right],\left[ T_i \sin \gamma_i \right]$ and $\left[ T_i \cos \gamma_i \right]$ we can reconstruct the vectors $e_{ji}, I_{ji}$ and the phase angles $\beta_i,\gamma_i$.

\begin{deluxetable*}{lcccccc}[b!]
\tablenum{1}
\tablecaption{Initial conditions of the Solar System propagation in Figure \ref{fig:4planets-equin}.}\label{t:secICs}
\tablewidth{0pt}
\tablehead{
\colhead{Planet} & \colhead{$a$ (au)} & \colhead{$e$} & \colhead{$i$ (deg)} & \colhead{$\Omega$ (deg)} & \colhead{$\omega$ (deg)} & \colhead{$M_0$ (deg)}  }
%\decimalcolnumbers
\startdata
Mercury & 0.39703           & 0.21337      &  6.936 & 48.264 & 31.991 & 52.745         \\ \hline
Venus   & 0.73096           & 0.012687     &  3.378 & 76.799 & 45.020 & 16.566        \\ \hline
Earth   & 1.0030            & 0.018402     &  0.001 & 154.979 & 296.322 & 8.654        \\ \hline
Mars    & 1.5177            & 0.093083     &  1.852 & 49.461 & 288.507 & 322.879          \\ \hline
Jupiter & 5.1904            & 0.047388     &  1.305 & 100.514 & 273.897 & 353.761         \\ \hline
Saturn  & 9.5499            & 0.05412      &  2.487 & 113.612 & 339.598 & 91.261         \\ \hline
Uranus  & 19.207            & 0.04628      &  0.772 & 73.997 & 96.864 & 189.506         \\ \hline
Neptune & 30.109            & 0.0091006    &  1.770 & 131.780 & 265.440 & 291.693          \\ \hline
\enddata
\tablecomments{From ephemeris DE431 at Epoch: JD$_0 = 2455562.5$ (2011 January 1) TDB}
\end{deluxetable*}

% \begin{table}[htb!]
% \centering
% \caption{Initial conditions of the propagation in figure \ref{fig:4planets-equin}. From ephemeris DE431 at Epoch: $JD_0 = 2455562.5$ (2011-Jan-01.0) TDB}\label{t:secICs}
% \begin{tabular}{l|l|l|l|l|l|l|}
% \cline{2-7}
%                               & {$a (au)$} & {$e$} & {$i$} & {$\Omega$} & {$\omega$} & {$M_0$} \\ \hline
% \multicolumn{1}{|l|}{Mercury} & 0.39703           & 0.21337      & 0.12105      & 0.84237           & 0.55834           & 0.92058        \\ \hline
% \multicolumn{1}{|l|}{Venus}   & 0.73096           & 0.012687     & 0.058964     & 1.3404            & 0.78574           & 0.28913        \\ \hline
% \multicolumn{1}{|l|}{Earth}   & 1.0030            & 0.018402     & 1.1989e-05   & 2.7049            & 5.1718            & 0.15104        \\ \hline
% \multicolumn{1}{|l|}{Mars}    & 1.5177            & 0.093083     & 0.032332     & 0.86325           & 5.0354            & 5.6353         \\ \hline
% \multicolumn{1}{|l|}{Jupiter} & 5.1904            & 0.047388     & 0.022771     & 1.7543            & 4.7804            & 6.1743         \\ \hline
% \multicolumn{1}{|l|}{Saturn}  & 9.5499            & 0.05412      & 0.043402     & 1.9829            & 5.9271            & 1.5928         \\ \hline
% \multicolumn{1}{|l|}{Uranus}  & 19.207            & 0.04628      & 0.013477     & 1.2915            & 1.6906            & 3.3075         \\ \hline
% \multicolumn{1}{|l|}{Neptune} & 30.109            & 0.0091006    & 0.030897     & 2.3000            & 4.6328            & 5.091          \\ \hline
% \end{tabular}
% \end{table}

\begin{figure}[!h]
	\centering
		\includegraphics[width=6in]{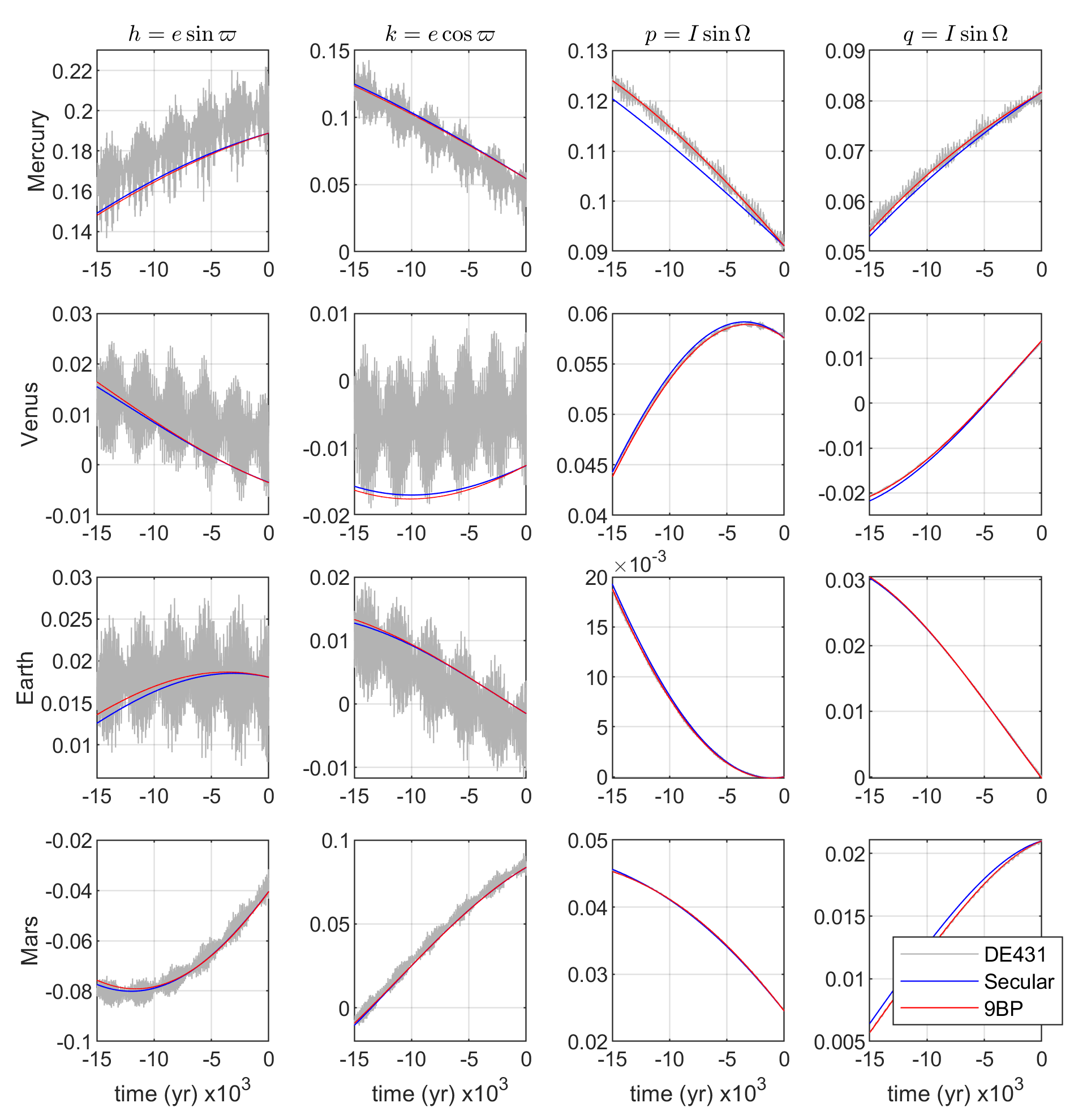}
	\caption{Semi-equinoctial elements of the inner Solar System planets obtained using three models. The analytical secular model derived is shown in blue, the integration of the 9BP of the main planets and the Sun in red and the ephemeris file DE431 is shown in grey. The initial conditions are obtained from the time average of the ephemeris file over the first orbital periods of the planets.}
	\label{fig:4planets-equin}
\end{figure}

Figure \ref{fig:4planets-equin} shows the solution of eq. \ref{eq:sec-sol} for Mercury, Venus, Earth and Mars as perturbed mutually and from the rest of planets of the Solar System. This model is compared to two other models for 15,000 years into the past. The first one is a numerical integration of the N-body problem taking into account the main 8 planets of the Solar System and the Sun. Then, we also compare to the planetary ephemerides DE431 \citep{Folkner2014}. While the complete ephemerides models show the short period effects, the secular component is modeled by the two simplified models.

{The initial conditions of the 9BP integration and the secular theory are obtained by averaging the full ephemeris model for two orbit periods. As a result, the initial conditions visually appear to be off from the mean of the full ephemeris solution, but they are the actual time average.} 

\newpage
The short-term components have a significant effect in the evolution of $h_j,k_j$. In the case of $p_j,q_j$, the secular component is the dominant effect of the evolution. {Because of the assumptions of small eccentricities and inclinations, the predicted frequencies are not perfectly accurate, as observed in the drift between the 9BP solution and the analytical theory of figure \ref{fig:4planets-equin}, especially in $p_j,q_j$.}

{A similar agreement in these elements is found for the gas giants. However, only the inner Solar System planets are shown as they are the bodies that are encountered by near-Earth objects. Thus, these are the planets for which we want to guarantee an accurate model of their secular dynamics. The analytical propagation of Mercury drifts the most from the full ephemeris solution, although it is the least relevant inner planet. Close encounters with Mercury are unfrequent and have a small effect, as Mercury is the least massive planet and it is encountered with very high relative velocity.}

As a result of the averaging of the perturbing potential, the semi-major axis of the bodies remains constant. The complete set of secular solutions includes the mean anomaly at epoch $\sigma_j$. Short term applications benefit from the improved characterization of the position of the bodies in their orbits. Solving for $\sigma_j$ is straightforward if we ignore the contribution of $R_{1,j}$, which has a small effect compared to $R_{0,j}$. The equation of motion for $\sigma_j$ becomes:

\begin{equation}
\dot{\sigma}_j=- \frac{\partial R_j}{\partial L_j} =- \frac{2}{n_{j} a_{j}} \frac{\partial R_{0,j}}{\partial a_j}
\end{equation}

%\newpage

\noindent
and the solution depends on whether the perturber is external or internal:

\begin{equation}
\dot{\sigma}_j = - \sum^{N}_{k=1,k\neq j} \frac{Gm_k}{n_{j} a^2_{j}} \bar{c}_{jk} D b^{(0)}_{1/2}
\end{equation}

\noindent
where $\bar{c}_{jk}=a_j/a^2_k$ in the case of an external perturber and $\bar{c}_{jk}=-1/a_j$ if the perturber is internal. The solution of the equation is simply a constant drift given by the rate $\dot{\sigma}$. This element completes the set of elements of the secular model. 

\subsection{Analytical secular dynamics of near-Earth asteroids}
\label{s:3c.secular1ast}

The secular dynamics of asteroids can be modelled as a particular case of the secular dynamics of multibody systems described above. In the present work we apply this solution to the evolution of the asteroid under the external perturbation of Jupiter. The solutions of equation \ref{eq:Nbodysec} simplify in the case of a system of 2 bodies with a massless internal body. We follow the same process to obtain the solution. Matrices $A_{jk}$ and $B_{jk}$ simplify to:

\begin{equation}
\begin{aligned}[c]
A_{jk} = \left(\begin{matrix}
B_{12} & A_{12}\\ 
0 & 0
\end{matrix}\right)
\end{aligned}
\qquad\qquad
\begin{aligned}[c]
B_{jk} = \left(\begin{matrix}
-B_{12} & B_{12}\\ 
0 & 0
\end{matrix}\right)
\end{aligned}
\end{equation}

\noindent
where the subindexes $1,2$ correspond respectively to the massless particle and the external perturber. The coefficients of the matrices are found as in equations \ref{eq:Ajk}-\ref{eq:Bjj} above. The solution to the eigenvalue problem yields the secular frequencies of the secular propagation $g_1=B_{12}$, $g_2=0$, $f_1=-B_{12}$, $f_2=0$. As expected, the elements of the perturber $h_2,k_2,p_2,k_2$ remain constant. The eigenvectors are the columns of the matrices:

\begin{equation}
\begin{aligned}[c]
\bar{e}_{jk} = \left(\begin{matrix}
1 & \kappa\\ 
0 & 1
\end{matrix}\right)
\end{aligned}
\qquad\qquad
\begin{aligned}[c]
\bar{i}_{jk} \left(\begin{matrix}
1 & \frac{\sqrt{2}}{2}\\ 
0 & \frac{\sqrt{2}}{2}
\end{matrix}\right)
\end{aligned}
\end{equation}

\noindent
where the constant $\kappa$ is found as the ratio between Laplace coefficients:

\begin{equation}
    \kappa = \frac{A_{12}}{-B_{12}} = \frac{b^{(2)}_{3/2}}{b^{(1)}_{3/2}}
\end{equation}

Note that the vector ($e_{12},e_{22}$) is not normalized. This is not necessary because in the process of obtaining the integration constants from the initial conditions the scaling of the eigenvectors is found. The solution of the elements of the massless particle becomes:

\begin{equation}  \label{eq:LLsol1}
\begin{aligned}[c]
h_1{(t)} &= S_1 \sin {(g_1 t + \beta_1)} + \kappa h_2 \\
k_1{(t)} &= S_1 \cos {(g_1 t + \beta_1)} + \kappa k_2 \\
\end{aligned}
\qquad\qquad
\begin{aligned}[c]
p_1{(t)} &= T_1 \sin {(f_1 t + \gamma_1)} + p_2 \\
q_1{(t)} &= T_1 \cos {(f_1 t + \gamma_1)} + q_2 \\
\end{aligned}
\end{equation}

\noindent
with constants of integration:

\begin{equation}  
\begin{aligned}[c]
S^2_1 &= e^2_{1,0} + \kappa^2 e^2_2 - 2 \kappa e_{1,0} e_2 \cos {(\varpi_{1,0} - \varpi_2)}  \\
T^2_1 &= i^2_{1,0} + i^2_2 - 2 i_{1,0} i_2 \cos {(\Omega_{1,0} - \Omega_2)}  \\
\end{aligned}
\qquad\qquad
\begin{aligned}[c]
\tan \beta_1  &= \frac{h_{1,0} - \kappa h_2}{k_{1,0} - \kappa k_2}  \\
\tan \gamma_1 &= \frac{p_{1,0} - p_2}{q_{1,0} - q_2}   \\
\end{aligned}
\end{equation}

\newpage
The time evolution of the Keplerian elements set can be obtained from the relationships with the semi-equinoctial set in equation \ref{eq:equinoc_def}. The solutions of $\varpi (t),\Omega (t)$ are the secular drift with frequencies $g_1,f_1$ that are equal with opposite signs. The solutions of $e(t),i(t)$ are oscillations with frequencies $g_1,f_1$ as obtained from the development of eccentricity $e_1 (t)=\sqrt{h^2_1 (t) + k^2_1 (t)}$ and inclination $i_1 (t) =\sqrt{p^2_1 (t) + q^2_1 (t)}$. The maximum and minimum values of eccentricity and inclination are:

\begin{equation}  
\begin{aligned}[c]
e^2_{1,min} &= S^2_1 + \kappa^2 e^2_2 - 2 S_1 \kappa e_2 \\
e^2_{1,max} &= S^2_1 + \kappa^2 e^2_2 + 2 S_1 \kappa e_2 \\
\end{aligned}
\qquad\qquad
\begin{aligned}[c]
i^2_{1,min} &= T^2_1 + i^2_2 - 2 T_1 i_2 \\
i^2_{1,max} &= T^2_1 + i^2_2 - 2 T_1 i_2 \\
\end{aligned}
\end{equation}

The secular model is computed for the fictitious asteroid of Case 1 of table \ref{t:cases} with the perturbation of Jupiter given by the elements of table \ref{t:secICs}. These cases are used later to demonstrate the propagation tool. For a nominal eccentricity of 0.15 the minimum eccentricity is 0.14946 and maximum is 0.17466. For a nominal inclination of 10 degrees, the minimum inclination is 7.41823 degrees and maximum is 10.02508 degrees. The characteristic period of the secular motion $T_{sec}$ is 154,116 years. 

This model assumes small eccentricities and inclinations. While these conditions are usually not fulfilled, it is important to remark that eccentricity and inclination are under frequent disturbance due to close encounters. Most importantly, the secular drift in $\Omega,\omega$ controls the evolution of the possible planetary encounters. 

%\subsection{Secular rates error}
%\newpage
The assumptions on the heliocentric orbit of the asteroid for the analytical secular perturbation solution are not always fulfilled among the NEO population. In this section we show that the analytical theory represents the dynamics of the perturbation by Jupiter. For this reason, we integrated the orbits of 4462 NEOs with $e<0.7$ and $i<0.5$ rad for 50,000 years. Note that in the solution of equation \ref{eq:LLsol1} if the terms of the external perturber are small the solution tends to a linear drift of the angles $\Omega, \varpi$. In addition, given the relationship between the frequencies $g$ and $f$, the relationship between the arguments rates is $\dot{\omega}=-2\dot{\Omega}$.

\begin{figure}[!h]
	\centering

		\includegraphics[width=5.5in]{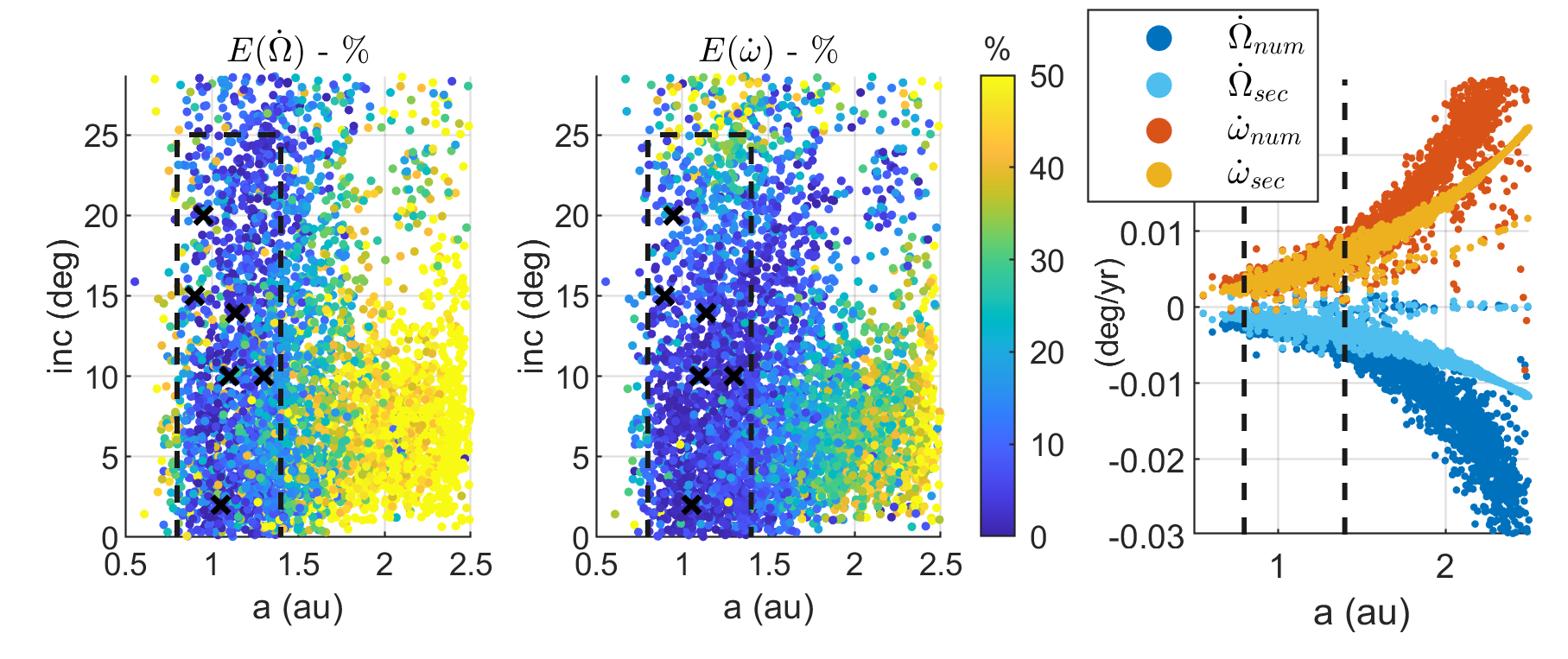}
	\caption{Error in the secular rates of near-Earth objects as obtained by linear regression of the propagation using Laplace-Lagrange secular theory and compared to numerical integration of the three-body problem in percent. Secular rates in ascending node (left) and argument of perihelion (center) as function of the initial conditions semi-major axis and inclination. {Secular rates in the angles as obtained with the two methods and function of the initial semi-major axis (right). The dashed lines indicate the region in which we compute the average errors. This region includes the initial conditions used throughout the paper, indicated as cross marks.}}
	\label{fig:3.errors-rates}
\end{figure}

Figure \ref{fig:3.errors-rates} shows the secular rates computed from linear regression of the time histories of $\Omega$,$\omega$. Note that from an initially larger list of NEOs, a significant fraction (12\%) was discarded because either eccentricity or inclination were larger than 0.5. An additional 11\% of the solutions were discarded because the error in the regression was too large or during the propagation close encounters with Jupiter moved the orbit of the NEO to a completely different location than the initial conditions. While the linear regression secular rates are not equivalent to the frequencies of the analytical solution, they serve as a comparison between the two dynamical models. The error is computed in percent relative to the rate measured from the regression of the numerically integrated trajectories, given by:

\begin{equation}
E(\dot{\omega}) = 100 \left| \frac{\dot{\omega}_{sec} - \dot{\omega}_{3BP}}{\dot{\omega}_{3BP}} \right|
\end{equation}

As observed in Figure \ref{fig:3.errors-rates} the rates obtained with the two methods agree towards the smaller end of semi-major axis. Since these are near-Earth objects, the condition of being in the vicinity of Earth means that eccentricity increases with semi-major axis. We can see that past 1.5-2 au the difference between the two models is increased, as well as the secular rates values themselves also increase. {This difference is also appreciated in the rates as function of semi-major axis, in which we show the agreement in the NEO region. Using the current model we find secular rates for 60\% of the population with an error less than 30\% in both $\dot{\omega}$ and $\dot{\varpi}$. If we limit the application of the secular theory to semi-major axes between 0.8-1.4 au (dashed region in Figure \ref{fig:3.errors-rates}) we find that this agreement improves to 88\%. It is important to note that the examples chosen to demonstrate the semi-analytical propagation tool fall within this region. Outside of this region we can verify if the secular rates found are reliable by using numerical integration. This test integration must be long enough to observe the secular rates, but still orders of magnitude shorter than the time-scales that we can more efficiently study using the semi-analytical propagation.}
%This difference is also appreciated in the correlation plot of the right, in which the correlation between the two rates persists but further from the predicted relationship of $\dot{\omega}=-2\dot{\Omega}$. 

At larger semi-major axis the effect of mean-motion resonances becomes important, and that Lidov-Kozai dynamics may better represent the dynamics for large eccentricities and inclinations \citep{Michel1996,Morbidelli2009}. {The implementation of additional analytical long-term dynamics models to model any generic asteroid is left as future work}.

%\newpage
\subsection{Finding the subsequent encounter} 
The analytical propagation of particles is interrupted when an encounter with a planet is detected. In principle it is not necessary to track the distance to planets at all times, since the regions in which encounters are possible are determined by the geometry of the asteroid around the central body. If the inclination relative to the planet is high, then the encounters are only possible in the vicinity of the ascending and descending node. However, the most generic approach is to track the distance between the asteroid and the crossing planets at all times. Thus, the results in this work follow the latter approach to find encounters within a closest approach distance of 0.1 au. When the two bodies are close, the unperturbed closest approach distance is found using a bisection method where the function is the derivative of the distance as obtained by finite differences. This process results in less evaluations of the relative distance function based on the heliocentric elements of the bodies.

The elements of the planets and the asteroid are propagated using the secular solution at the date of start of the encounter, which is defined below. The transition between models consists in the conversion between the sets of elements, obtaining the necessary Keplerian elements in the process. These are semi-equinoctial elements for the analytical perturbed propagation as in equation \ref{eq:equinoc_def} and Delaunay elements for the quadrature of the Lagrange planetary equations.% as in equation \ref{eq:delaunayelems}.

%\newpage
\subsection{Evaluation of planetary encounters}\label{s:III-Elems-var}

Close encounters are commonly solved using the analytical \"{O}pik theory  \citep{opik1976interplanetary}. While this theory requires the least computational resources, its accuracy is limited to specific circumstances. The quadrature of Lagrange Planetary Equations can be used to solve close encounters \citep{Alessi2015}. In this work we derive a solution using this method for generic close encounters using Delaunay elements. The two methods are compared to the integration of the three body problem from the same date and during the same period of time. 

%From the methods available in the literature to evaluate planetary encounters we selected two. 
%These are the analytical \"{O}pik theory and the proposed quadrature of Lagrange Planetary Equations. 

%\newpage
\subsubsection{\"{O}pik theory of close encounters}

An analytical solution to the planetary close encounter problem was derived by \"{O}pik \citep{opik1976interplanetary}. This solution was extended and studied in detail by \cite{Valsecchi2003,Valsecchi2015}. The encounter solution uses a linearized mapping from orbital elements to a planetocentric Cartesian frame, that is later expressed in B-plane coordinates. Then, the encounter is assumed instantaneous and the incoming asymptote and B-plane both rotate. The new B-plane coordinates are mapped back to the orbit elements space.

The analytical solution is derived for a hyperbolic flyby around a point secondary mass. This mapping between B-plane coordinates and orbit elements is linearized in the impact parameter. Thus, the encounter must be close for the method to be reliable. Additionally, if the inclination is small the relative velocity coordinates become undefined. A possible way to avoid this is by using a method sometimes referred as pseudo-\"{O}pik \citep{Greenberg1988}. In this case the relative velocity vector is computed directly and defines the turn angle $\gamma$ at the time of closest approach:

\begin{equation}
    \tan{\frac{\gamma}{2}} = \frac{m}{bU^2}
\end{equation}

\noindent
{where $m$ is the mass of the planet in units of the mass of the Sun, $b$ is the impact parameter and $U^2$ is the relative velocity in units of the circular velocity of the planet. Here we use the unperturbed trajectory of the planet and asteroid to find these quantities. That is, the impact parameter and relative velocity are found as the planetocentric distance and velocity at closest approach.}

%======== Description Imported from section II.Errors
% The algorithm to describe a flyby consists in mapping from orbit elements to pre-encounter Cartesian coordinates, then to pre-encounter \"{O}pik variables. Post-encounter \"{O}pik variables are then mapped back to orbit elements. The set of \"{O}pik variables are the B-plane coordinates \{$\zeta,\xi$\}, relative velocity vector in modulus and angles \{$U,\theta,\phi$\} and time of B-plane piercing $t_b$. The coordinate $\zeta$ is closely related to the timing of the flyby, and $\xi$ is related to the minimum orbit intersection distance (MOID). 

%In this work the mapping from encounter variables to orbit elements is used to generate the initial conditions of the flybys . 
%===========================

\begin{figure}[t]
	\centering
		\includegraphics[width=5.5in]{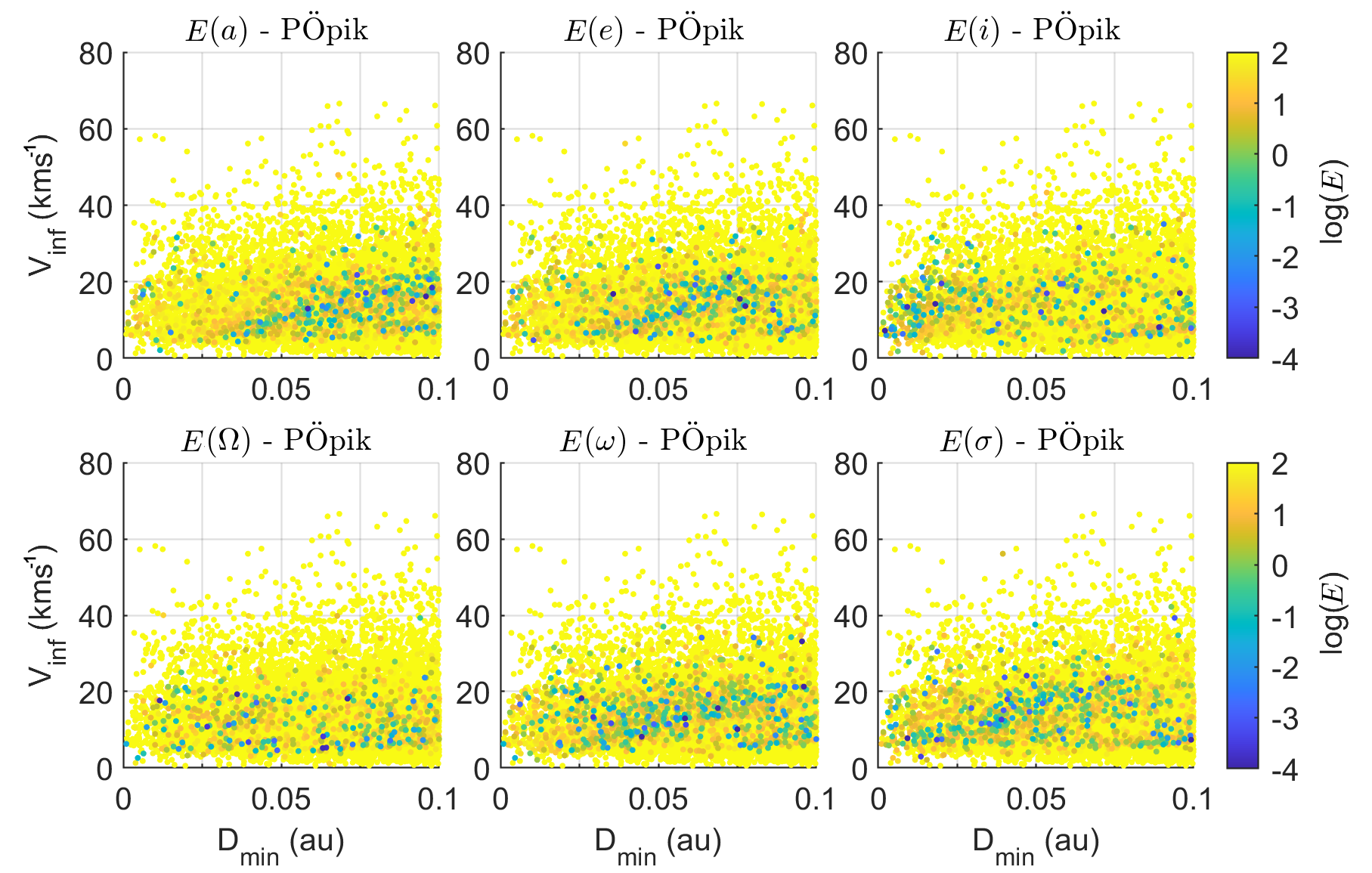}		
		\includegraphics[width=5.5in]{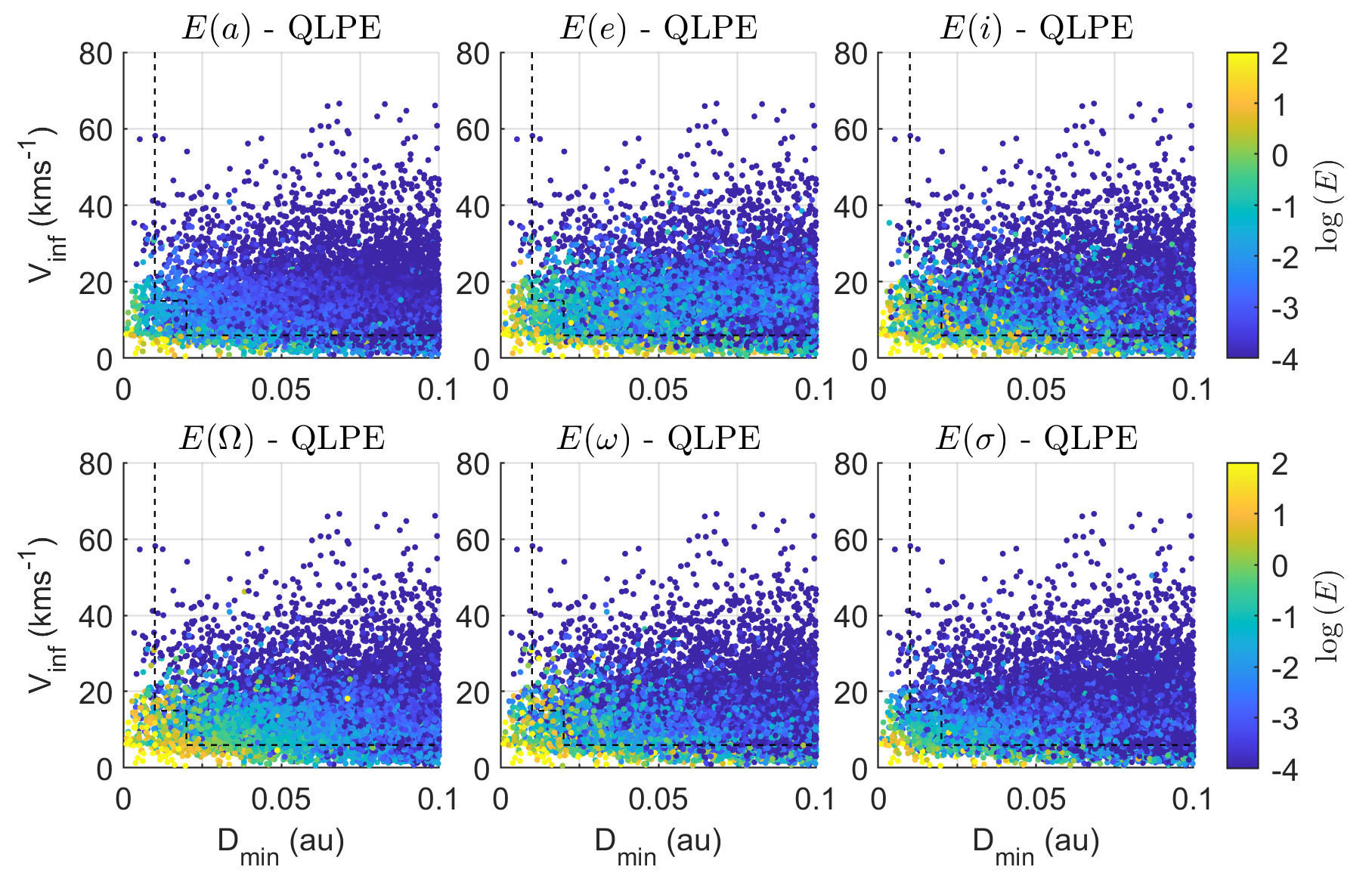}
	\caption{Logarithm of the errors in the computation of the final Keplerian elements given by pseudo-\"{O}pik theory (P\"{O}pik - first and second rows) and the quadrature of LPE (QLPE - third and fourth rows). The list of flybys used is shown in figure \ref{fig:3.CAD-dca-vinf} and generated as described in the text. The flybys are represented in the plane of relative velocity at closest approach $V_{inf}$ (km s$^{-1}$) and distance of closest approach (au). {The dashed area in the QLPE error plots separates the region in which the encounters are computed using an integration of the 3BP during the semi-analytical propagation.}}
	\label{fig:3.errors-CAD-dca-vinf}
\end{figure}

%\newpage
\subsubsection{Lagrange Planetary Equations}

The proposed computation of the encounter effect is computed as follows. The variation in elements over the encounter event is obtained from a quadrature of the Lagrange Planetary Equations assuming the geometry of the unperturbed flyby. The elements used are obtained from the secular propagation of the asteroid. The Lagrange Planetary Equations describe the evolution of orbit elements due to a perturbing potential. The derivation of Lagrange Planetary Equations can be found in some references for different sets of orbit elements \citep{Brouwer1961,Roy2004}. In general, they have the form of:

\begin{equation}\label{eq:lpeD}
    \bm{\dot{D}} = \left[L (\bm{D})\right] \frac{\partial R}{\partial \bm{D}}(\bm{D},t) 
\end{equation}

\noindent
where $\bm{D}$ is the set of elements of choice and $L(\bm{D})$ is a function of the elements that depends on the chosen elements. In this case the perturbing potential $R$ is the gravitational potential of the encountered planet. Note that without further simplifying assumptions the partials of the perturbing potential are function of the elements and function of time. {For the elements of choice we take the partial derivatives that relate the set of elements to Keplerian elements $\bm{K} = [a,e,i,\Omega,\omega,\sigma]$ and Cartesian coordinates. From the orbital elements representations available to choose, the current implementation uses the Delaunay elements:}

\begin{equation}
\begin{aligned}[c]
L&=\sqrt{\mu a}\\
G&=L\sqrt{1-e^2}\\
H&=G \cos{i}
\end{aligned}
\qquad\qquad
\begin{aligned}[c]
l&=\sigma\\
g&=\omega\\
h&=\Omega
\end{aligned}
\end{equation}

{The Lagrange Planetary Equations with the perturbing potential of Equation \ref{eq:Pot_direct} with $j$ being the asteroid and $k$ the encountered planet can be written as:}

\begin{equation}
\begin{aligned}[c]
\frac{\mathrm{d} L}{\mathrm{d} t}&=\frac{\partial R}{\partial {r}} 
\frac{\partial {r}}{\partial \bm{K}} 
\frac{\partial \bm{K}}{\partial l}\\
\frac{\mathrm{d} G}{\mathrm{d} t}&=\frac{\partial R}{\partial {r}} 
\frac{\partial {r}}{\partial \bm{K}} 
\frac{\partial \bm{K}}{\partial g}\\
\frac{\mathrm{d} H}{\mathrm{d} t}&=\frac{\partial R}{\partial {r}} 
\frac{\partial {r}}{\partial \bm{K}} 
\frac{\partial \bm{K}}{\partial h}
\end{aligned}
\qquad\qquad
\begin{aligned}[c]
\frac{\mathrm{d} l}{\mathrm{d} t}&=-\frac{\partial R}{\partial {r}} 
\frac{\partial {r}}{\partial \bm{K}} 
\frac{\partial \bm{K}}{\partial L}\\
\frac{\mathrm{d} g}{\mathrm{d} t}&=-\frac{\partial R}{\partial {r}} 
\frac{\partial {r}}{\partial \bm{K}} 
\frac{\partial \bm{K}}{\partial G}\\
\frac{\mathrm{d} h}{\mathrm{d} t}&=-\frac{\partial R}{\partial {r}} 
\frac{\partial {r}}{\partial \bm{K}} 
\frac{\partial \bm{K}}{\partial H}
\end{aligned}
\end{equation}

The proposed solution is the integration of these differential equations around the encounter date $t_{e}$ and assuming the unperturbed geometry of the flyby. Hence, the asteroid coordinates are obtained from the heliocentric elements secularly propagated until the start of integration date $D_0$ and the quadrature is only a function of time: 

\begin{equation}
\Delta{\bm{D}}=\int_{t_{e}-\delta t}^{t_{e}+\delta t}\bm{\dot{D}}(\bm{D}_0,t)dt
\end{equation}

\newpage
The integration is conducted for a fraction of the orbit period around the closest approach distance. This fraction is a constant set large enough such that the effect of the encounter is captured completely. In this work we use a self-coded fast quadrature function based on the midpoint rule and a total integration time of 20\% of the orbital period. This method avoids the frame transformation to the center of the planet, since it considers the planet as an external perturber of the asteroid motion around the Sun. For this reason, it is not possible to obtain a closed form solution of the integral. Nonetheless, this approach does not imply further assumptions that limit its range of applicability. Future work will be done in finding the optimal set for this application. This approach is accurate for the vast majority of encounters, but it is less accurate for the closest ones, as we explore in the following section.

%\subsubsection{Error in the evaluation of planetary encounters}
%\label{s:flybyvalidation} 

Using the list of flybys generated in figure \ref{fig:3.CAD-dca-vinf} we computed the errors of \"{O}pik theory and the quadrature of LPE compared to the solution of the encounter using the three-body problem. The  error $E(K)$ is relative to the variation and in percent, given by:

\begin{equation}
E(K) = 100 \frac{\Delta K_{QLPE} - \Delta K_{3BP}}{\Delta K_{3BP}}
\end{equation}

The results of this evaluation are described in figure \ref{fig:3.errors-CAD-dca-vinf}. Using pseudo-\"{O}pik theory there is a region in the space of relative velocity and closest approach distance in which flybys can be computed accurately. However, this region is not constant for all Keplerian elements. In addition, most flybys are not computed correctly using this method. Slow flybys break the assumption in \"{O}pik theory that the behavior during the flyby is modeled by the two-body hyperbolic interaction. Many of the faster flybys occur with Venus and Mercury. The two-body hyperbolic flyby model fails to characterize the effect of flybys with these less massive planets even if they are faster.

%\newpage
Using the quadrature of the Lagrange Planetary Equations 99\% of the flybys list are computed with less than 3\% of error, and more than 88\% with less than 0.1\% error. The flybys that are not computed accurately with this method are very close and with a slow relative velocity. These flybys can break the assumption of the unperturbed geometry of the flyby. Given that these encounters also cause significant variations in the elements, these infrequent encounters are solved using a three-body problem integration in Cartesian coordinates. The criteria to solve these encounters using the alternative method is by defining three threshold regions in the encounter parameters: with very small $V_\infty$, very small closest approach distance, and a combination of both close to zero. This process simplifies the detection of collisions with the planets during the numerical integration in Cartesian space in the heliocentric frame.  

\subsection{{Semi-analytical propagation vs. numerical integration}}

{In the previous sections we validate the individual pieces of the semi-analytical propagation tool. Once combined, we want to compare the resulting trajectories to trajectories obtained using numerical integration. With this purpose we generate a fictitious NEO population and propagate their orbits using both methods.}

{The fictitious NEO population we define consists in normal distributions for the perihelion distance, eccentricity, inclination. The distributions are centered respectively around $0.8$au, $0.2, 10\deg$ and with standard deviations of $0.05$au, $0.05, 3\deg$. Arguments of the node, perihelion and initial mean anomalies are defined as uniform distributions in the $[0,360]$ degrees range. We sample 1000 particles from these distributions as our test set for the comparison between the two methodologies. We setup the numerical integration of the asteroid orbits considering the planets as third body perturbers. The model we use for the orbits of the planets is the secular theory developed in section \ref{s:3c-propagation}.}

\begin{figure}[!h]
	\centering
		\includegraphics[width=5.8in]{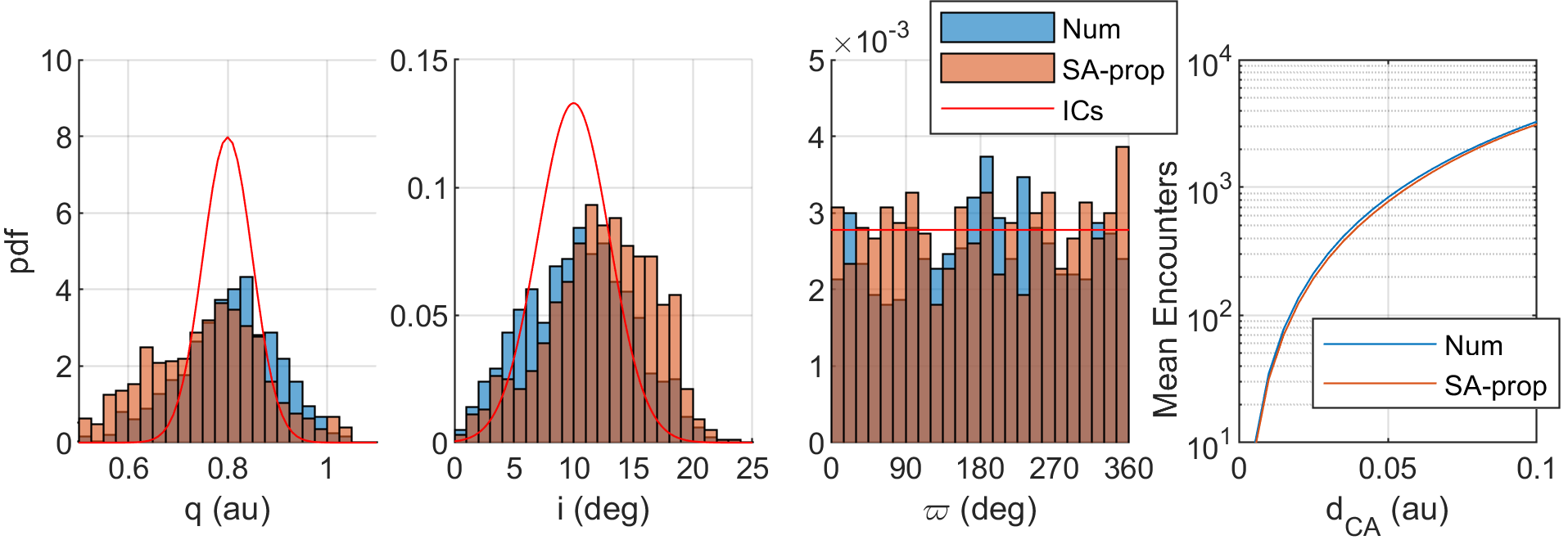}
	\caption{{Comparison between the 1000 trajectories obtained using numerical integration (blue) and semi-analytical propagation (orange) of a fictitious population of NEOs after 200,000 years. The probability density function is shown for the perihelion distance, inclination and argument of perihelion. The analytical probability density function of the initial conditions is shown as a continuous red line. On the right, the mean number of encounters found as function of the closest approach distance for both methods.}}
	\label{fig:num-vs-saprop}
\end{figure}

{Figure \ref{fig:num-vs-saprop} shows the results of the propagation using both methods as well as the distributions that defined the initial conditions. In addition, we find that the cumulative distribution of the mean number of encounters versus closest approach distance is matched very closely. After 200,000 years, the presence of planetary encounters causes a dispersion of the initial distributions. The distribution obtained through semi-analytical propagation is able to track this drift.}

{The main difference between the results using the two methods is that the numerically integrated distribution shows a small drift in the center of the distribution, but very similar dispersions. In terms of the longitude of the perihelion, the resulting distributions after 200,000 years remain uniform using either of the two propagation methods. The other significant difference between the two methods is in the required the computational time, which is discussed next.}

%{We compare the results of the numerical integration of (35107) 1991 VH shown in section \ref{s:2background} with the results of the semi-analytical propagation for 500,000 years. Figure \ref{fig:num-vs-saprop} shows the distributions in elements using both methods. In semi-major axis, eccentricity and inclination, the final distributions of the trajectories obtained with numerical integration are more dispersed. This is mainly caused by the presence of short-period perturbations.}

%{After 500,000 years the arguments of perihelion have become uniformly distributed. This result is found using both methods. In the case of the arguments of the node, this distribution is not yet completely uniform but it already spreads along a significant portion of the space.}

%In section \ref{s:2background} we show how an initially very close distribution becomes stochastic. 

\newpage
\subsection{Computational time of the semi-analytical propagation tool} \label{s:6speed}

%\remove[OFM]{Besides the insight on the dynamics that the secular model can provide,}
The semi-analytical propagation of near-Earth asteroids reduces the computational time required to obtain long-term trajectories. The use of numerical techniques is limited almost exclusively to the computation of planetary encounters, that represent a fraction of the simulation time.

In order to estimate the speedup of the propagation we generate a fictitious population of NEOs and propagate them for 100,000 years with the semi-analytical propagation tool and with numerical integration. The semi-analytical propagation tool uses the secular solution of the Solar System to compute the orbits of the planets over long periods of time. Then, planetary encounters can occur with any planet. The secular propagation of the asteroid is derived as described in the text accounting only for Jupiter, which accurately represents the asteroid motion in the absence of resonances. The semi-analytical propagation of 100,000 years with this setup is computed in around 5 seconds. In order to account for these effects in numerical integrations we use the N-Body problem integrator IAS15 \citep{Rein2014} with all the planets and the asteroid.

The simulation is setup using high-level programming code that runs libraries in more efficient low-level code. This is the case for both numerical integrations and semi-analytical propagation, running in the same 2.5GHz Intel Core i7 processor. The result is a speed-up of x500-x1000 of the semi-analytical propagation tool as compared to the numerical integration. The current implementation allows room for significant speed-up that is left for future work. The perturbed long-term propagation could be extended to use other suitable models of interest. The computational cost is not expected to increase while we use analytical solutions of these long-term perturbations.

%========================================================
% PROPAGATION RESULTS
%========================================================
\newpage
\section{Semi-analytical propagation results} \label{s:4longterm-prop}

In this section we demonstrate the semi-analytical propagation tool in a variety of scenarios. First, we want to compare the semi-analytical model with trajectories obtained through numerical integration. Matching very accurately trajectories obtained with more complex models is outside the scope of the comparison. Even if the models were identical, trajectories under encounters are very sensitive to the initial conditions and under small perturbations they diverge into different paths. This effect was visualized in figure \ref{fig:2.chaos5k} using only numerical integration. For this reason, long term simulations may focus on the statistical analysis of the dynamical evolution rather than individual trajectories. Throughout the section, the simulations include Jupiter as the only planet secularly perturbing the asteroids. All the inner Solar System bodies are considered to evaluate planetary encounters. These are secularly evolving due to mutual perturbations and the perturbations of the outer Solar System planets, as described in section 3.a..

\subsection{Short-term propagation of near-Earth objects using different models}

\begin{figure}[!h]
	\centering
		\includegraphics[width=5.5in]{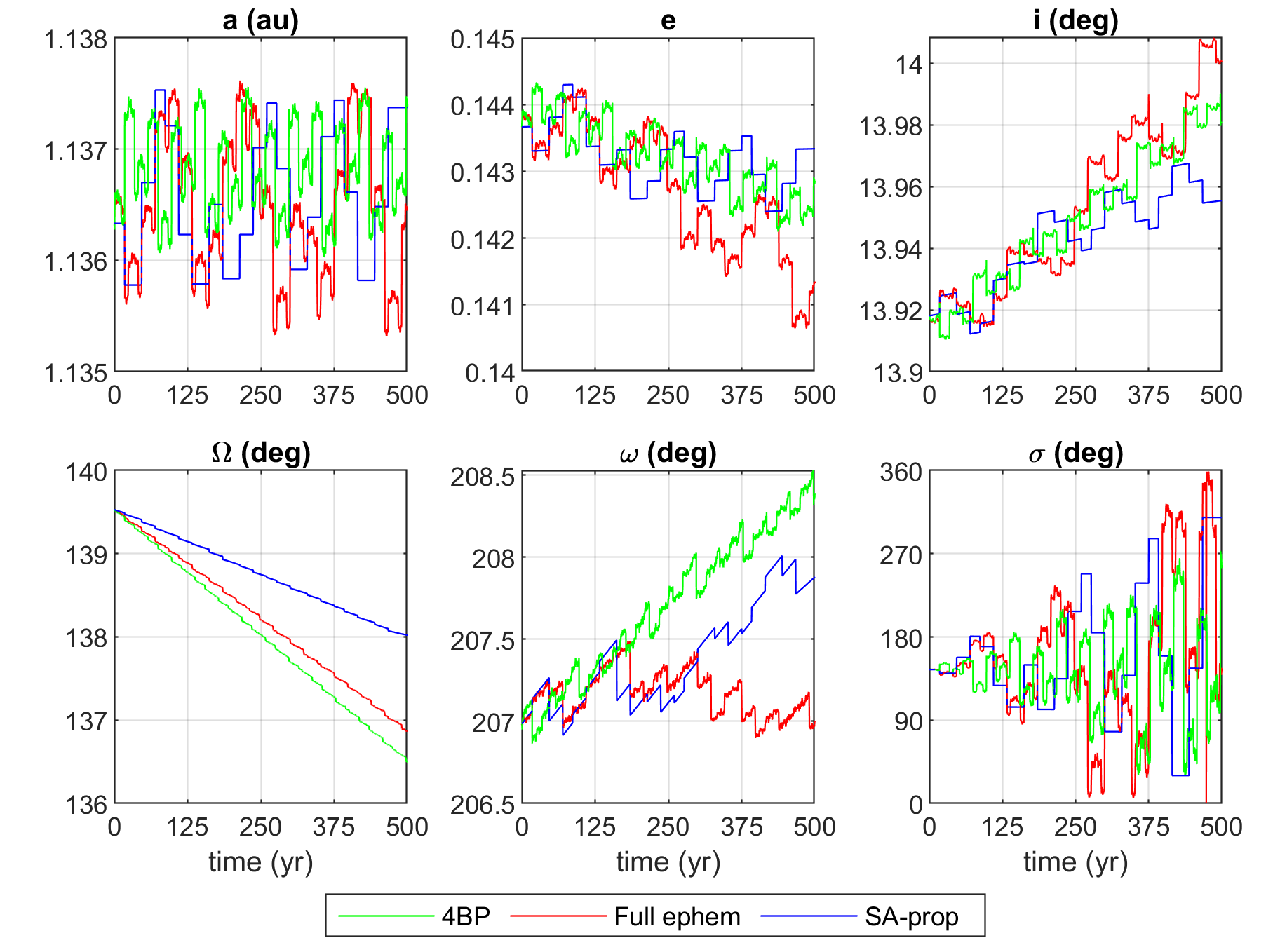}
            \caption{Trajectories of a particle that mirrors the binary asteroid (35107) 1991 VH for 500 years. Comparison of three methods. First, numerical integration of the heliocentric orbit of the asteroid with two external perturbers in constant sets of elements: Earth and Jupiter (4BP, in red).  Full Ephemeris as accessed in HORIZONS \citep{Giorgini} (green), and the semi-analytical propagation method (SA-prop, blue).}
        \label{fig:1e2yrs-oe}
\end{figure}

The asteroid chosen for the tool demonstrations is the binary (35107) 1991 VH. The first reference trajectory is obtained from the HORIZONS system of JPL \citep{Giorgini}. The second model is the numerical integration of the asteroid motion under the influence of the Sun, Earth and Jupiter. Jupiter is the main driver of the secular motion, which is observed as a linear drift in the argument of perihelion and argument of the ascending node. Given the current orbit of (35107) 1991 VH, it only experiences planetary encounters with Earth in the next few centuries. In Figure \ref{fig:1e2yrs-oe}, we compare these two models from numerical integration with the present semi-analytical propagation tool. 

In the three trajectories we observe similar behavior, although it manifests differently in every element. First, there is a close agreement in the encounter dates of the Sun-Earth-Jupiter integration and the encounters found by the propagation tool that shows in all elements. {The encounter dates can be distinguished as the discontinuities in the trajectories, especially in the semi-analytical propagation trajectory}. The variation of the semi-major axis is characterized by a random walk  from the planetary encounters. In both eccentricity and inclination there is a relevant role of the encounters with an additional secular component that the secular model is able to model.

The dynamics of the argument of the ascending node are dominated by the secular drift. There is not a significant effect that can be perceived by the planetary encounters and this is expected when the encounters occur with a unique planet and close to the node. What we observe is that the secular dynamics including the complete effects of  Earth and Jupiter are very similar, and the analytical secular drift is off by about a degree after the 500 years of the propagation.  The secular drift rate in the argument of perihelion is not as trivial to compare since it must be observed between encounters, although good agreement is found too. Last, the mean anomaly at epoch evolves over time with an increasing amplitude present in all three models.

\begin{figure}[!h]
	\centering
		\includegraphics[width=5.5in]{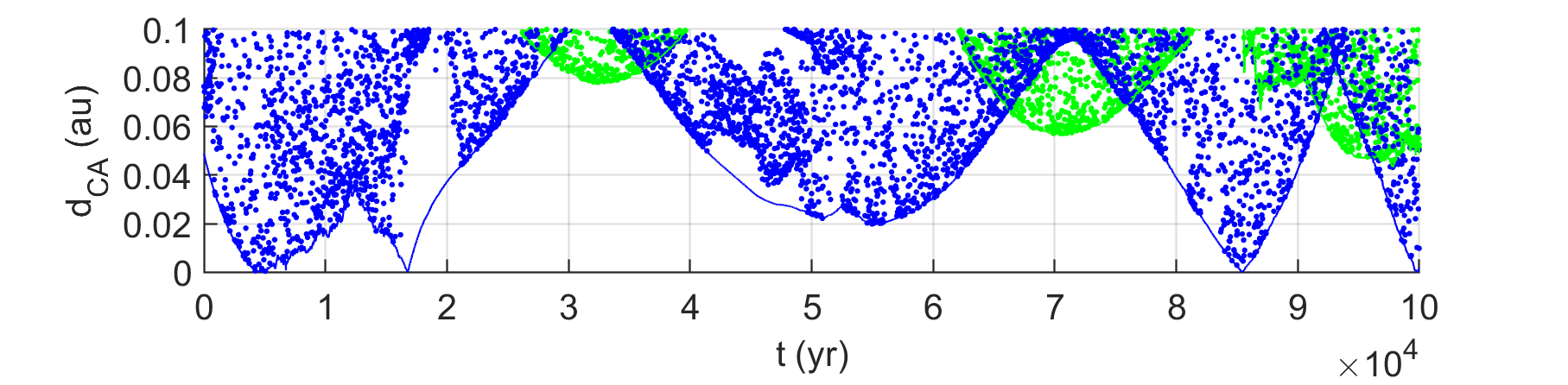}
		\includegraphics[width=5.5in]{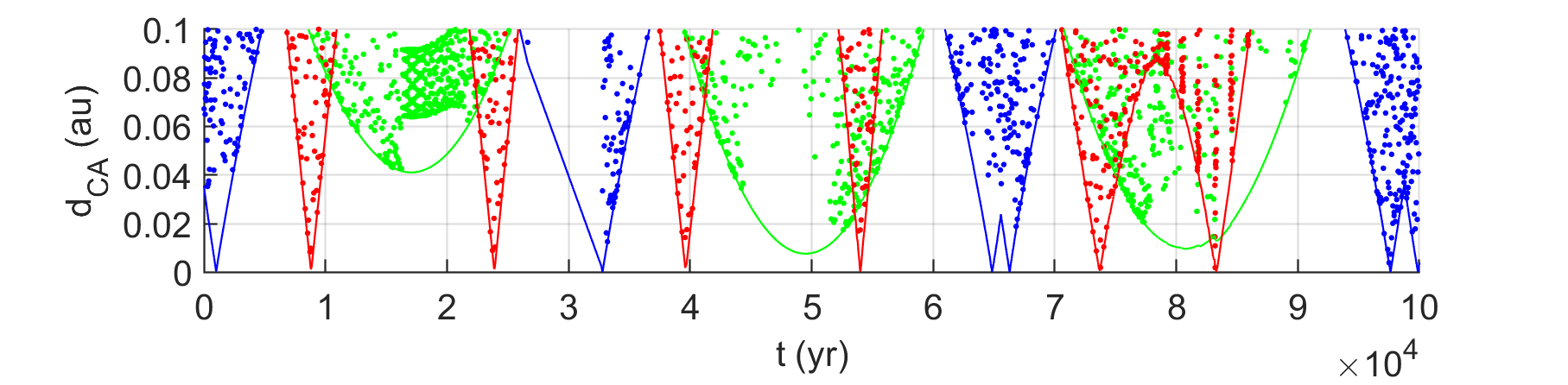}
		\includegraphics[width=5.5in]{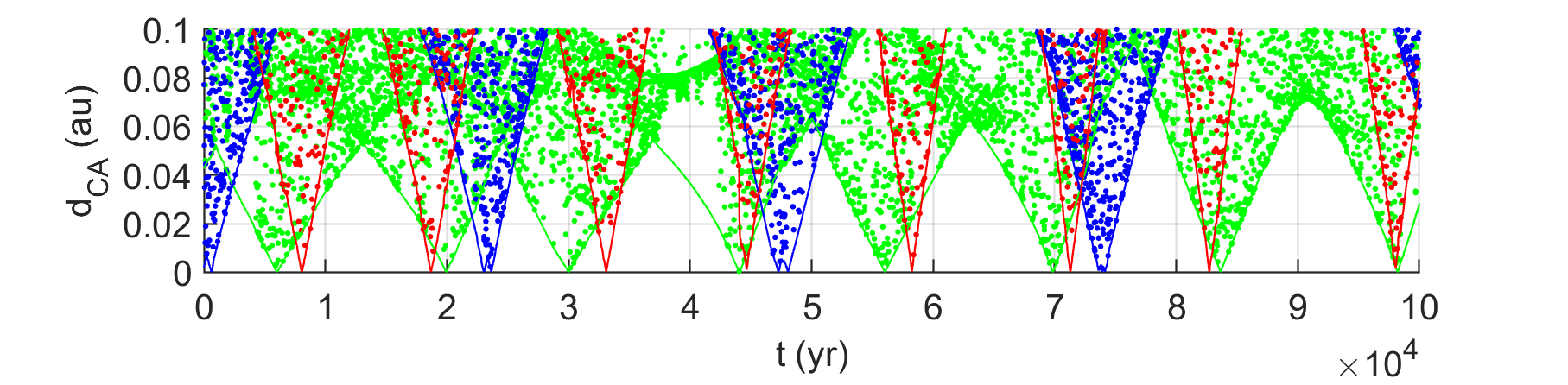}
            \caption{Close encounters in the semi-analytical propagation of the fictitious NEOs Cases 1-3 with initial conditions given in table \ref{t:cases}. Closest approach distance $d_{CA}$ (au) of the encounters and MOID with the planets. {Dots indicate close encounters, solid lines indicate the MOID.} Color code indicates the planets: (Green) Venus, (Blue) Earth, (Red) Mars.}
	\label{fig:4sapropex-1}
\end{figure}

\begin{deluxetable*}{lccccccc}[h!]
\tablenum{2}
\tablecaption{Near-Earth objects used as example cases for the demonstration of the semi-analytical propagation tool.}\label{t:cases}
\tablewidth{0pt}
\tablehead{
\colhead{Asteroid} & \colhead{$a (au)$} & \colhead{$e$} & \colhead{$i$ (deg)} & \colhead{$\Omega$ (deg)} & \colhead{$\omega$ (deg)} & \colhead{$M_0$ (deg)} & \colhead{JD (TBD)} }
%\decimalcolnumbers
\startdata
Case 1   & 1.1                           & 0.15                           & 10                             & 90                                  & 90                                  & 90     & 2451545.0  \\ \hline
Case 2   & 1.2                           & 0.35                           & 40                             & 90                                  & 90                                  & 90  & 2451545.0   \\ \hline
Case 3   & 1.3                           & 0.5                            & 10                             & 90                                  & 270                                 & 90    & 2451545.0  \\ \hline
Case 4 & 0.95   & 0.07   & 20    & 90  & 90   & 90 &  2451545.0   \\ \hline
Case 5 & 0.9    & 0.25   & 15    & 90  & 90   & 90 &  2451545.0   \\ \hline
1991 VH & 1.1373    & 0.14426   & 13.912    & 139.37  & 206.88   & 302.39 &  2456902.5   \\ \hline
1996 FG3 & 1.0543   & 0.34987     & 1.9903      & 299.88                              & 23.930                              & 147.277  & 2454796.5   \\ \hline
\enddata
\tablecomments{The elements of asteroids 1991 VH and 1996 FG3 were retrieved from HORIZONS \citep{Giorgini}. Using DE431 and SB431-N16. 1991 VH: Orbit solution date 2021 April 15, 1996 FG3: Orbit solution date of 2021 April 26.}
\end{deluxetable*}

\newpage
\subsection{Long-term propagation and the MOID} %\note[OFM]{Separated in subsection}

%It is of relevance to the system to show the with Earth. 
The minimum orbit intersection distance (MOID) indicates what the minimum distance between any two points of the two heliocentric Keplerian orbits is. In this case we focus on the orbit of Earth and the orbit of the asteroid. The MOID is also used as one of the criteria to define an asteroid as a potentially hazardous asteroid. There are many algorithms available in the literature to compute the MOID \citep{Gronchi2005,Wisniowski2013,Armellin2010}. In this paper we use the tool derived in \cite{Hedo2018,Hedo2020} based on an asymptotic approach.

The MOID constrains the minimum distance of a possible close encounter. In other words, the periods with a large MOID are absent of close encounters. Three examples are used to visualize time histories of close encounters and the evolution of the MOID for 100,000 years. These are obtained for high and low eccentricities and inclinations, and the initial conditions are found in table \ref{t:cases}.

\begin{figure}[!h]
	\centering
		\includegraphics[width=5.5in]{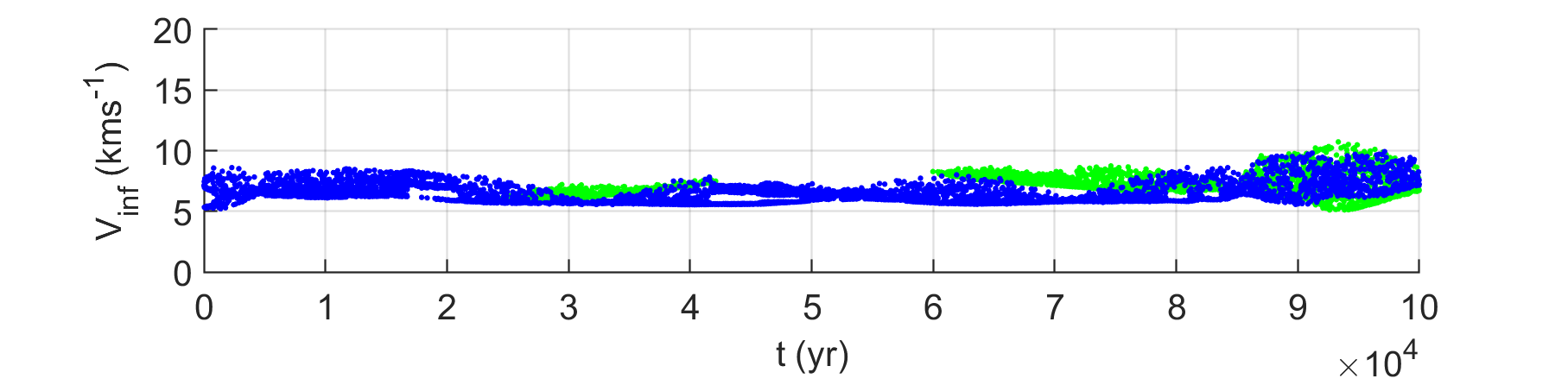}
		\includegraphics[width=5.5in]{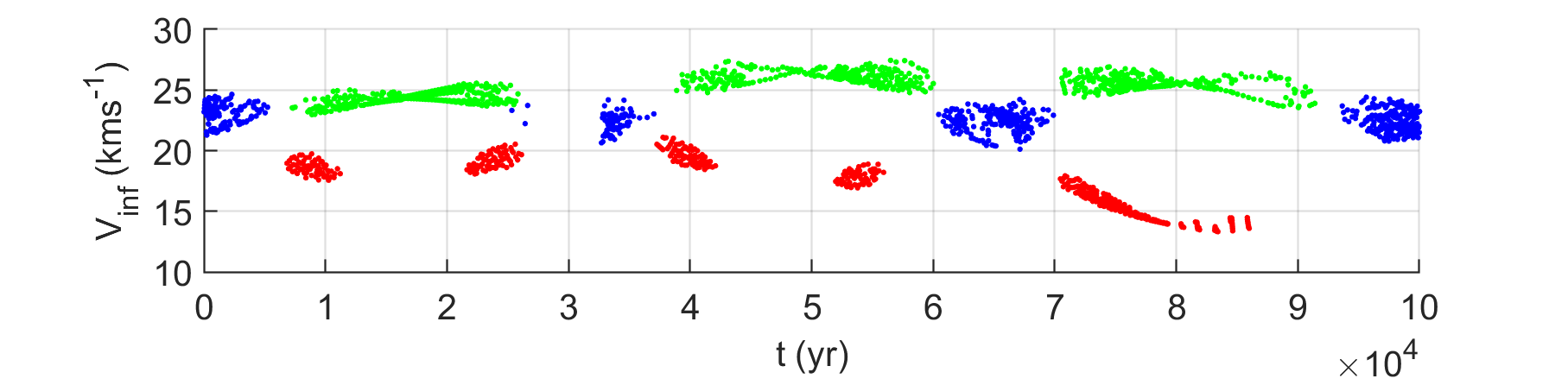}
		\includegraphics[width=5.5in]{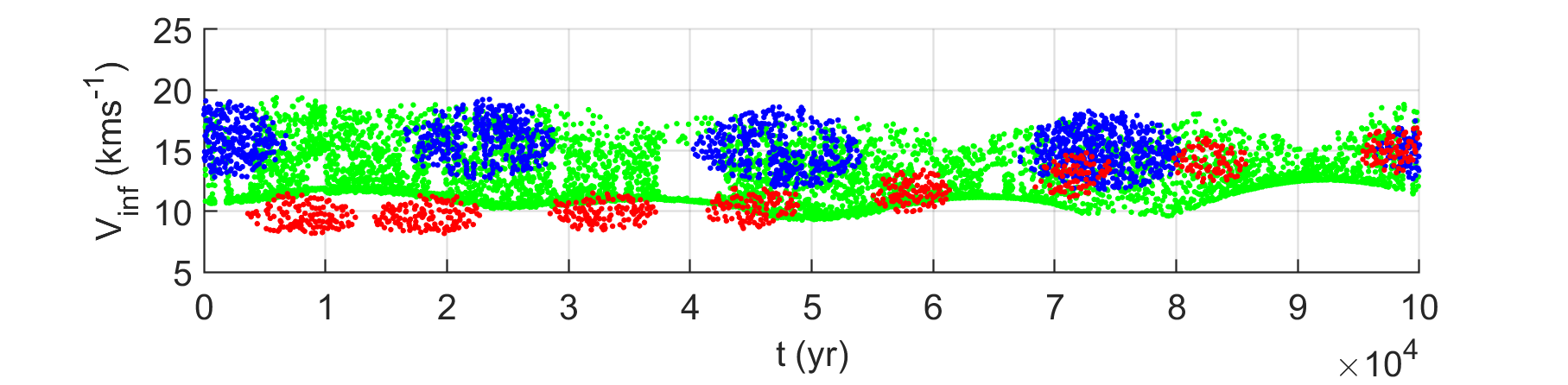}
            \caption{Close encounters in the semi-analytical propagation of the fictitious NEO Cases 1-3 with initial conditions given in table \ref{t:cases}. $V_\infty$ at closest approach of the encounter (km s$^{-1}$). Color code indicates the planets: (Green) Venus, (Blue) Earth, (Red) Mars.}
	\label{fig:4sapropex-2}
\end{figure}

% \newpage
The distributions of closest approach distances are shown in figure \ref{fig:4sapropex-1} and the unperturbed relative velocity $V_\infty$ at those encounters is found in figure \ref{fig:4sapropex-2}. Case 1 is an example of a NEO with relatively low eccentricity and inclination. In  these conditions, close encounters are only possible with Earth and at a low relative velocity. The MOID oscillates secularly with long periods of low MOID. Case 2 is an example of an opposite scenario in which both eccentricity and inclination are large. In the secular evolution of the MOID this translates in short periods of low MOID and long periods absent from encounters. This is scenario in which the semi-analytical propagation of the asteroid allows a rapid propagation until the next period of frequent encounters. Case 2 faces high velocity close encounters with Venus, Earth and Mars. 
%Observing the evolution of the MOID before the complete dispersion of the cloud of points, the evolution is clearly secular with a characteristic oscillation.

Case 3 is an example of a NEO with high eccentricity and low inclination. This combination of factors leads to a large frequency of close encounters with the inner planets. In this case, encounters are very frequent with Venus, Earth and Mars. The close encounters experienced by Case 3 are with a relative velocity smaller than in Case 2 given the reduced inclination. Even under the elevated frequency of close encounters, the secular signature of the MOID is maintained. The structure shows until the event of an energetic close encounter. The occurrence of such encounters is just a matter of probability of having the right timing during the low-MOID intervals of the secular propagation.

\newpage
\subsection{Statistics of long-term propagation}%\label{s:stochs}

The chaotic nature of the dynamics implies that the study of the orbital evolution over long timescales should be done statistically. Given the uncertainty in the orbit solution of an asteroid, we can sample a large number of particles and study the dynamical paths that the different particles take. Because of the sensitivity to initial conditions in planetary encounters, very well determined distributions diverge in a few centuries to widely different paths. We demonstrate these effects by propagation of 500 particles that sample uncertainty distributions around a nominal asteroid orbit for 500,000 years. We inspected 7 examples: first, in detail, orbit solution of (35107) 1991 VH; then, the orbit solution of (175706) 1996 FG3; last, the previous Cases 1-5 of table \ref{t:cases} with artificial orbit uncertainties as described in appendix \ref{app:uncerts}.

\begin{figure}[h!]
	\centering
		\includegraphics[width=5.5in]{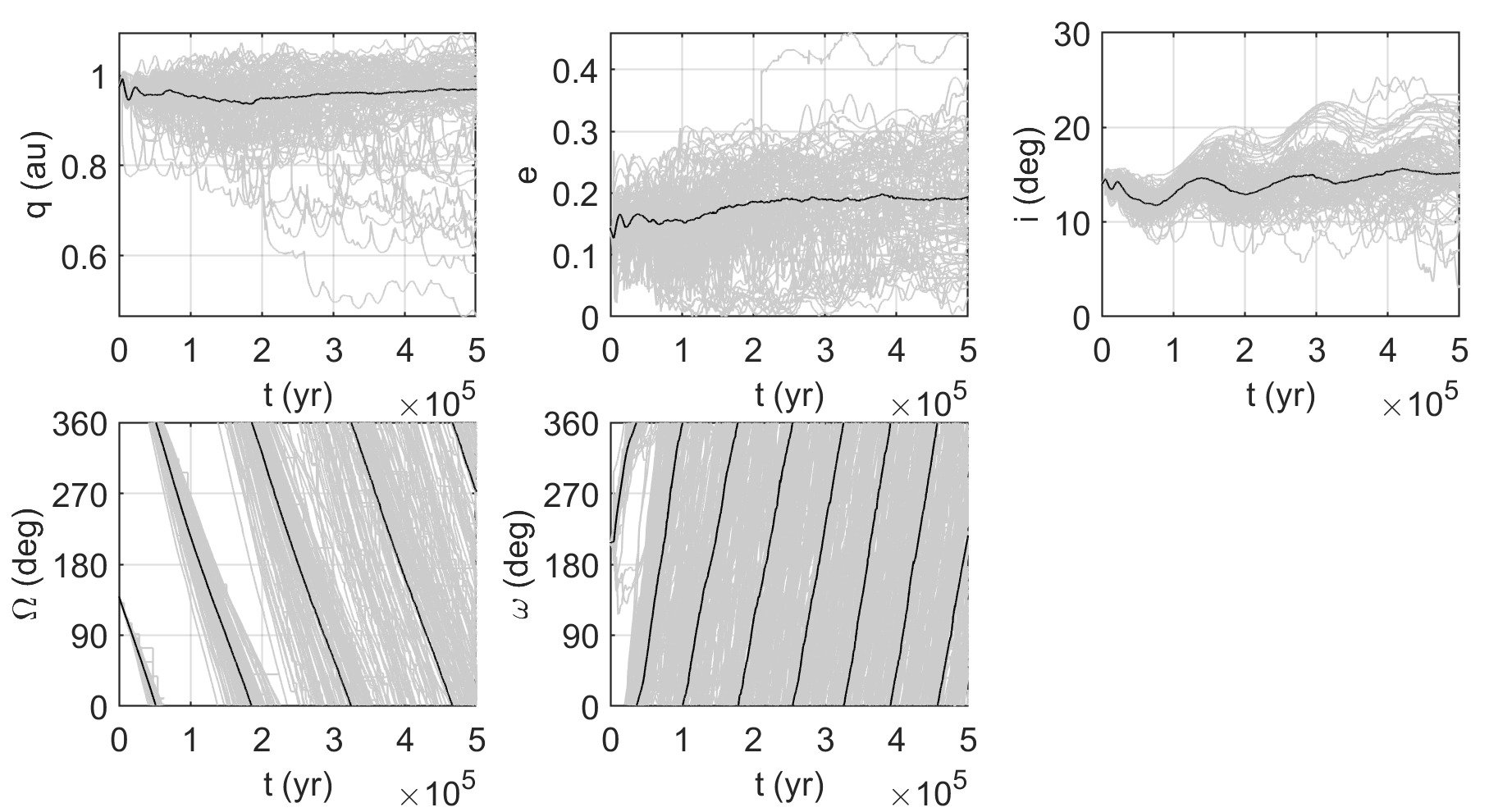}
		\includegraphics[width=5in]{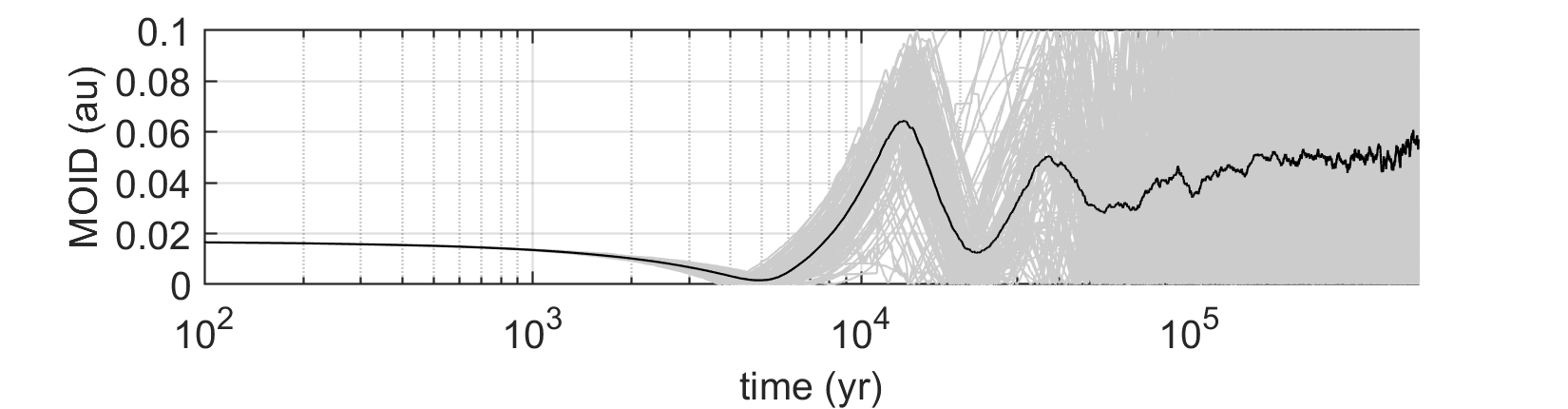}
            \caption{500,000 year Monte Carlo semi-analytical propagation of asteroid (35107) 1991 VH. Initial conditions are given in table \ref{t:cases} as obtained from HORIZONS \citep{Giorgini} {and uncertainties in the distribution are obtained from JPL’s SSD/CNEOS Small-Body DataBase \citep{SBDB}} as  described in Appendix \ref{app:uncerts}. Elements shown are perihelion, eccentricity, inclination, argument of the ascending node, argument of perihelion and minimum orbit intersection distance (MOID). Grey lines show individual simulations, black lines are the median of the 500 simulations of each parameter shown.}
             
            %  {\color{red} Noticed that some lines appear to have 'jumps', will review plotting and secular interpolation between datapoints.}
            
            %\caption{1 million year Monte Carlo semi-analytical propagation of asteroid (35107) 1991 VH. Initial conditions are generated form the uncertainty in the orbit solution as given in HORIZONS\cite{Giorgini}. Elements shown are perihelion, eccentricity, inclination, argument of the ascending node and argument of perihelion. \remove[OFM]{ Additionally, the MOID between the asteroids and Earth is shown.} Grey lines show individual simulations, black lines are the median of the distribution of each parameter shown.}
	\label{fig:5e5yrs}
\end{figure}

\begin{figure}[h!]
    \centering
    \includegraphics[width=5.5in]{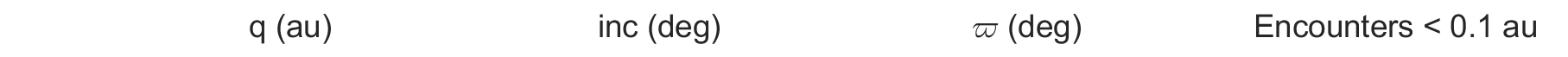}
    \includegraphics[width=5.5in]{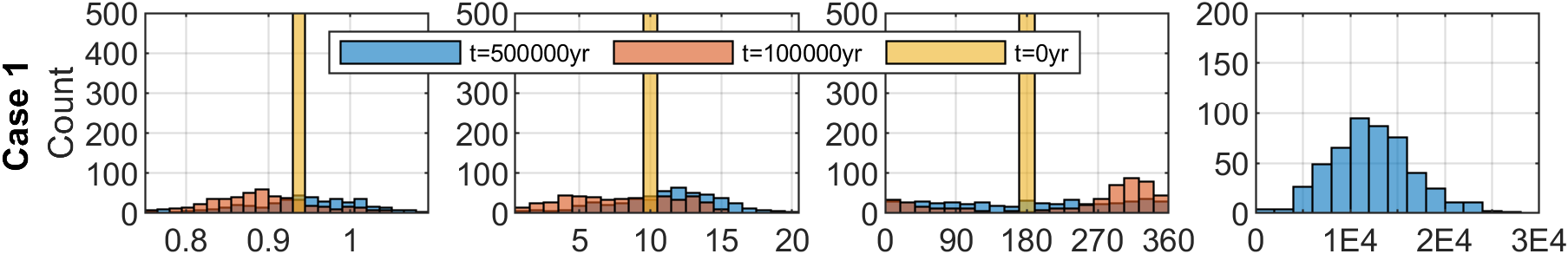}
    \includegraphics[width=5.5in]{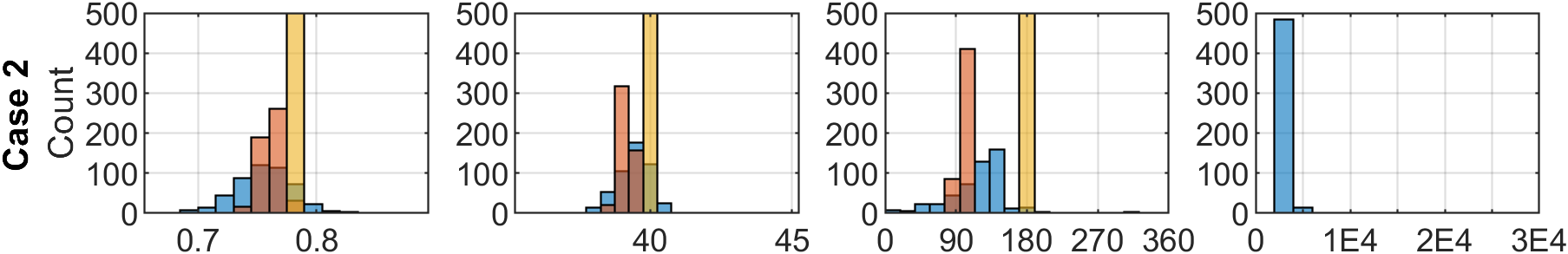}
    \includegraphics[width=5.5in]{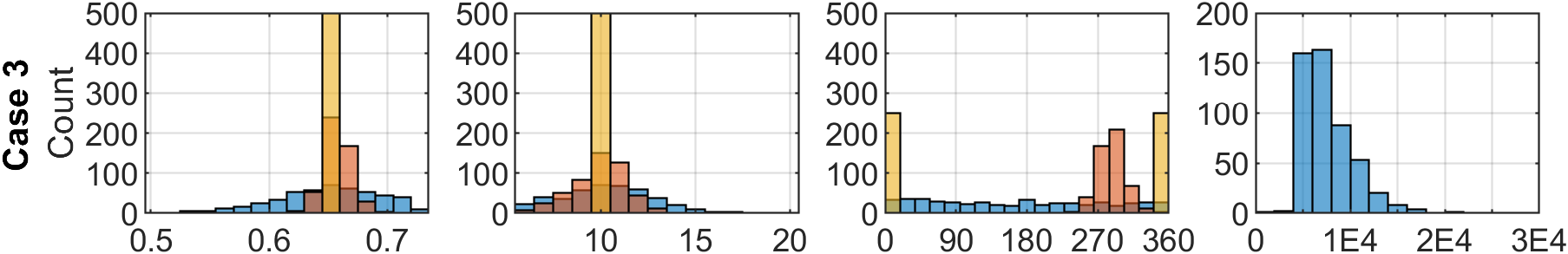}
    \includegraphics[width=5.5in]{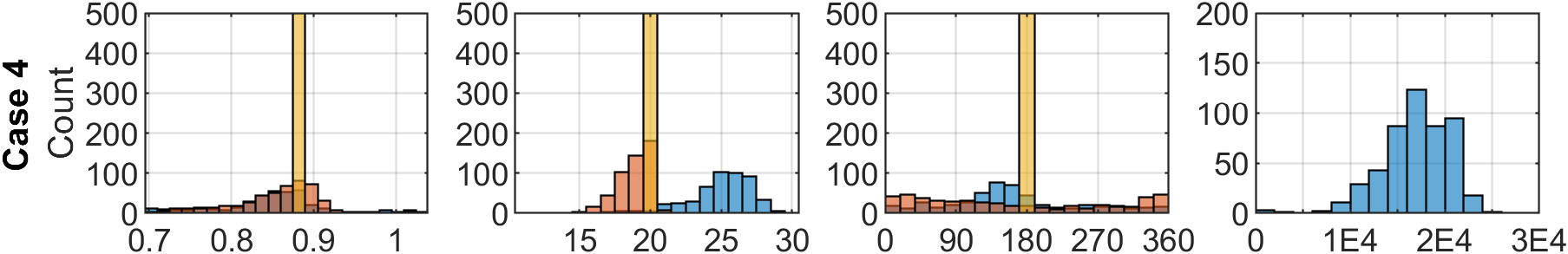}
    \includegraphics[width=5.5in]{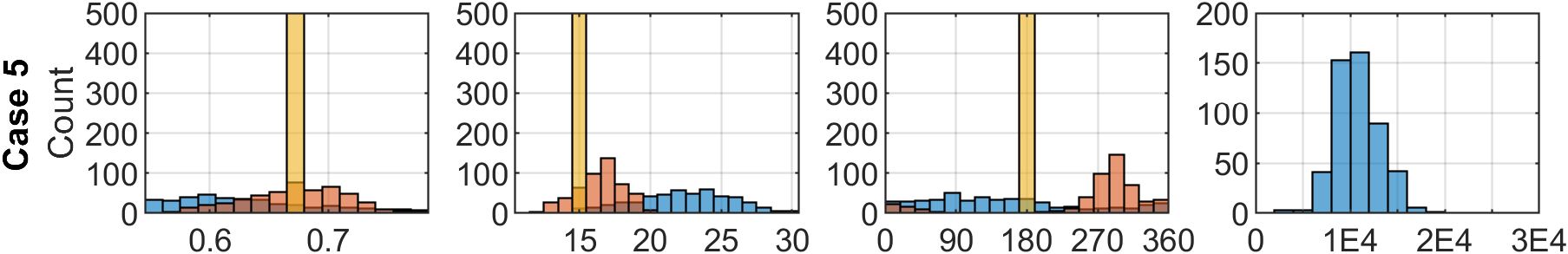}
    \includegraphics[width=5.5in]{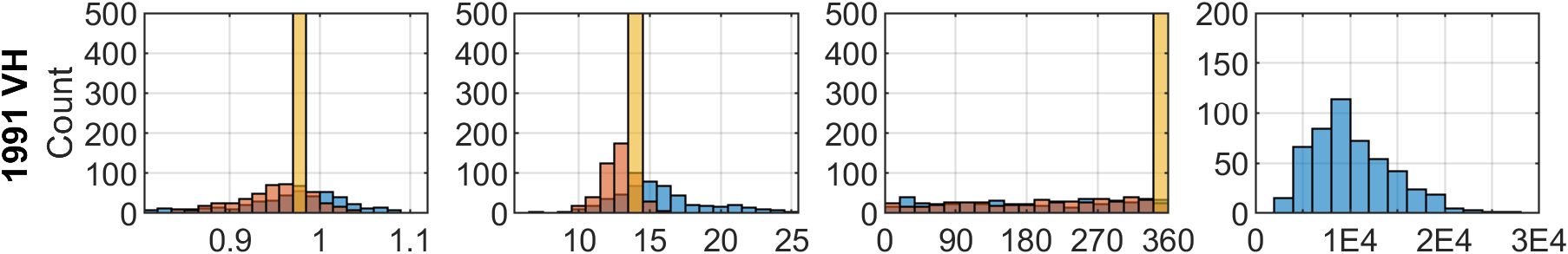}
    \includegraphics[width=5.5in]{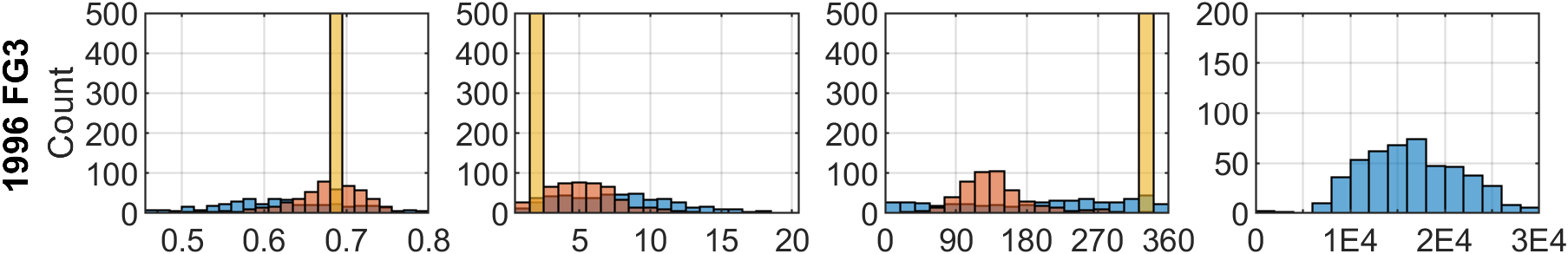}
    \includegraphics[width=5.5in]{hists-labels.png}
    
    \caption{Semi-analytical propagation of  asteroid Cases 1-5, (35107) 1991 VH, (175706) 1996 FG3 for 500,000 years. Histograms at the initial time, after 100,000 years and at the final time. Initial conditions are given in table \ref{t:cases} as obtained from HORIZONS \citep{Giorgini} {and uncertainties in the distribution are obtained from JPL’s SSD/CNEOS Small-Body DataBase \citep{SBDB}} as  described in Appendix \ref{app:uncerts}.}
    \label{fig:hists-cases}
\end{figure}

% \begin{figure}[h!]
%     \centering
%     \includegraphics[width=5.5in]{hists-Case1.png}
%     \includegraphics[width=5.5in]{hists-Case2.png}
%     \includegraphics[width=5.5in]{hists-Case3.png}
%     \includegraphics[width=5.5in]{hists-VH.png}
%     \includegraphics[width=5.5in]{hists-FG3.png}
    
%     \caption{Semi-analytical propagation of asteroids Cases 1-3, (35107) 1991 VH, (175706) 1996 FG3 for 500,000 years. Histograms at the initial time, after 100,000 years and at the final time. Initial conditions are given in table \ref{t:cases} as obtained from HORIZONS \citep{Giorgini} and uncertainties in the distribution are obtained from JPL’s SSD/CNEOS Small-Body DataBase \citep{SBDB}) as  described in Appendix \ref{app:uncerts}.}
%     \label{fig:hists-cases}
% \end{figure}

Figure \ref{fig:5e5yrs} shows the time histories of the individual runs of the cloud of points originally neighboring (35107) 1991 VH, showing that the cloud of particles distributes over a wide region of near-Earth space. % Figure \ref{fig:elems-moid} shows snapshots of the distribution after 1000, 10,000 and 100,000 years. 
 On the order of hundreds of thousand years, the dispersion is accomplished by the less frequent very close encounters. Eccentricity and inclination show the secular component but are dispersed by the presence of encounters. The dynamics of argument of ascending node and argument of perihelion remain mainly secular with a degree of dispersion because of the presence of encounters. Note that the initial uncertainty on the orbit of (35107) 1991 VH is very small as shown while demonstrating the chaotic dynamics nature in Figure \ref{fig:2.chaos5k} of the background section.  As it was observed in the detailed analysis of a shorter simulation in figure \ref{fig:1e2yrs-oe}, the mean anomaly at epoch changes completely with small changes in the semi-major axis. This fact reflects in the long-term simulations as a complete uniformization after just a few centuries.

%\newpage
The binary (35107) 1991 VH currently presents a MOID that is decreasing. This means that after a few millenia the probability of experiencing very close encounters increases. In the statistical analysis, this probability shows in that a fraction of the fictitious asteroids experience such encounters. We observe that towards the end of the simulation there is a large dispersion in inclination and perihelion distance. By the end of the simulation, the angles $\Omega-\omega$ are dispersed along a linear drift as caused by numerous close encounters that not only change these angles, but modify the secular dynamic frequencies.% The main behavior allows us to make predictions of the dynamical evolution of the asteroid, as we explore next.

%[Figure moved to the end!] To illustrate this phenomenon we show two instants of time of the statistical distribution of the asteroids in figure \ref{fig:elems-moid} color-coded by the MOID.

% Observing the distribution of the arguments we note a the fraction of asteroids that experienced close encounters and increased their MOID separating from the larger cloud of particles. This phenomenon was also revealed studying the statistics of (35107) 1991 VH using numerical integration in figure \ref{fig:2.chaos5k}. 
% Note - These used to be before sentence: "We observe that towards..."

\begin{figure}[b!]
    \centering
    \includegraphics[width=4.5in]{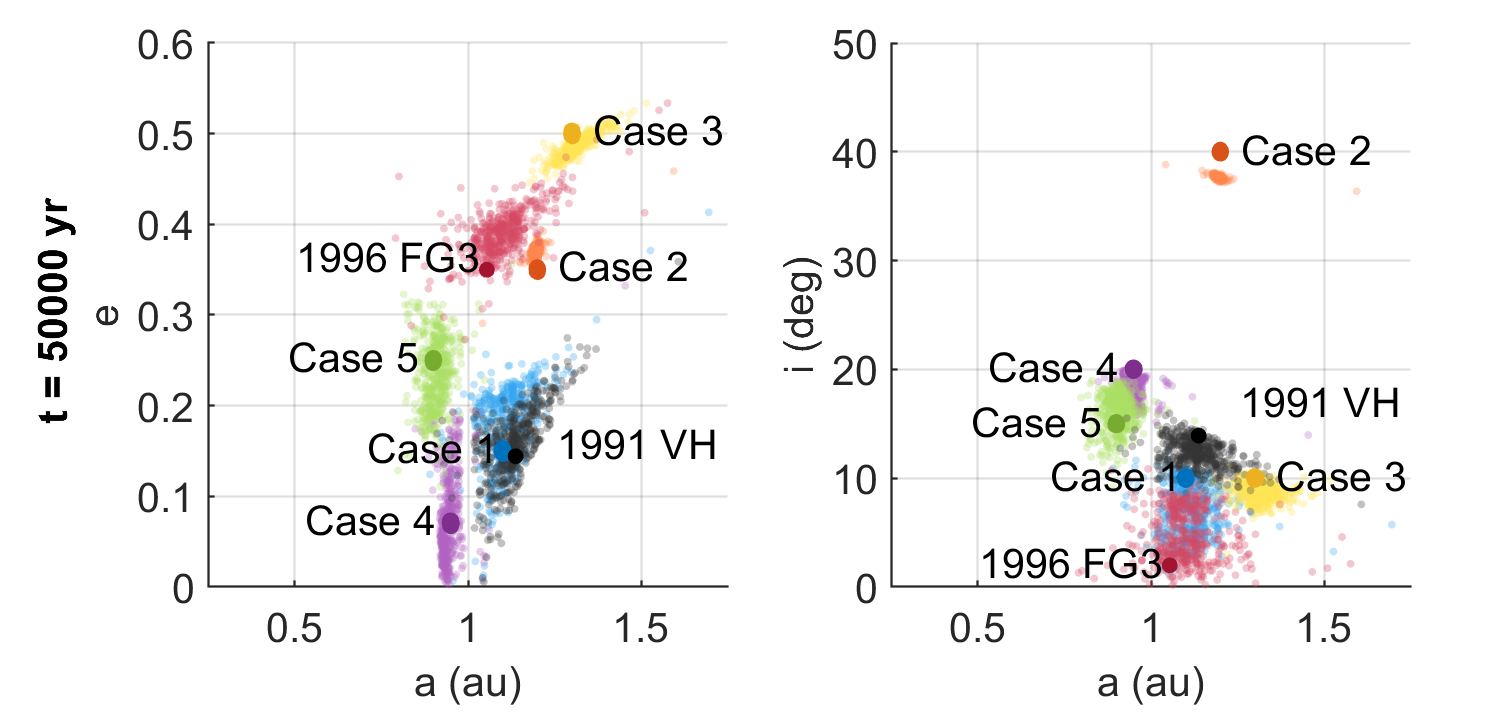}
    \includegraphics[width=4.5in]{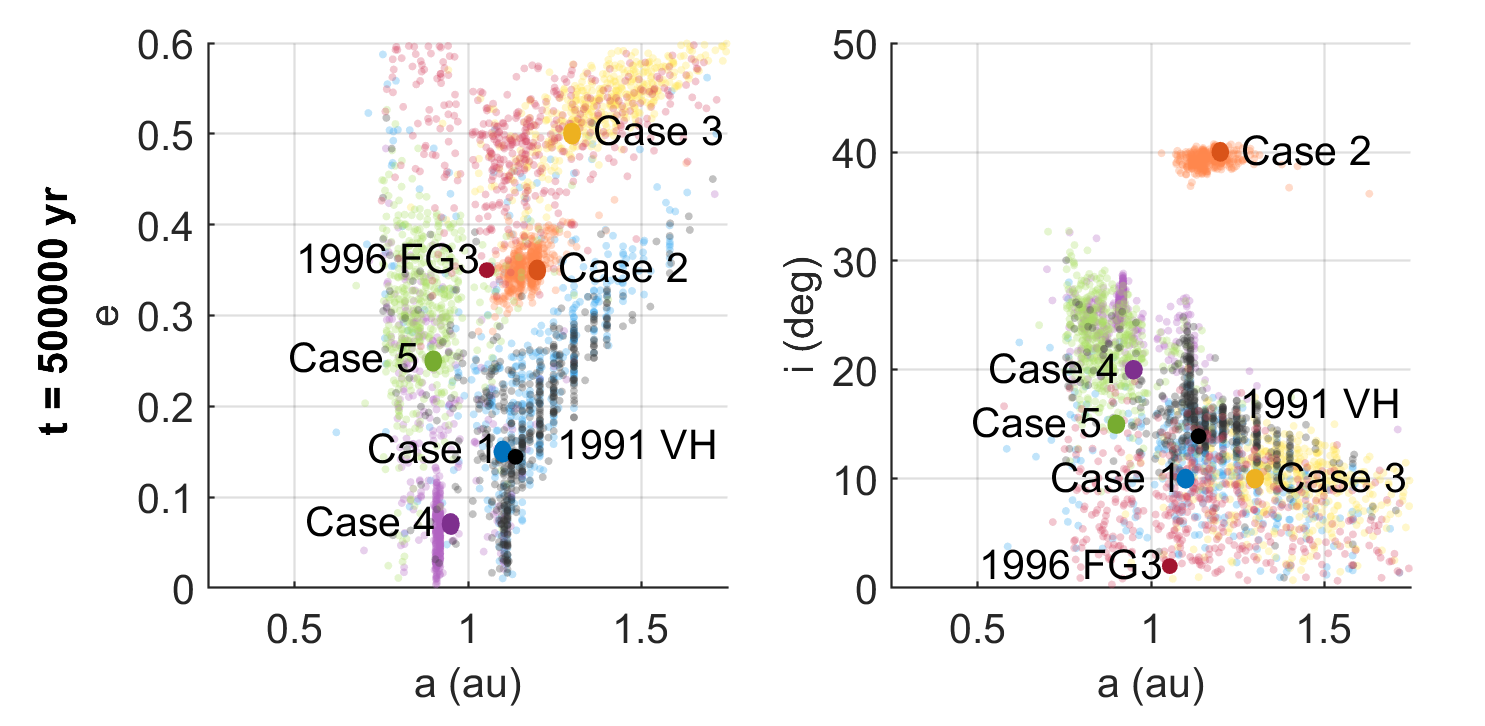}

    \caption{Dispersion in the distributions of the 5 asteroid in table \ref{t:cases},  (175706) 1996 FG3 and (35107) 1991 VH after 100,000 years (top) and 500,000 years (bottom). Semi-major axis vs. eccentricity (left) and semi-major axis vs. inclination (right). Initial uncertainty distributions detailed in appendix \ref{app:uncerts} and shown in red. See text for the explanation of the resonances found in the semi-analytical propagation.}
    \label{fig:aei-cases}
\end{figure}

\begin{figure}[h!]
    \centering
    \includegraphics[width=5.5in]{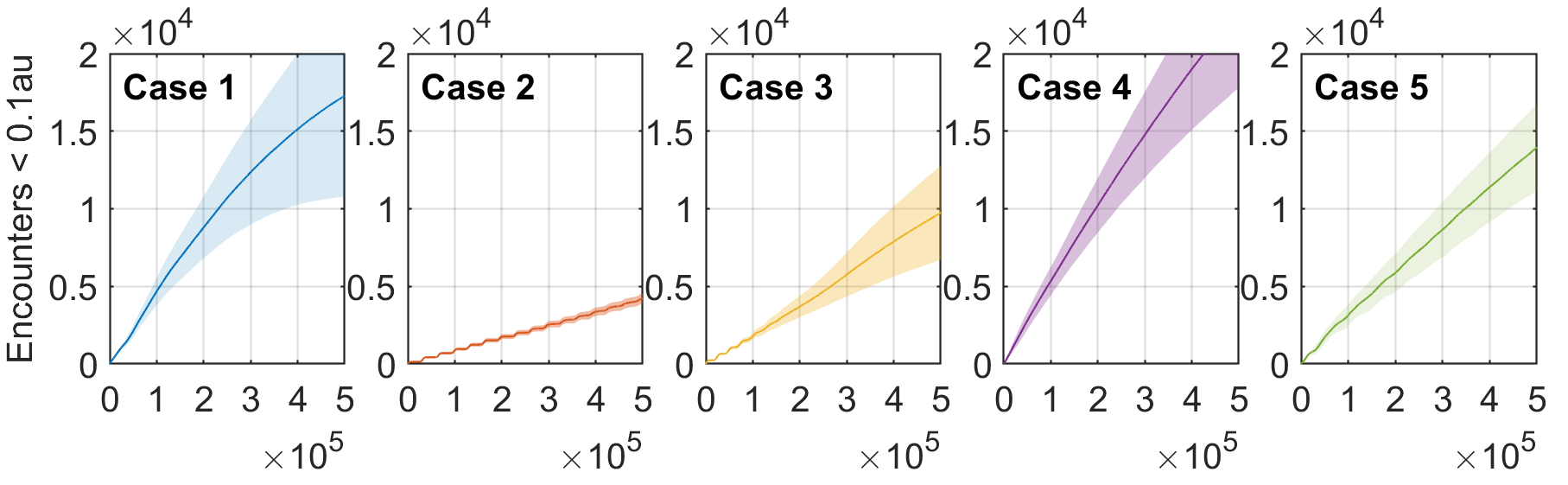}
    \includegraphics[width=4in]{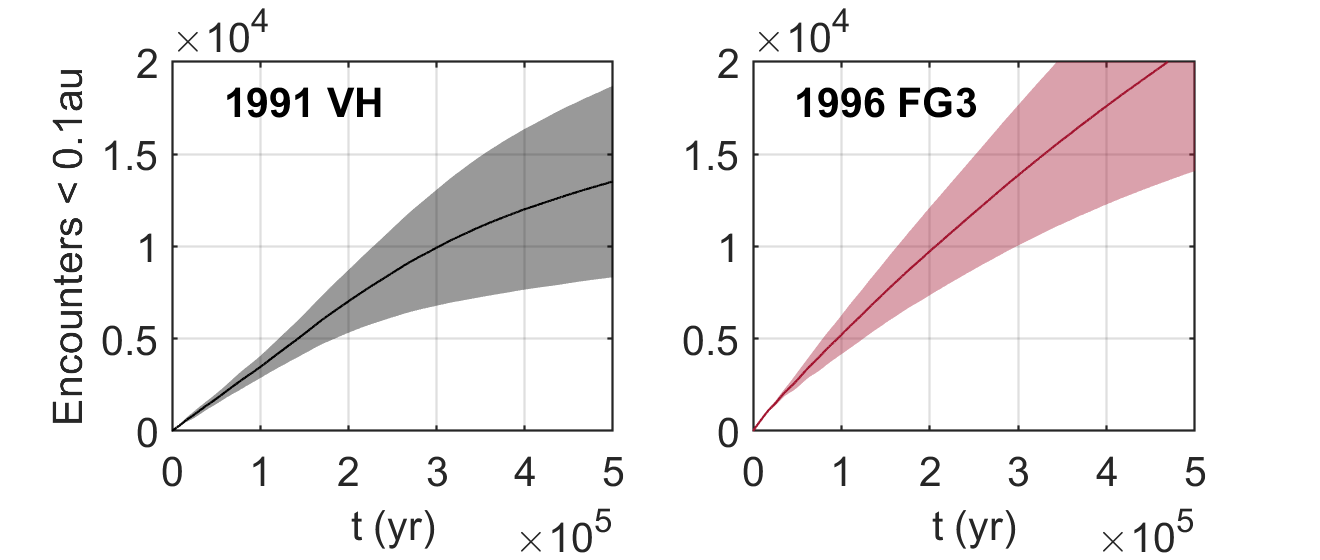}
    
    \caption{Number of encounters experienced by the 5 test cases given in table \ref{t:cases}, (175706) 1996 FG3 and (35107) 1991 VH during 500,000 years. Mean of the distributions of 500 particles (solid lines) and 1-$\sigma$ bounds (shadowed area).}
    \label{fig:nenc-time-cases}
\end{figure}

\begin{figure}[b!]
    \centering
    \includegraphics[width=5.5in]{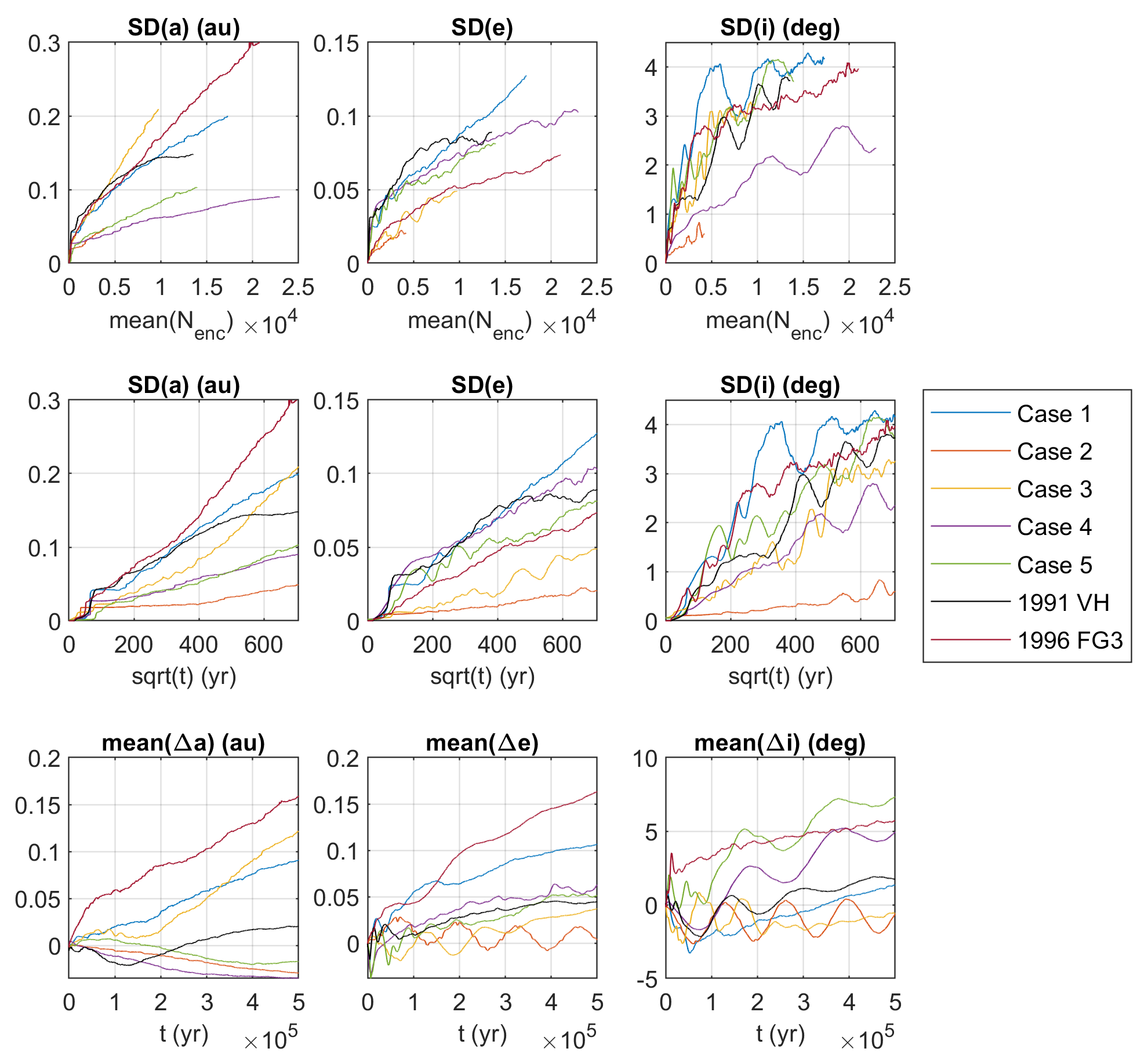}
    
    \caption{Statistical evolution of the distributions of the 5 test cases given in table \ref{t:cases}, (175706) 1996 FG3 and (35107) 1991 VH during 500,000 years. Standard deviation of semi-major axis, eccentricity, inclination as function of the number of encounters (top row), square root of time (center row). Variation of the mean value over time with respect of the initial value of semi-major axis, eccentricity and inclination (bottom row).}
    \label{fig:std-cases}
\end{figure}

The long-term dynamics of 6 more examples are integrated for 500,000 years. These are the Cases that we used to illustrate the long-term dynamics with initial conditions in Table \ref{t:cases}. The case of the binary (175706) 1996 FG3 is also added to the discussion, as it is the other target of exploration of the Janus mission \citep{Scheeres2020Janus}. The uncertainties of (175706) 1996 FG3 and (35107) 1991 VH are sampled based on their publicly available orbit solutions. In the case of the fictitious asteroids we used an arbitrary distribution. Both approaches are explained in detail in the appendix \ref{app:uncerts}. The statistical distributions after 100,000 and 500,000 years are shown in Figure \ref{fig:hists-cases}. The combined final and initial distributions are shown in Figure \ref{fig:aei-cases}. The recorded number of encounters are shown in Figure \ref{fig:nenc-time-cases} and more details on the statistical distributions over time are shown in Figure \ref{fig:std-cases}. Next, we describe the dynamical evolution of these test cases.

%\newpage
In Figure \ref{fig:aei-cases}, the 500 virtual asteroids that are generated at the initial time per case are in the same bar of the histogram. The effect of repeated encounters causes the distributions to spread along near-Earth space. This dispersion is clearly shown in the perihelion distance in all cases, with a general trend of a decrease in the distance. Figure \ref{fig:aei-cases} additionally shows this spread of all the cases together. 

The presence of mean motion resonances in the inner Solar System can protect asteroids from close encounters \cite{Milani1989}. In the semi-analytical propagation of near-Earth asteroids the orbits may drift to these regions, as observed in Figure \ref{fig:aei-cases}. The resonance regions are found for semi-major axes larger than $a=1$ au. In these cases encounters with the Earth stop occurring for a period of time. However, this clustering of particles is not found in numerically integrated populations or in the discovered population of near-Earth asteroids. Thus, we investigate the trajectories obtained through NBP dynamics.

%\newpage
The process of obtaining the secular long-term perturbation eliminates the short-period perturbations. The latter perturbations cause an oscillation in the orbit elements of non-negligible amplitude. In order to measure the influence of this effect, we included an analytical oscillation in the semi-major axis with the frequency of the orbital period and an amplitude of the order of 0.01 au. This extension completely eliminates the artificial mean motion resonance regions. The analytical characterization of the short-period perturbation is left as future work, as its contribution must be considered for the complete set of orbital elements and is not expected to significantly modify the obtained distributions.

In general, we observe that asteroids encountering the planets more frequently disperse their distributions faster. This is the case for the Janus targets and Case 1, that present the largest standard deviations increase in the simulation time (Fig. \ref{fig:std-cases}). This dispersion is not only shown in the elements, but also in the number of encounters (Fig. \ref{fig:nenc-time-cases}). 

%\newpage
The evolution of the distribution as caused by encounters could be modelled as a random walk. If this hypothesis is true, then the standard deviation in the population increases linearly with the square root of time. Figure \ref{fig:std-cases} tests graphically this hypothesis for semi-major axis, eccentricity and inclination. Initially in all cases there is a fast increase in the standard deviations. After the few first millennia, some of the distributions follow the hypothesis of the linear relationship $\sigma \propto \sqrt{t}$, especially in the semi-major axis.

In eccentricity and inclination, we observe that there is a secular component in the evolution of the distribution. The secular component is observed in both the evolution of the standard deviation and the mean of the variations (Fig. \ref{fig:std-cases}). These are shown with respect to the initial values to have a common reference in the comparison of cases. 

The dispersion in semi-major axis modifies the secular rates of the drift in argument if perihelion and argument of the node. This effect combined with the direct variation of the angles during planetary encounters leads to the uniformization of the distribution in $\omega, \Omega$. In the duration of our simulations of 500,000 most cases approach this uniformization as we showed in figure \ref{fig:hists-cases}. 

\begin{figure}[h!]
    \centering
    \includegraphics[width=5in]{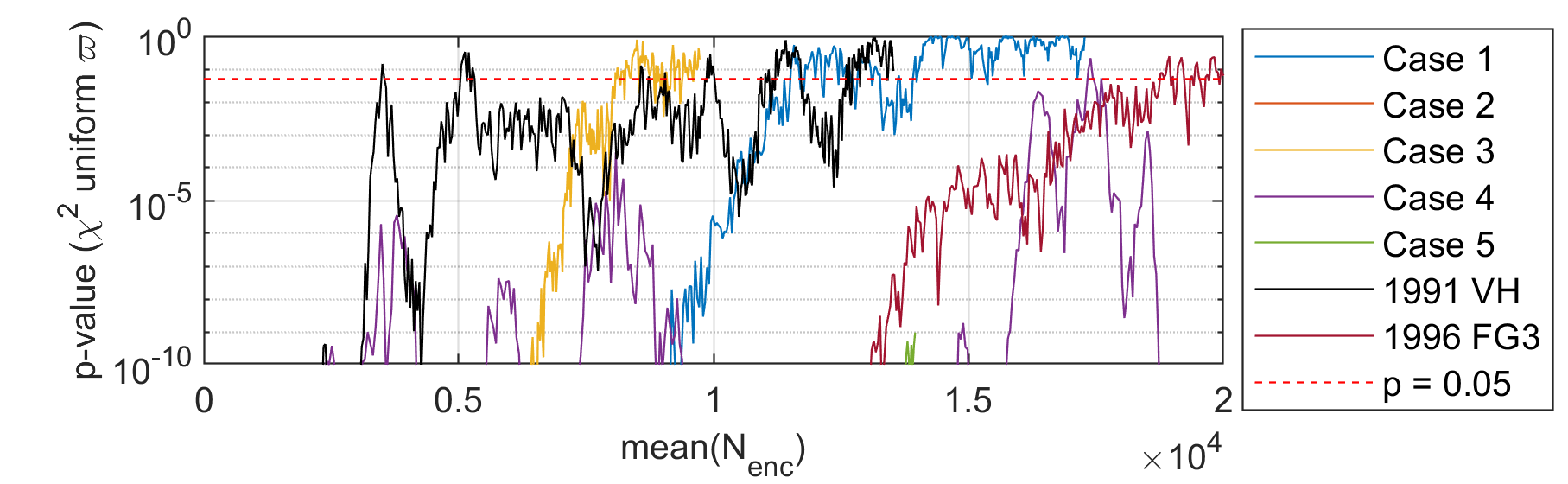}
    \includegraphics[width=5in]{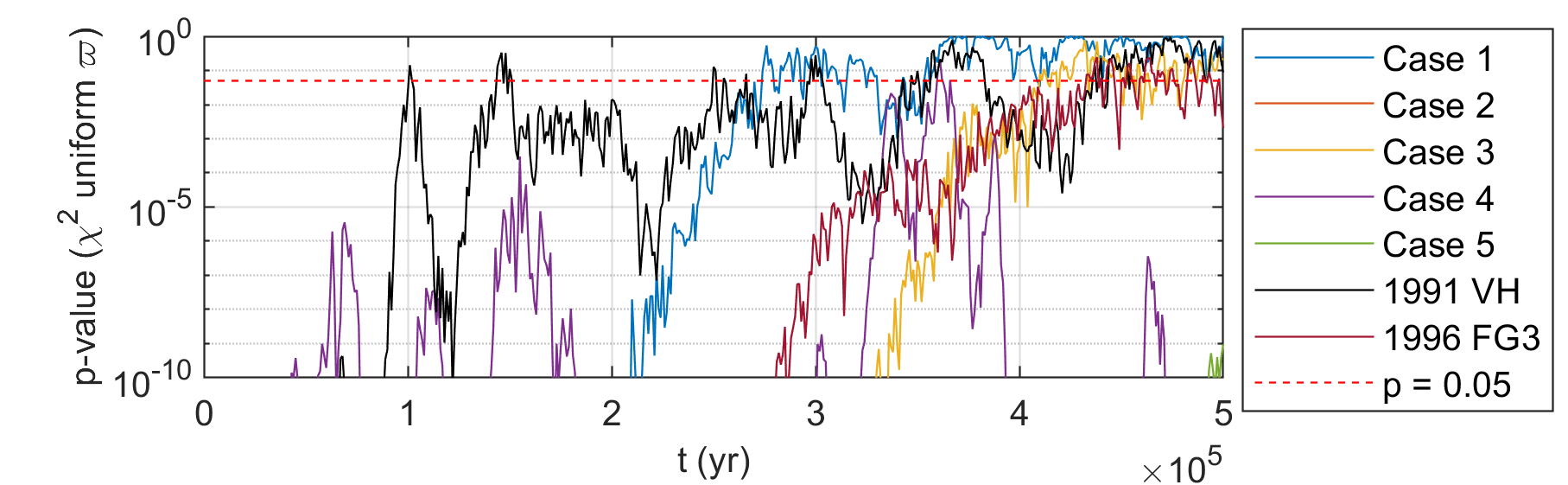}
    
    \caption{P-value of the chi-squared test of the uniform distribution of the longitude of perihelion $\varpi = \omega + \Omega$ for the 5 test cases given in table \ref{t:cases}, (175706) 1996 FG3 and (35107) 1991 VH during 500,000 years. P-value is shown as function of the mean number of encounters (top) and function of time (bottom).}
    \label{fig:pval-cases}
\end{figure}

We can compute more rigorously whether the distributions can be considered uniform by conducting the chi-squared test on the longitude of perihelion $\varpi = \omega + \Omega$. Figure \ref{fig:pval-cases} shows the result of the chi-squared test of a uniform distribution over the simulation time for all cases. If the p-value is larger than our threshold of $p=0.05$, we can consider that our null hypothesis of the uniform distribution of $\varpi$ is true. Cases 1, 3, 4, (175706) 1996 FG3 and (35107) 1991 VH reach this threshold, while Cases 2 and 5 do not approach the significance by the end of the simulation time. It is interesting to show the evolution of the p-value compared to the mean number of encounters. In figure \ref{fig:pval-cases} we show how the distribution of Case 3 tends to the uniformization in $\varpi$ with less encounters than other cases that achieve this distribution earlier in the simulation time. This is expected since this case experiences more frequent encounters with the most massive planets and with a slower relative velocity, which means that the impact of these encounters in the dispersion of their distributions is larger.

From the general trends that we observed, the only case that is very different is Case 2, that experiences much fewer encounters. As we illustrate in Figure \ref{fig:4sapropex-1}, Case 2 experiences close encounters much less frequently than the other cases. The relative velocity is also larger in this case, which means that the effects of the encounters are not as strong. The case of (175706) 1996 FG3 is more difficult to fit in the general description of the dynamics, as the close encounters do not cause such a fast dispersion of the distribution. This binary asteroid is studied in more detail in section \ref{s:janustargs}.

%as the variation of mean and standard deviation of the distributions is not as accurately described by a random walk in semi-major axis and the combination with secular oscillations in eccentricity and inclination. Interestingly, the angles become uniformized faster than other cases, but requiring a much larger number of encounters on average.

% Notes of the section description
% •	To random walk or not to random walk…
% •	In inclinations we observe the combination of a secular component that translates the distribution and the dispersion caused by the encounters. 
% •	The rotation of the orbit in the heliocentric frame is shown by the longitude of the perihelion $\varpi=\omega+\Omega$. While after 100,000 years the main component is secular, the distribution becomes uniform by the end of the simulation time.

%========================================================
% Section about short-period components
%========================================================

%\newpage
%\subsection{Mean motion resonances and short-period perturbations}\label{s:mmr}

% \begin{itemize}
%     \item Usual mean-motion resonances effects
%     \item Mean motion in our study
%     \item Fixing by adding oscillation in semi-major axis
%     \item Will study this component also in the other elements
% \end{itemize}

%========================================================
% DISCUSSION
%========================================================

\newpage
\section{Discussion}\label{s:5disc}

% - Dynamics that are observed/known that the semi-analytical model shows
%     - Presence of encounters
%     - Uniformization of the angles
% - Limits of the model, possibility and plans to extend the model
% See case 2. With Lidov-Kozai
% - Dynamics that allow further computation, and how we will do it

The semi-analytical propagation tool shows the main dynamical effects observed in long-term numerical integration of the inner Solar System. The secular drifts caused by Jupiter move the asteroids between the vicinities of the different planets of the inner Solar System. This effect causes a seasonal presence of strong close encounters that can disturb both the orbit of the asteroid and its physical properties. While the time-scales of these events is of millions of years \citep{Fang2012}, if we sample a large enough number of particles we can measure the probabilities of collisions or the potential disruption of other physical properties of  asteroids. The measurement of the collision probabilities was outside the scope of this paper, but in this work a few collisions were detected in the uncertainty sampling of the asteroids.

The analytical modelling of the dynamics far from the planets was done using the Laplace-Lagrange theory, which works well in a large fraction of the NEO population. {For this reason we defined a region in near-Earth space in which the secular model works best, as shown in Figure \ref{fig:3.errors-rates}}. However, we could extend the modelling in the regions of large eccentricity and inclination. In the previous section we describe the low frequency of encounters that is characteristic of asteroids with high inclination, specifically with Case 2. An asteroid of these characteristics would be likely to be dominated by the Lidov-Kozai effect, in which there is an exchange between high inclination-low eccentricity periods and low inclination-high excentricity periods. This would mean that Case 2 evolves to become a case closer to Case 3, in which encounters are more frequent. The use of analytical models of the Lidov-Kozai model \citep{kinoshita2007general} for the perturbed propagation is left for future work.

Using the semi-analytical propagation tool we observe the stochastic nature of the dynamics. However, the effect is different on each of the elements. While in semi-major axis we observe what could be described as a random-walk process, the angles $\Omega$ and $\omega$ become uniformly distributed. Eccentricity and inclination show a mixed effect between a random-walk that adds dispersion to the distribution and the oscillations driven by the secular theory. 

%========================================================
% 1991 VH and 1996 FG3
%========================================================

\newpage
\section{Orbit histories of the Janus mission targets}\label{s:janustargs}

The binary asteroids (175706) 1996 FG3 and (35107) 1991 VH are the targets of the exploration mission Janus \citep{Scheeres2020Janus}. The long-term orbital dynamics of the asteroids are studied in this section with two goals. First, characterizing the stochastic nature of their long-term dynamics. Then, estimating the probability that they have been potentially disrupted by a very close encounter. With these purposes, we study their orbital origins by sampling 1000 particles and propagating them for 1Myr into the past using the semi-analytical propagation tool.

\begin{figure}[h!]
    \centering
    \includegraphics[width=5.5in]{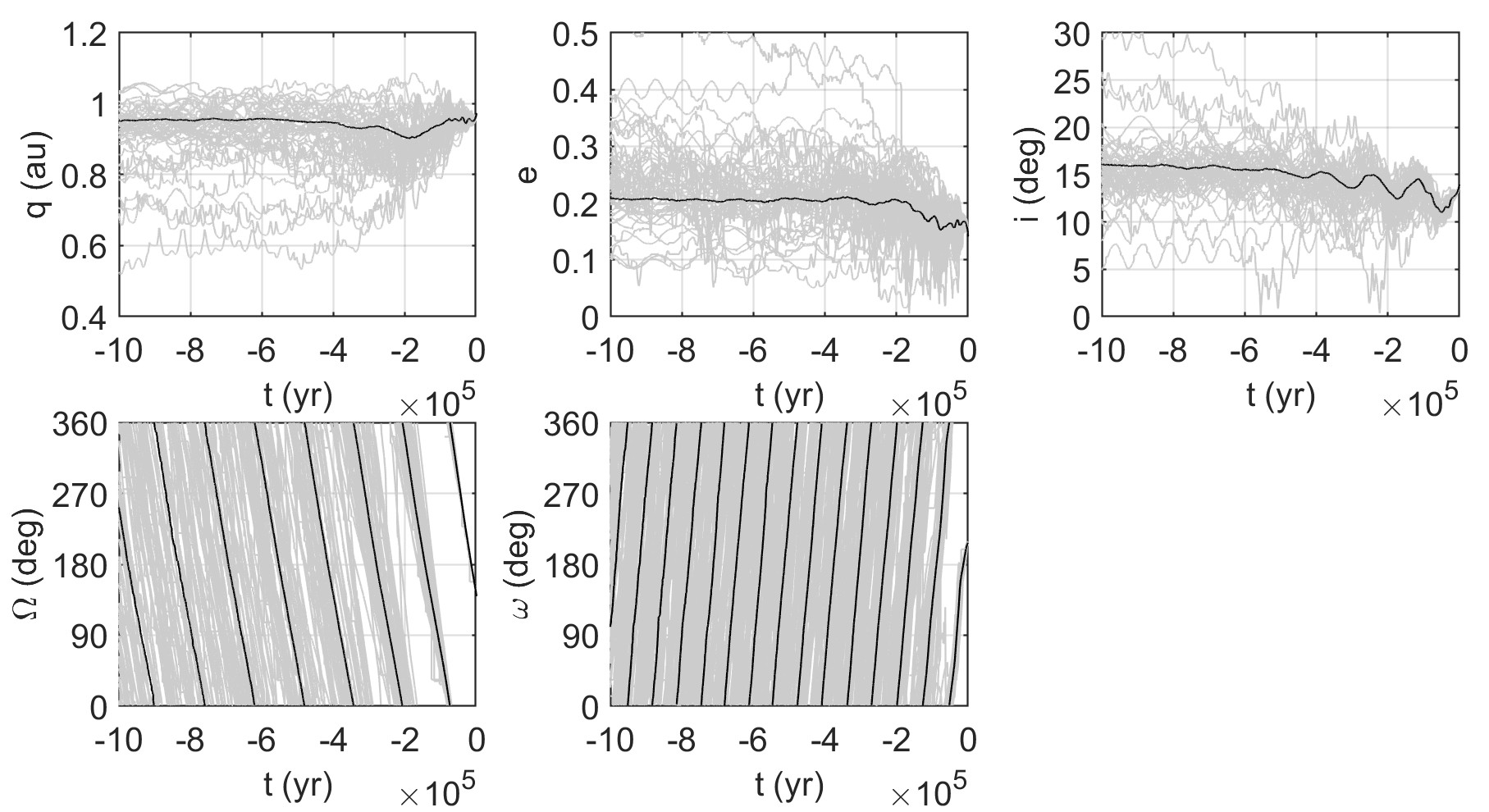}
    \includegraphics[width=5in]{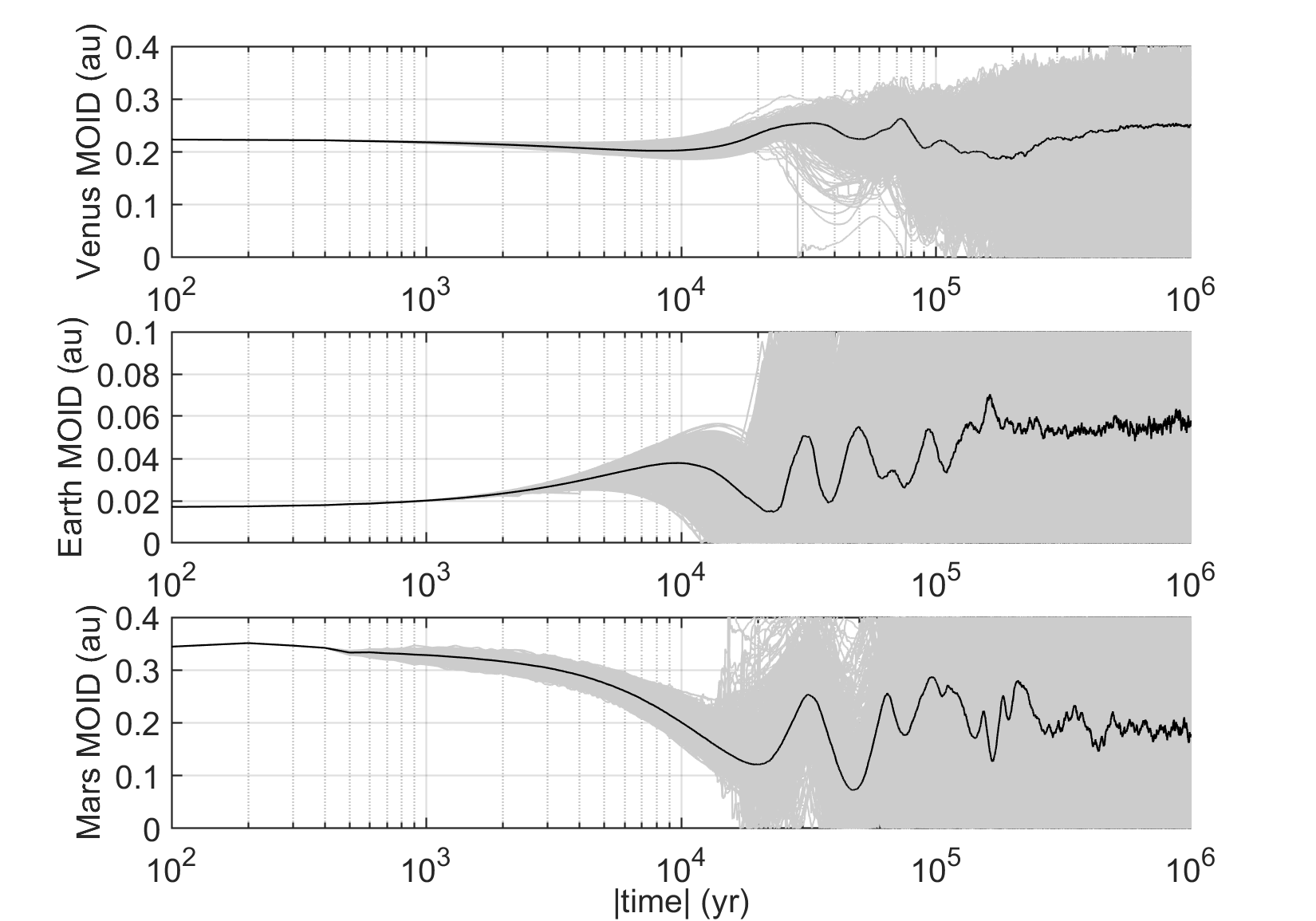}
    \caption{Orbit history of (35107) 1991 VH in the last million years. Initial conditions are given in table \ref{t:cases} as obtained from HORIZONS \citep{Giorgini} {and uncertainties in the distribution are obtained from JPL’s SSD/CNEOS Small-Body DataBase \citep{SBDB}} as  described in Appendix \ref{app:uncerts}. Elements shown are perihelion distance, eccentricity, inclination, argument of the ascending node, argument of perihelion and minimum orbit intersection distance (MOID) with Venus, Earth and Mars. Grey lines show individual simulations, black lines are the median of the 1000 simulations of each parameter shown. Only 50 example runs and the median of the full distribution are shown in orbit elements (Top 2 rows).}
    \label{fig:oe-1e6-VH}
\end{figure}

% \begin{figure}[h!]
%     \centering
%     \includegraphics[width=5.8in]{hists-VH-1e6.png}
%     \caption{Histograms of the orbit history of (35107) 1991 VH at the initial time, 100,000 years ago and 1Myr ago. Initial conditions are given in table \ref{t:cases} as obtained from HORIZONS \citep{Giorgini}) and uncertainties in the distribution are obtained from JPL’s SSD/CNEOS Small-Body DataBase \citep{SBDB}) as  described in Appendix \ref{app:uncerts}.}
%     \label{fig:hists-FG3}
% \end{figure}

\begin{figure}[h!]
    \centering
    \includegraphics[width=5.5in]{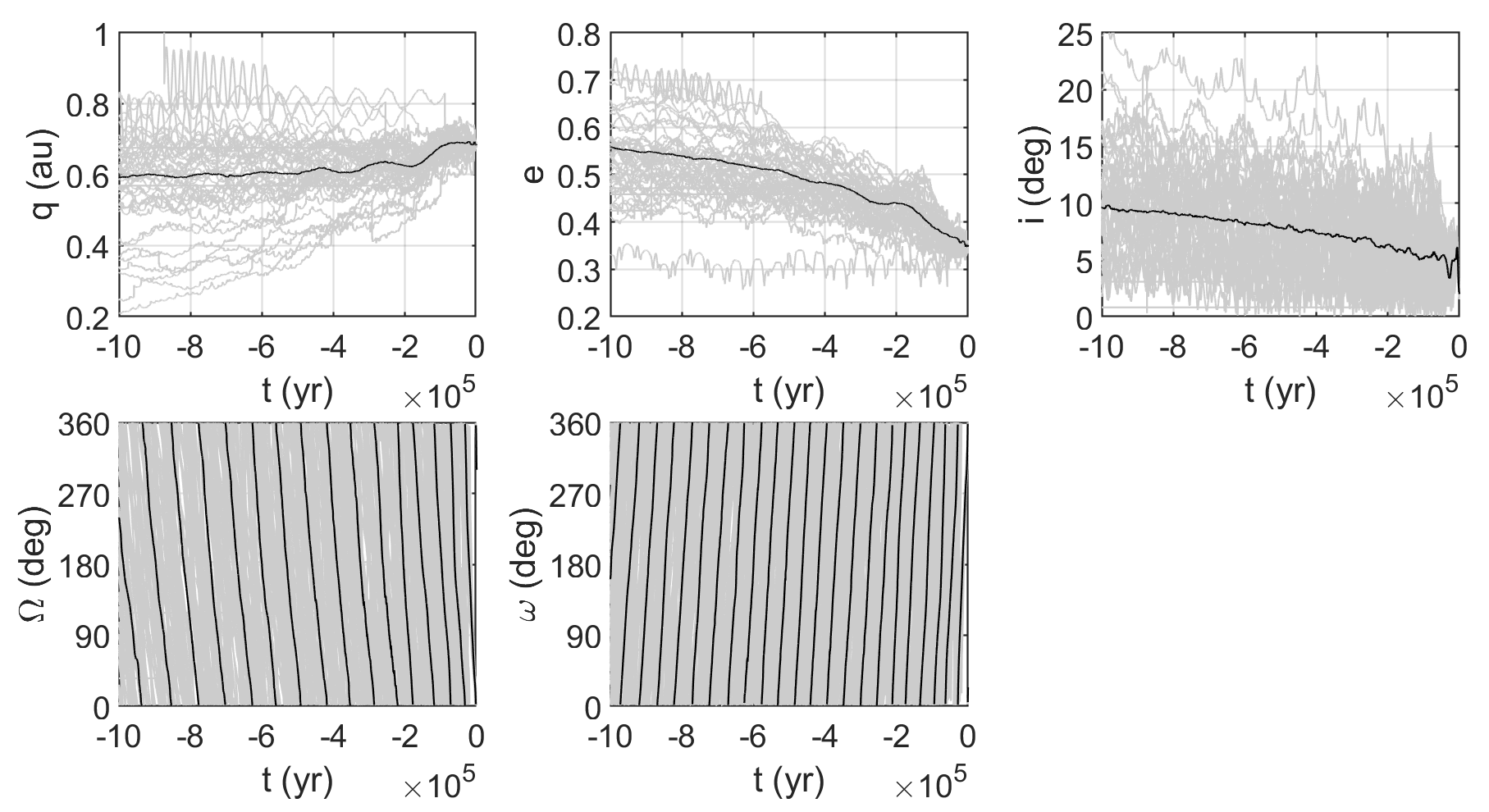}
    \includegraphics[width=5in]{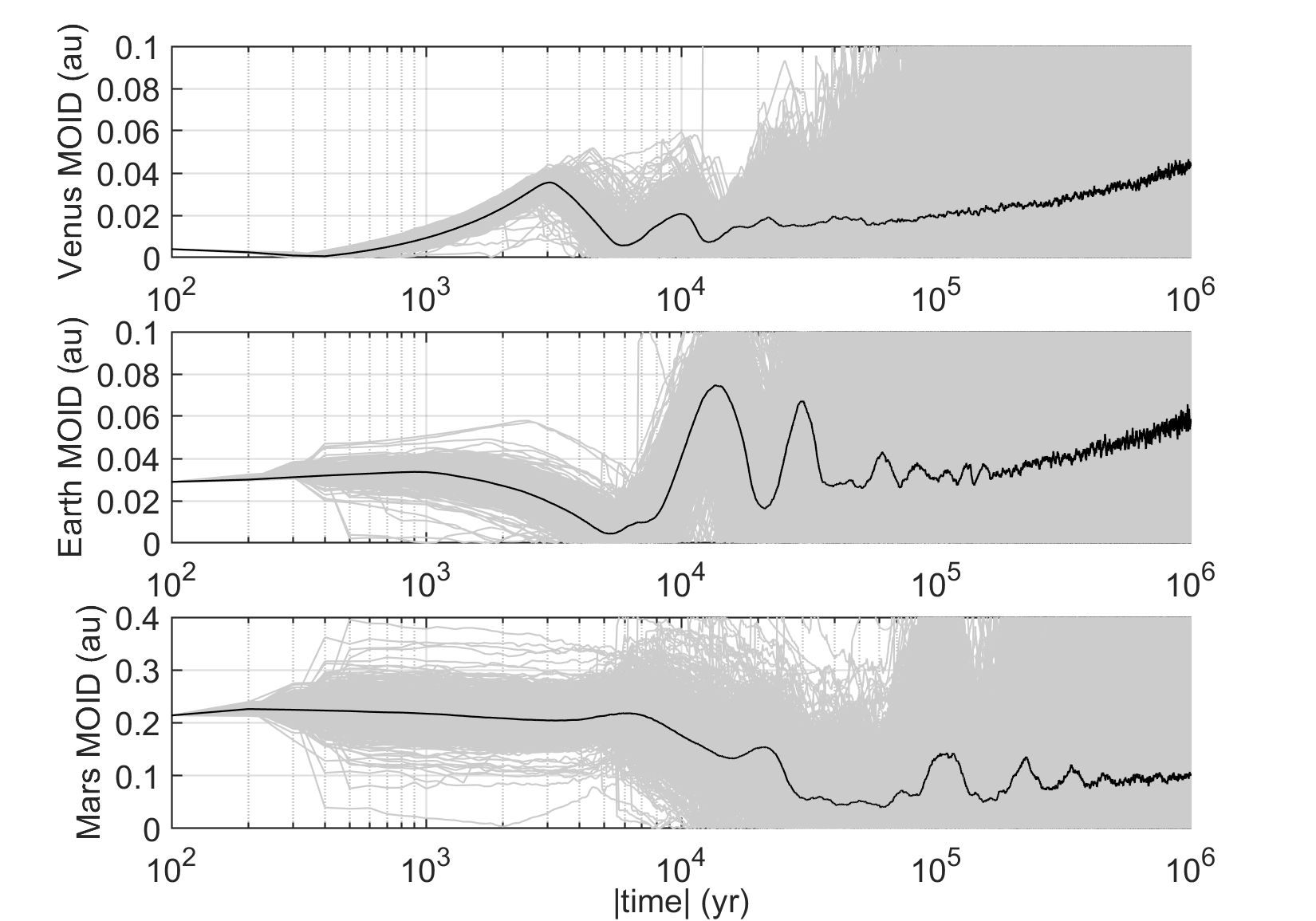}
    \caption{Orbit history of (175706) 1996 FG3 in the last million years. Initial conditions are given in table \ref{t:cases} as obtained from HORIZONS \citep{Giorgini} {and uncertainties in the distribution are obtained from JPL’s SSD/CNEOS Small-Body DataBase \citep{SBDB}} as  described in Appendix \ref{app:uncerts}. Elements shown are perihelion distance, eccentricity, inclination, argument of the ascending node, argument of perihelion and minimum orbit intersection distance (MOID) with Venus, Earth and Mars. Grey lines show individual simulations, black lines are the median of the 1000 simulations of each parameter shown. Only 100 example runs and the median of the full distribution are shown in orbit elements (Top 2 rows).}
    \label{fig:oe-1e6-FG3}
\end{figure}

The orbit time histories of (35107) 1991 VH and (175706) 1996 FG3 are shown respectively in figure \ref{fig:oe-1e6-VH} and figure \ref{fig:oe-1e6-FG3}. For clarity, we show only a subset of the runs and the median of the full distribution of 1000 runs. The minimum orbit intersection distance (MOID) is shown for the inner Solar System planets Venus, Earth and Mars. These metrics show when close encounters with these planets are possible. The presence of frequent close encounters causes the dispersion of the orbit histories. This feature manifests in the orbit history of (175706) 1996 FG3, in which the period of very low Venus MOID corresponds with a dispersion in the overall statistical representation of the orbit.

% \newpage
% \subsection{Orbit history of (175706) 1996 FG3}

\begin{figure}[h!]
    \centering
    \includegraphics[width=5.8in]{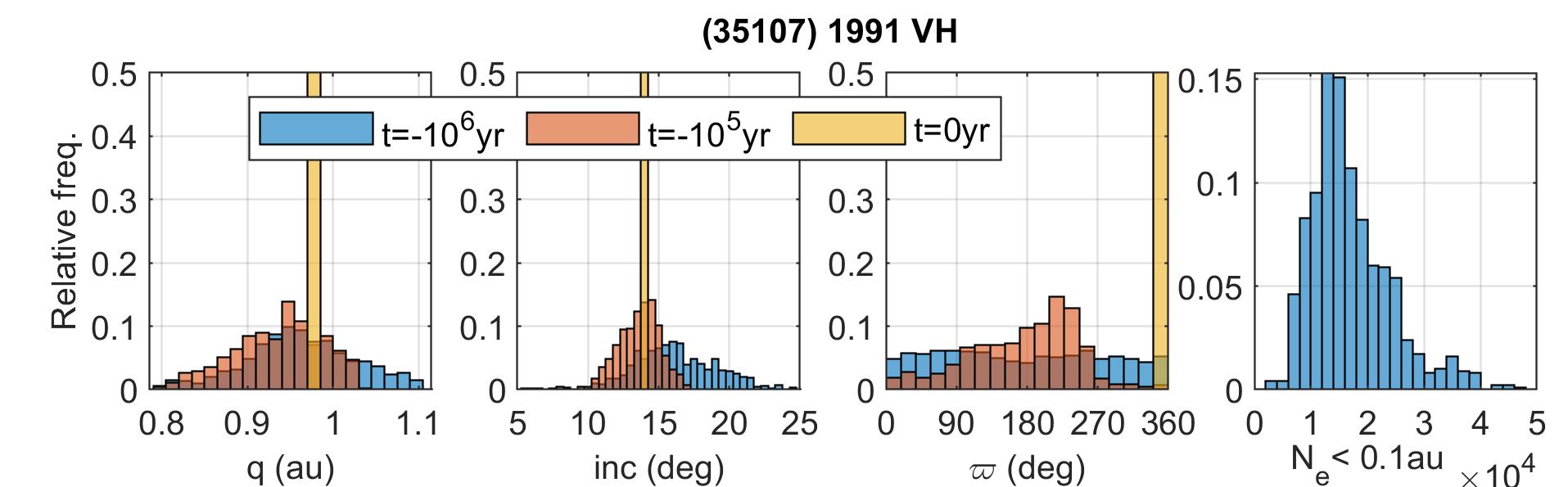}
    \includegraphics[width=5.8in]{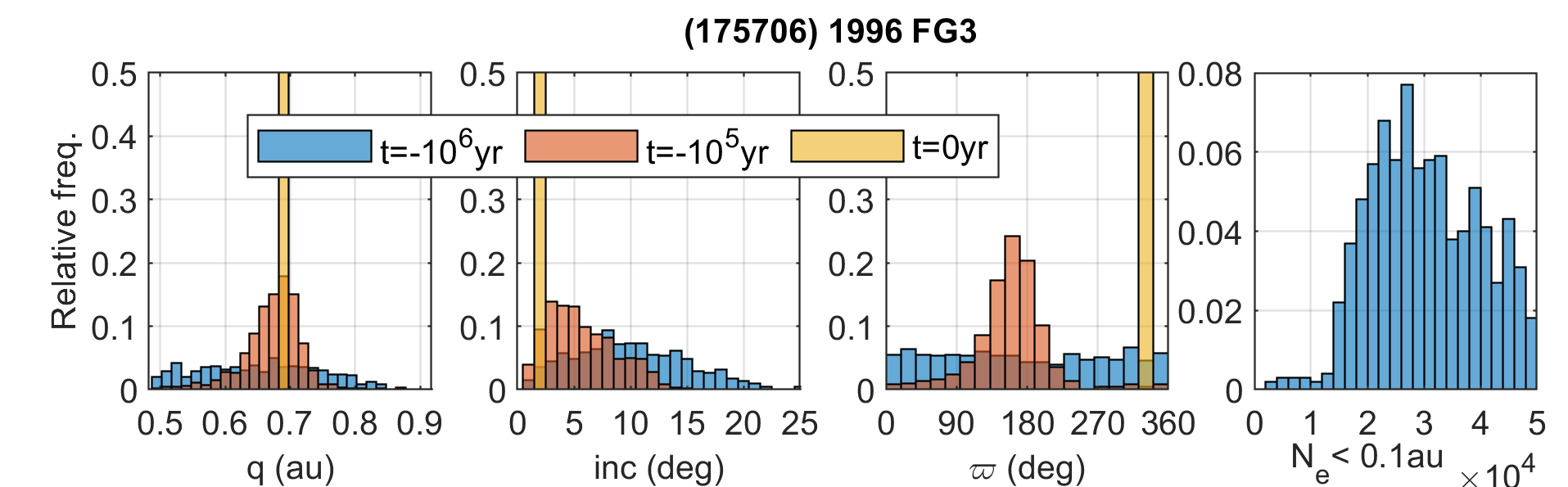}
    \caption{Histograms of the orbit history of (35107) 1991 VH and (175706) 1996 FG3 at the initial time, 100,000 years ago and 1Myr ago. Initial conditions are given in table \ref{t:cases} as obtained from HORIZONS \citep{Giorgini} {and uncertainties in the distribution are obtained from JPL’s SSD/CNEOS Small-Body DataBase \citep{SBDB}} as  described in Appendix \ref{app:uncerts}.}
    \label{fig:hists-FG3}
\end{figure}

Similarly to the long-term dynamics into the future studied in section \ref{s:4longterm-prop}, the initially very close distribution becomes a wide statistical distribution when propagated far into the past. In Figure \ref{fig:hists-FG3} we show histograms of the orbit elements and the number of encounters recorded below a closest approach distance threshold of 0.1 au. The orbit evolution of (35107) 1991 VH is mostly a spread around the initial conditions. However, (175706) 1996 FG3 is in a particular initial orbit with low inclination. On average, the very low inclination and high eccentricities drift toward a smaller eccentricity and higher inclination that are more frequent in the secular cycle. In both cases, the longitude of the perihelion becomes uniformly distributed. In the next section we characterize this uniformization process.

\newpage
\subsection{Stochastic modelling of the long-term dynamics}\label{s:stochs}

In section \ref{s:4longterm-prop} we show that we can model the long-term dynamics with a random walk in semi-major axis, eccentricity and inclination. In addition, the latter two present also the influence of the oscillations of the secular theory. We also want to study the uniformization in the longitude of the perihelion, as this process occurs with time but also with a repeated number of close encounters.

\begin{figure}[h!]
    \centering
    \includegraphics[width=5.5in]{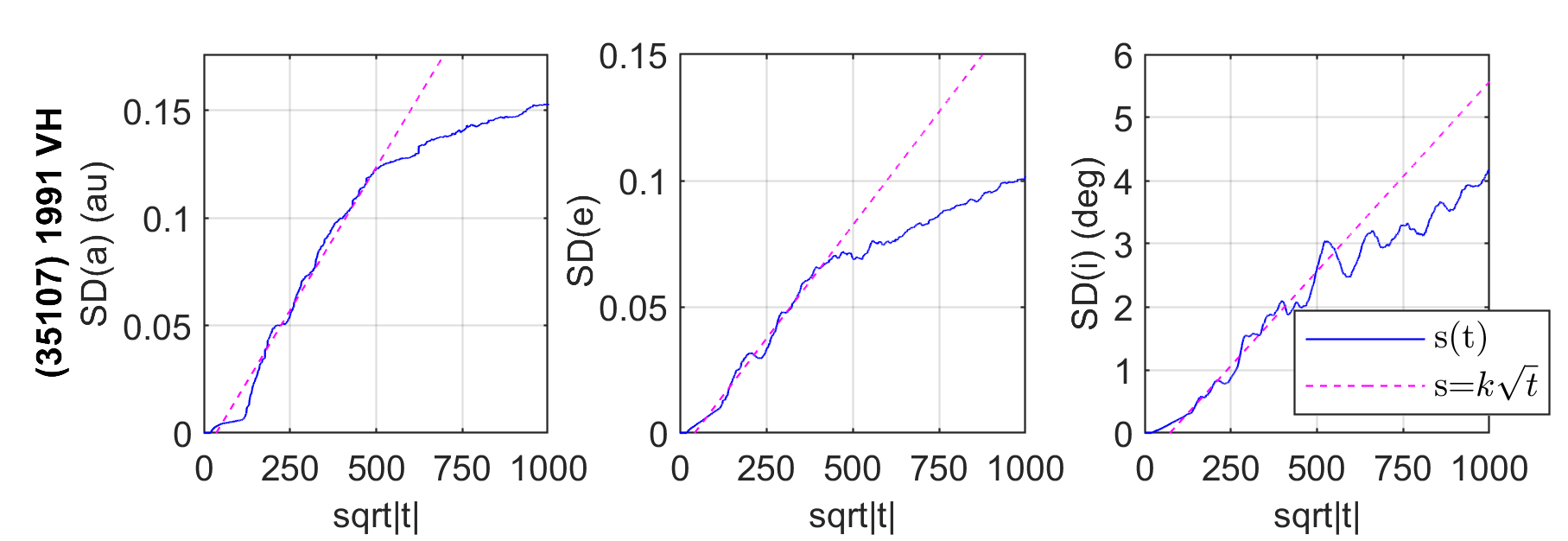}
    \includegraphics[width=5.5in]{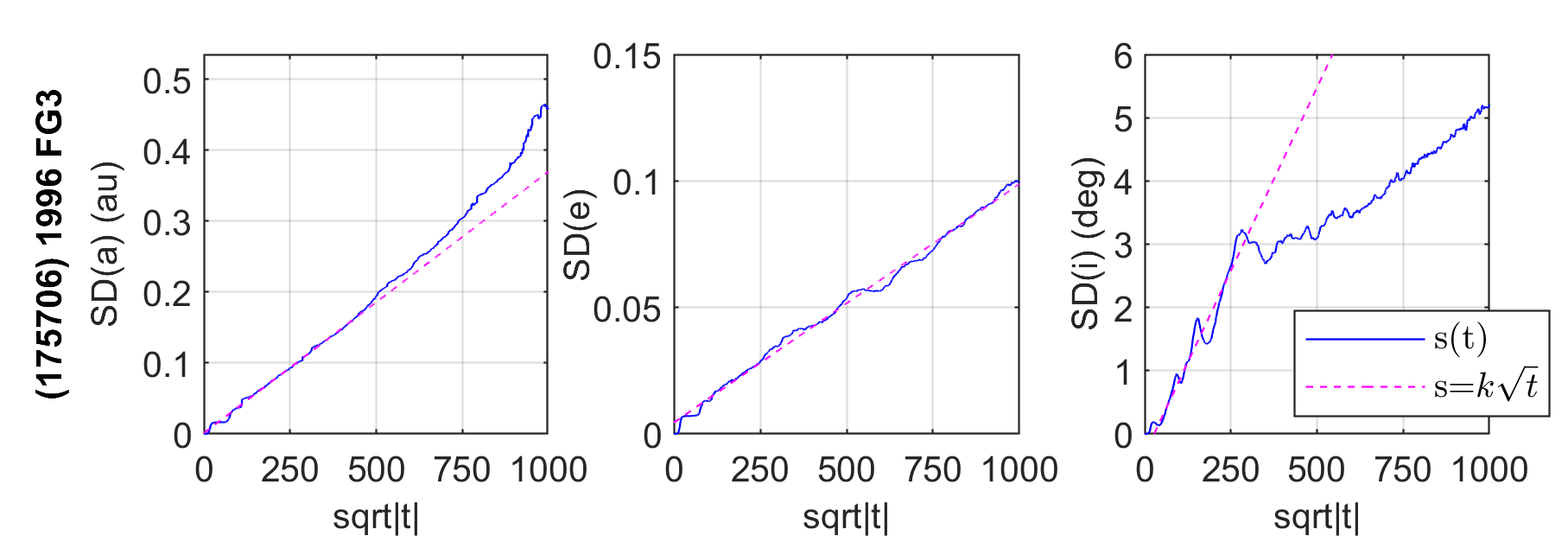}
    \caption{Random walk statistical modelling of the evolution of semi-major axis, eccentricity, inclination of (35107) 1991 VH (top) and (175706) 1996 FG3 (bottom). Standard deviation of the 1000 Monte Carlo runs as function of the square root of time and random walk model fit.}
    \label{fig:sqrt-VH}
\end{figure}

The random-walk model can be characterized by a linear increase in standard deviation with the square root of time. Figure \ref{fig:sqrt-VH} shows the standard deviation of the 1000 Monte Carlo experiment that we conducted into the past of the two Janus targets (35107) 1991 VH and (175706) 1996 FG3. We fit a linear model to the standard deviation evolution, and measure the slopes to compare the evolution of the two targets. In the case of (35107) 1991 VH we avoid the use of the full simulation time, as the standard deviation bends from the initial linear increase. {The slower increase after this period occurs when the distribution migrates from a configuration with slower encounters. The opposite case occurs with inclinations, in which the rapid increase of (175706) 1996 FG3 from the low-inclination initial regime slows down after the initial growth.}

The measured slopes are reported in table \ref{tab:janustock}, showing that the random walk of (175706) 1996 FG3 is faster in semi-major axis and inclination. However, because of the bends in the progression after a few hundred thousand years, the final standard deviations are not substantially larger than the ones of (35107) 1991 VH after the million years into the past in eccentricity and inclination.

\begin{deluxetable*}{cccccc}[htb!]
\tablenum{3}
\tablecaption{Stochastic modelling of the long-term dynamics of the Janus targets\label{tab:janustock}}
\tablewidth{0pt}
\tablehead{
\colhead{Target} & \multicolumn3c{Random walk $k_{x}$ constant ($s_{x}=k_{x}\sqrt{t}$)} & \multicolumn2c{$\varpi$ Uniformization} \\
\colhead{Name} & \colhead{$k_{a}$} & \colhead{$k_{e}$} & \colhead{$k_{i}$} &
\colhead{Time} & \colhead{Encounters $< 0.1$ au}  \\
\colhead{} & \colhead{(au/$\sqrt{yr}$) $\cdot 10^{-3}$} & \colhead{(1/$\sqrt{yr}$) $\cdot 10^{-3}$} & \colhead{(deg/$\sqrt{yr}$) $\cdot 10^{-3}$ } &
\colhead{$yr\cdot 10^{3}$} & \colhead{Number}}
% \colhead{Number} & \colhead{Number} & \nocolhead{Name} & \colhead{Type} &
% \multicolumn2c{(kpc)} & \colhead{Constellation} }
%\decimalcolnumbers
\startdata
(35107) 1991 VH   & 0.2661 & 0.1799 & 6.0055 & -434 & 11800  \\ \hline
(175706) 1996 FG3 & 0.3688 & 0.0944 & 11.598 & -797 & 29800  \\
\enddata
%\tablecomments{cumantaris varis.}
\end{deluxetable*}

\begin{figure}[h!]
    \centering
    \includegraphics[width=5.8in]{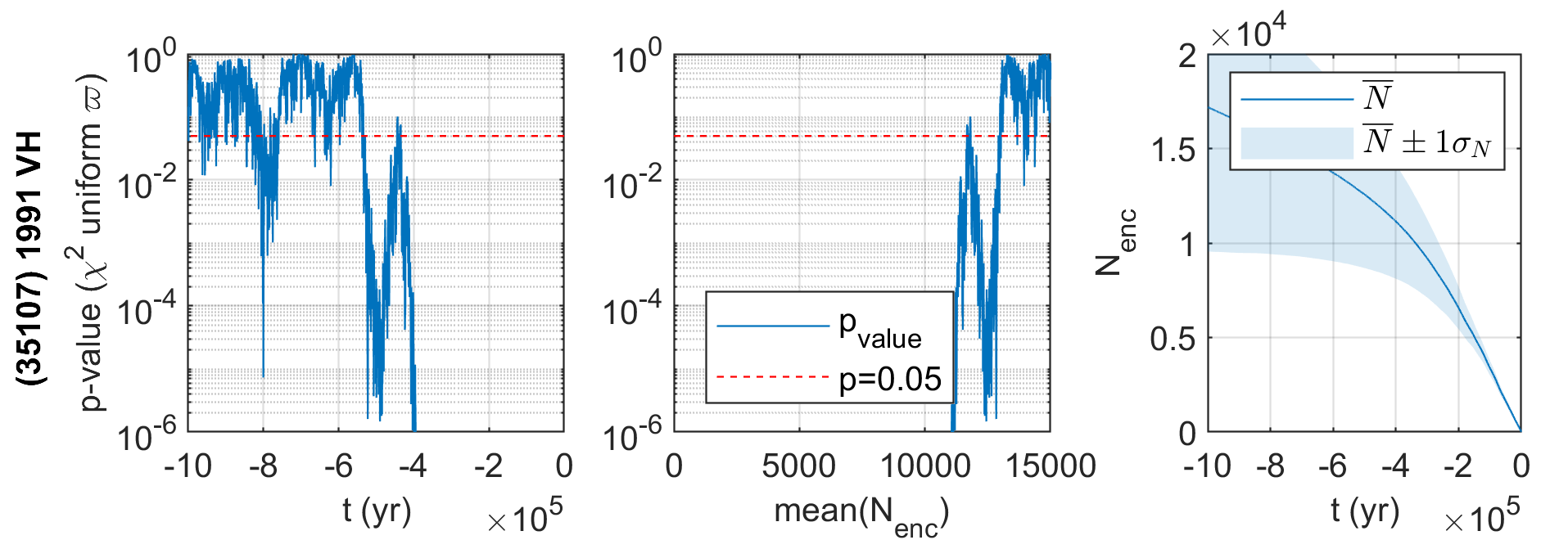}
    \includegraphics[width=5.8in]{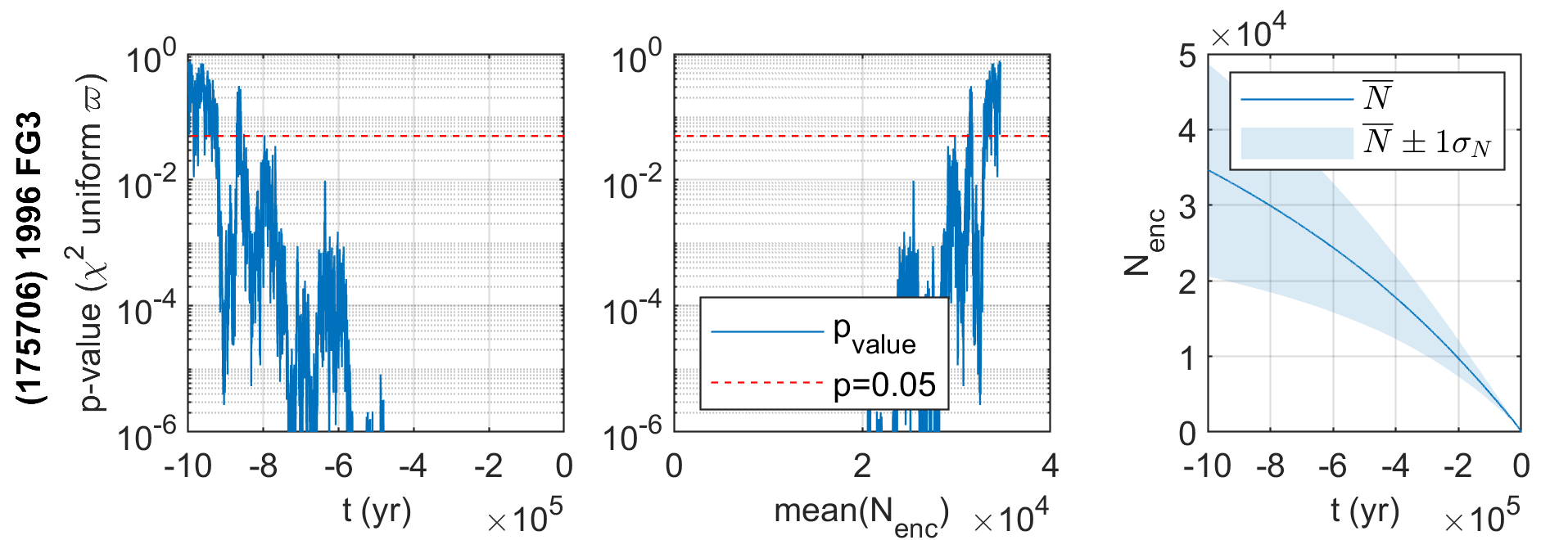}
    
    \caption{P-value of the chi-squared test of the uniform distribution of the longitude of perihelion $\varpi = \omega + \Omega$ of (35107) 1991 VH (top) and (175706) 1996 FG3 (bottom). P-value is shown over the past 1Myr (left) and as function of the mean number of encounters (center). The mean number of encounters of the 1000 Monte Carlo runs is shown over time the corresponding with 1-$\sigma$ bounds (right).}
    \label{fig:pval-VH}
\end{figure}

The process of uniformization of the longitude of perihelion is shown in figure \ref{fig:pval-VH}. We conduct the chi-squared test of the uniform distribution over the 1Myr simulation, to find when the hypothesis of the uniform distribution is significant. In table \ref{tab:janustock} we show the first time in which this criterion is satisfied, both in time and mean number of encounters: -434,000 years and a mean of 11800 encounters for (35107) 1991 VH, and -797,000 years and a mean of 29800 encounters for (175706) 1996 FG3. 

The uniformization of (35107) 1991 VH is faster than the uniformization of (35107) 1991 VH in both time and mean number of encounters. It is remarkable that {(35107) 1991 VH} takes a much lower mean number of encounters. This is explained by the faster relative velocities of the encounters of (175706) 1996 FG3 and a larger fraction occuring with Mars, a less massive planet. The relative velocities of a few of the recorded flybys are shown in figures \ref{fig:disrupt-VH} and \ref{fig:disrupt-FG3} in the context of studying the probability that a close encounter could potentially disrupt the binaries. 

{The comparison between the two binary systems highlights how the effect of the encounters depends on the relative velocities and the mass of the planets. In general, slow encounters and with larger planets are more efficient at causing the distributions to become uniform. However, depending on the heliocentric orbit these encounters may be more or less frequent. Thus, leveraging both effects is required to obtain a stochastic representation of the long-term dynamics of NEOs under frequent encounters.}

% \begin{figure}[h!]
%     \centering
%     \includegraphics[width=5.5in]{sqrtt-J2.png}
%     \caption{Random walk statistical modelling of the evolution of semi-major axis, eccentricity, inclination of (175706) 1996 FG3. Standard deviation of the 1000 Monte Carlo runs as function of the square root of time and random walk model fit.}
%     \label{fig:sqrt-FG3}
% \end{figure}

% \begin{figure}[h!]
%     \centering
%     \includegraphics[width=5in]{pval-J2.png}
%     \caption{P-value of the chi-squared test of the uniform distribution of the longitude of perihelion $\varpi = \omega + \Omega$ of (175706) 1996 FG3. P-value is shown over the past 1Myr (left) and as function of the mean number of encounters (right).}
%     \label{fig:pval-FG3}
% \end{figure}

\newpage
\subsection{Potentially disruptive planetary encounters}

The two binary targets of the Janus mission present different relative orbits as observed by radar and photometry \citep{pravec2016binary,2021DDA....5240506M}, showing that (175706) 1996 FG3 is in a stable orbital state and (35107) 1991 VH is in a chaotic state. The perturbed state of (35107) 1991 VH could be explained by a recent very close encounter with the inner Solar System planets \citep{HeggieRasio1996}. Thus, it is of interest to characterize the frequency of such encounters in the orbital history of asteroid binaries.

{Using the semi-analytical propagation tool we obtain the history of flybys over a long time period of time. The perturbation in the orbit of a binary system during a planetary close encounter is studied in detail as described in \citep{MEYER2021114554}. In this section we combine both results to predict the frequency of a disrupting flyby.}

{The effect of the close encounter on the binary can be modelled as an impulsive variation in the binary Keplerian elements. In \cite{MEYER2021114554}  the effect of close encounters to singly synchronous binary asteroids is studied. The variation in semi-major axis, eccentricity, and inclination obtained with numerical methods was compared to analytical expressions for the impulsive variation in binary Keplerian elements \citep{fang2011binary}. We used these analytical expressions as they provide an estimate of the variation as function of the relative velocity and distance of closest approach.}

\begin{figure}[b!]
    \centering
    % %\includegraphics[width=4.5in]{dCA-disr-Earth4.png}
    % \includegraphics[width=4.5in]{fbs-vh-earth.png}
    % %\includegraphics[width=4.5in]{dCA-disr-Earth.png}
    % %\includegraphics[width=4.5in]{dCA-disr-Mars2.png}
    %\includegraphics[width=4.5in]{fbs-vh-mars.png}
    
    \includegraphics[width=4.5in]{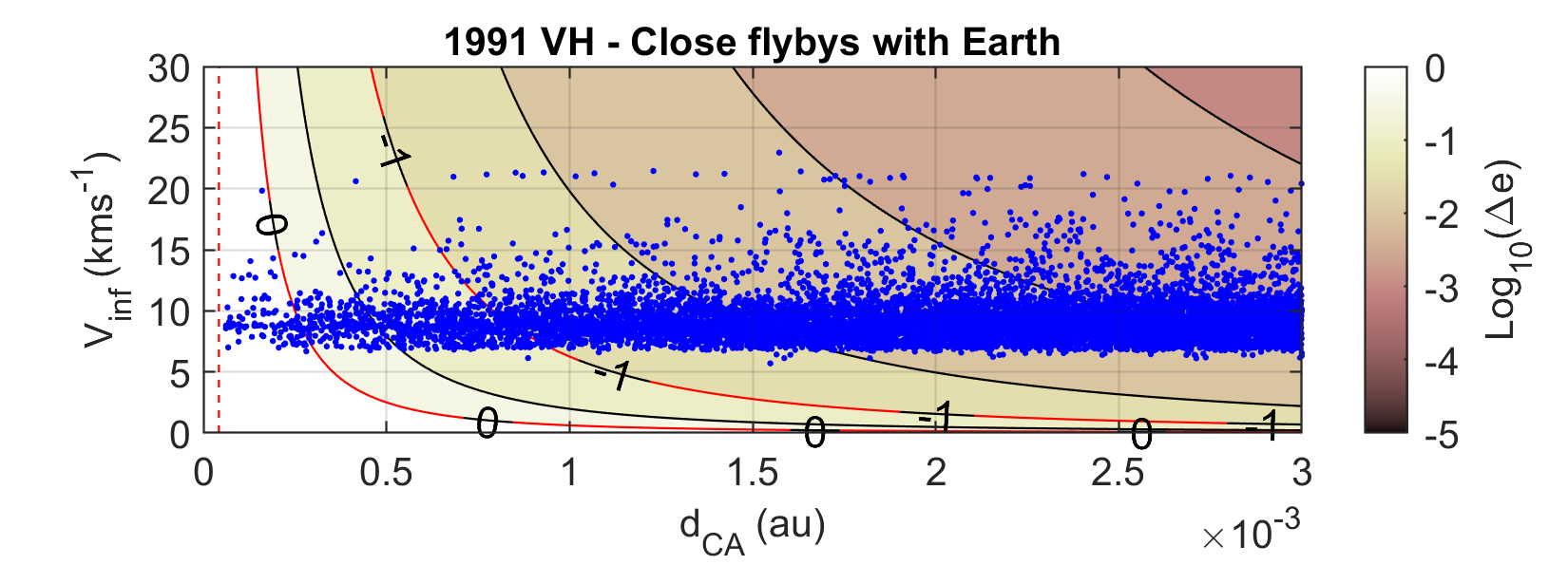}
    \includegraphics[width=4.5in]{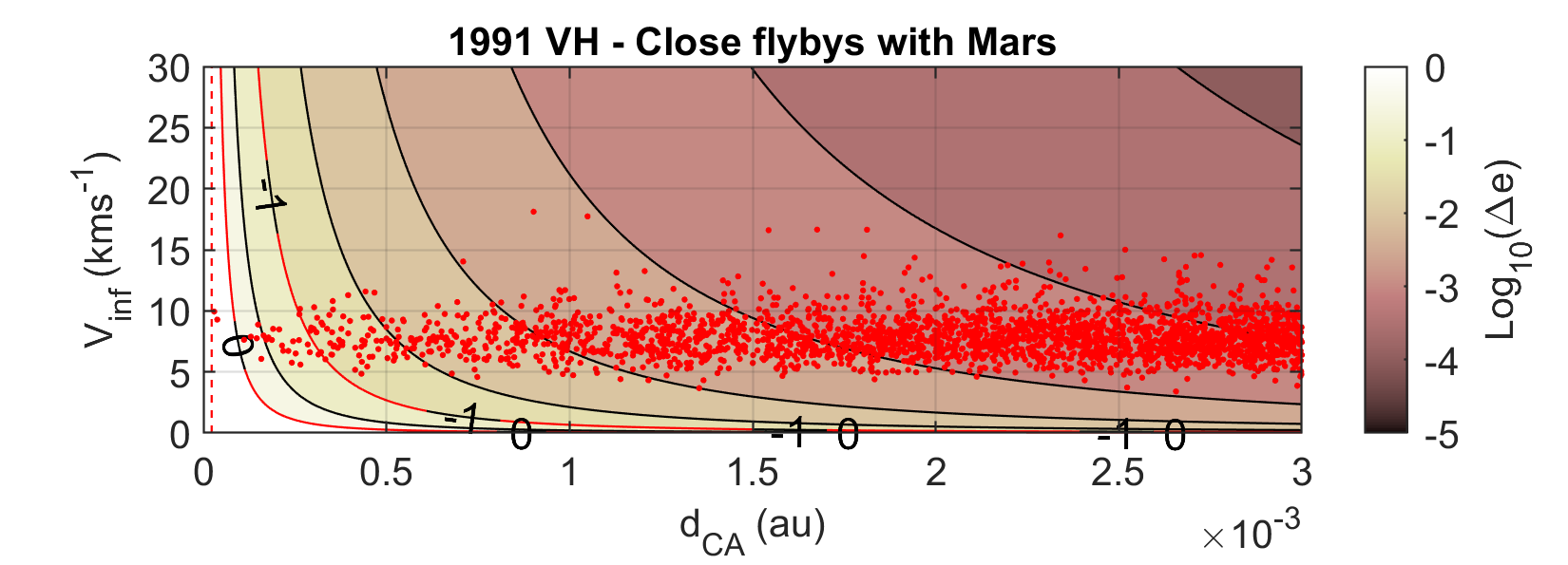}
    
    \caption{Potentially disruptive encounters recorded by the 1000 virtual (35107) 1991 VH binaries during a million years into the past. The background contour represents the logarithm of the mean variation in binary eccentricity during a close encounter with a planet: Earth (blue, top) or Mars (red, bottom). The encounters found during the propagation in this threshold are shown by their closest approach distance (au) and $V_\infty$ (km s$^{-1}$). The radius of the planet is shown as a dashed line. The initial uncertainty distribution of (35107) 1991 VH is detailed in appendix \ref{app:uncerts}.}
    \label{fig:disrupt-VH}
\end{figure}

\begin{figure}[t!]
    \centering
    \includegraphics[width=4.5in]{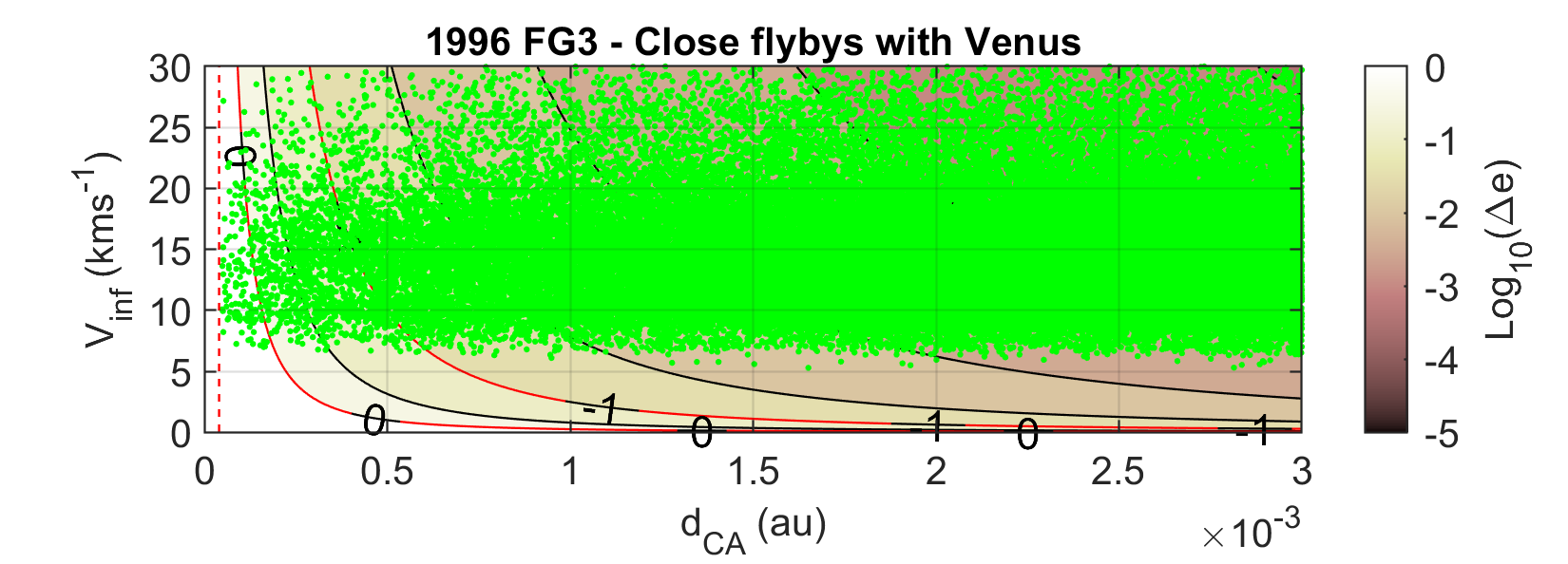}
    \includegraphics[width=4.5in]{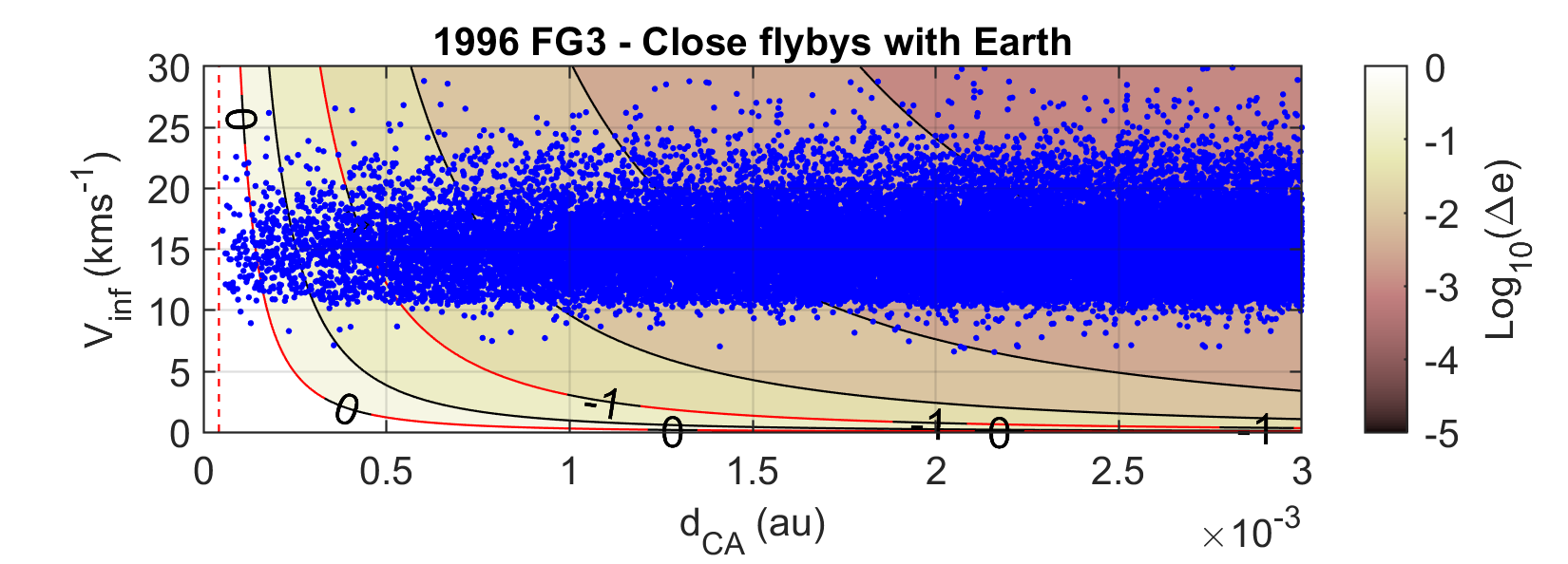}
    \includegraphics[width=4.5in]{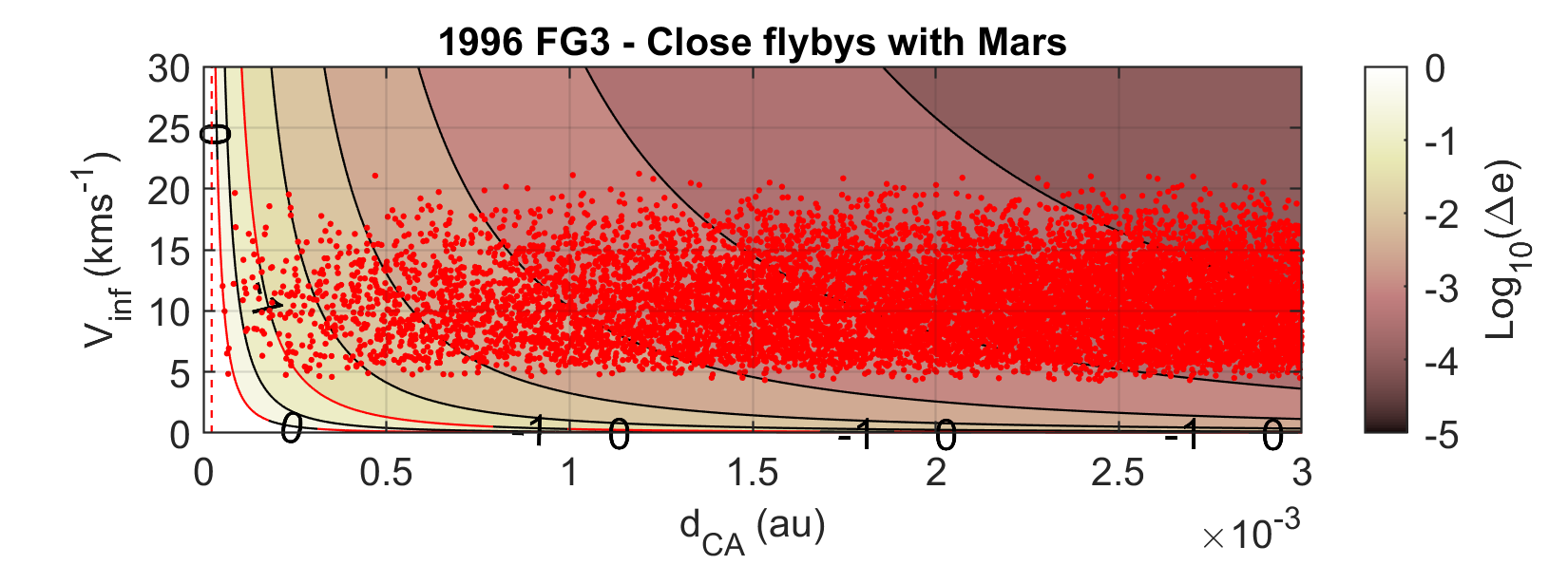}
    \caption{Potentially disruptive encounters recorded by the 1000 virtual (175706) 1996 FG3 binaries during a million years into the past. The background contour represents the logarithm of the mean variation in binary eccentricity during a close encounter with a planet: Venus (green, top), Earth (blue, center) or Mars (red, bottom). The encounters found during the propagation in this threshold are shown by their closest approach distance (au) and $V_\infty$ (km s$^{-1}$). The radius of the planet is shown as a dashed line. The initial uncertainty distribution of (175706) 1996 FG3 is detailed in appendix \ref{app:uncerts}.}
    \label{fig:disrupt-FG3}
\end{figure}

For every binary and encountered planet we can generate contours of the  variation of the eccentricity. In Figures \ref{fig:disrupt-VH} and \ref{fig:disrupt-FG3} we show these contours respectively for (35107) 1991 VH and (175706) 1996 FG3. In every figure we highlight two levels, a variation of 0.1 in eccentricity, and a variation of 1, which would mean that the binary is completely separated. The probability of disruption in the binary orbit increases as the relative velocity and closest approach distance are reduced.

Using the semi-analytical propagation tool we track the close encounters below the threshold of 0.003 au, above which the variation in the binary Keplerian elements becomes negligible. For each binary we generate 1000 trajectories for a million years into the past. All the encounters that are found in this threshold are plotted in figures \ref{fig:disrupt-VH} and \ref{fig:disrupt-FG3} and separated by encountered planet.

The encounters potentially disruptive recorded for (35107) 1991 VH with Earth and Mars are shown in figure \ref{fig:disrupt-VH}. Less than 1\% of the recorded encounters are with Venus so the map with this planet is not included. The relative velocities of the flybys are mostly found between 5 and 15 km s$^{-1}$. In the case of (175706) 1996 FG3 this range of possible relative velocities is larger in all the planets. In addition, many encounters are found with quite slower $V_\infty$, which means that even if they are not as close, they can still potentially cause a disruption.

As we observed in figure \ref{fig:nenc-time-cases}, (175706) 1996 FG3 experiences more frequent encounters. However, the regions in which the encounters are potentially disruptive depend on the current orbital configurations of the binaries. In this case, (175706) 1996 FG3 requires closer and slower encounters to obtain the same mean variation in binary elements.

\begin{figure}[h!]
    \centering
    % \includegraphics[width=5.8in]{ddis-janus.png}
    % % \includegraphics[width=5.8in]{prob-disr-VH-FG3-ci.png}
    % \includegraphics[width=5.8in]{pdis-janus.png}
    \includegraphics[width=5.8in]{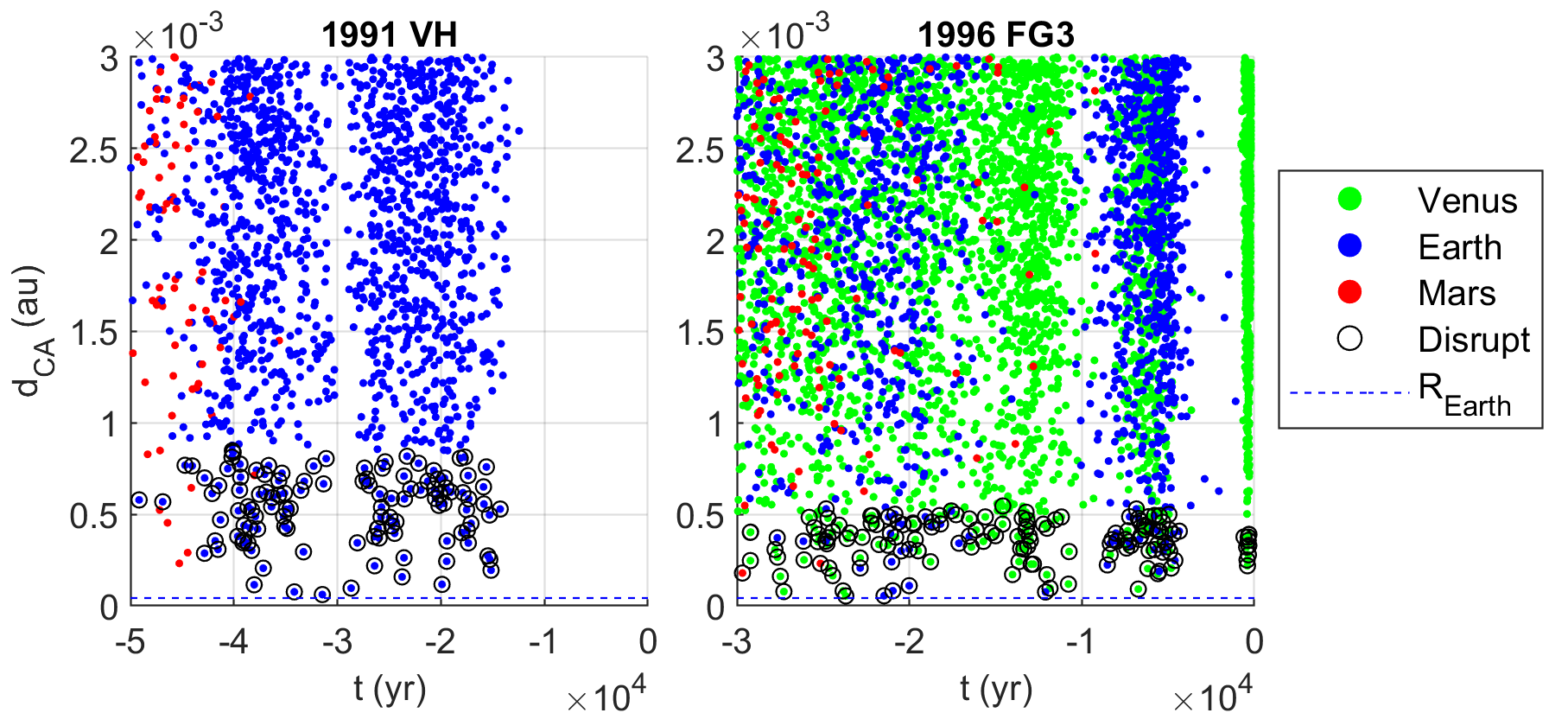}
    \includegraphics[width=5.8in]{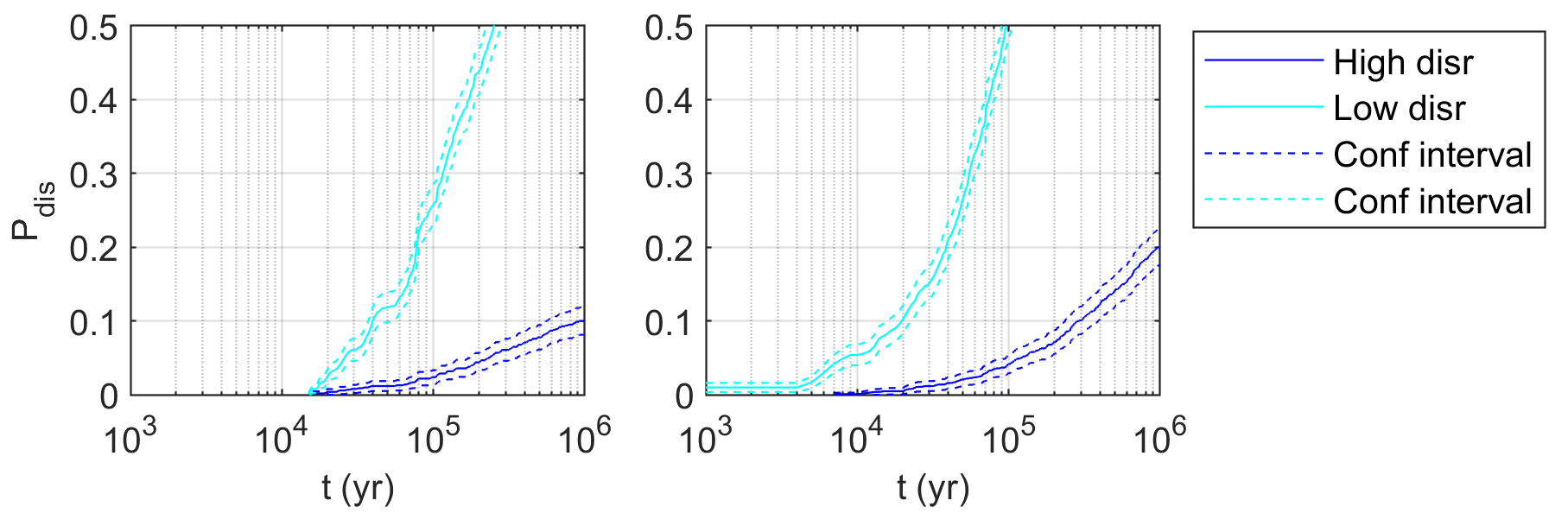}
    
    \caption{(Top) Potentially disruptive encounters time history in the most recent periods of close encounters of (35107) 1991 VH and (175706) 1996 FG3. In the most recent 50,000 years, (35107) 1991 VH experiences Earth and Mars encounters. In the latest 10,000 years, (175706) 1996 FG3 experiences Venus and Earth encounters. The close encounters below the line of mean variation in binary eccentricity of 0.1 are highlighted with a black circle and the dashed line marks the radius of Earth. (Bottom) Probability of disruption based on the number of encounters found below thresholds of mean $\Delta e=0.1$ for low disruption and $\Delta e=1$ for high disruption. Confidence intervals on the probability of disruption prediction are shown in dashed lines.}
    \label{fig:disrupt-times}
\end{figure}

%\newpage
Considering the orbit history in a million years, a non-negligible probability exists that both binaries have been potentially disrupted at some point. However, it is possible that the signature of these potential disruptions vanish if there is energy dissipation in the system. Thus it is relevant to study the probability that the binary orbits have potentially been disrupted recently in the orbit histories. We can study the first fraction of the long-term secular periods, in which the particles still have not mixed. Figure \ref{fig:disrupt-times} shows the history of potentially disruptive encounters in the recent periods of frequent encounters.

\newpage
The last period of possible very close encounters that we find for (35107) 1991 VH starts beyond the last 10,000 years. As determined by a mean variation in binary eccentricity of 0.1, we find that 61 of the 1000 test runs experience a potential disruption in the period of time up until the last 30,000 years. When we incorporate the next period of close encounters, the probability of a potential disruption increases to 131 out of 1000 runs in the last 60,000 years.

Given the current orbit state of (175706) 1996 FG3, we find very close encounters with Venus in the last millennia. Because of the more restrictive closest approach distance for a potential disruption to happen, only 10 out of the 1000 test runs experienced this potential disruption in this period of frequent encounters. If we consider the last 10,000 years, then the probability increases to 54 cases in which at least one potentially disruptive encounter was found.

If we keep increasing the time in which we consider all the potentially disruptive encounters, we can estimate the probability that a potentially disruptive encounter occurs. These probabilities are shown in figure \ref{fig:disrupt-times} with the corresponding 95\% confidence intervals. The probability of suffering a disruptive encounter of (175706) 1996 FG3 increases faster than the probability of (35107) 1991 VH.  This is explained by the significantly higher number of recorded close encounters. Thus, it is not possible to explain the chaotic state of (35107) 1991 VH only from the long-term probability of experiencing such encounters. However, a low probability in recent times combined with the incapability to dissipate the perturbation in a long time could explain the chaotic state of (35107) 1991 VH. Thus, future work will be done in the lines of characterizing the timescales of the dissipation of perturbations due to close encounters.

% Note that dissipation of it will be needed to explain it's current states!

% \begin{itemize}
%     \item VH: Venus not plotted, some conclusions
%     \item FG3: some conclusions, Range in relative velocities, more restrictive in required flybys
%     \item Overall, in 1Myr, very likely. But, secular periods make us wonder, what about recently?
%     \item Recent disruption study
% \end{itemize}

%\subsection{(35107) 1991 VH}
%Note that previous figure was into the future, now figure is into the past, AND a million years, and 1000 runs.
%Venus not plotted (only X flybys)

%\subsection{(175706) 1996 FG3}

%\subsection{Recent disruption of (35107) 1991 VH and (175706) 1996 FG3}

% {\color{red}    
% Future work ideas that we have from past brainstorming:
% \begin{itemize}
%     \item Variety of analytical solutions: Lidov-Kozai, mean-motion resonances and select among them based on elements
%     \item Compute proper elements as initial conditions for analysis. How does it compare to representing an asteroid by stochastics?
%     \item Use simplified search of encounters as function of MOID map. We skipped this work to make it more general, but this study is also relevant to the computation of the probabilities of collision.
% \end{itemize}}

%========================================================
% FINAL SECTIONS
%========================================================

\newpage
\section{Conclusions} \label{s:6conclusions}

%{\color{red} Update conclusions with new sections!}

In this paper we present a rapid semi-analytical propagation tool for asteroids in the inner Solar System. The tool combines an analytical solution for the secular dynamics and the evaluation of planetary encounters. The derived solution of planetary encounters proves to accurately model the effect of the  majority of flybys that asteroids experience in the inner Solar System. 

The long-term effect of the perturbation by Jupiter is captured by the analytical secular solutions in a large fraction of the NEO population. The combination with detection and evaluation of close encounters recreates the full dynamics as we demonstrate for the case of (35107) 1991 VH. 
%The secular drift causes near-Earth asteroids to experience periods of time without close encounters. 

%The analytical modelling of the asteroid allows the prediction of dynamical properties without the precise propagation.  Thus we use the secular propagation to compute how common it is that close encounters within a certain threshold are possible. We find that the system trends to be well described by the ergodic definition.

%Since the secular drift causes the asteroid to completely visit the space of angles we can assume the trajectory to be ergodic. We use this assumption to predict long-term properties sparing the use of computationally expensive Monte-Carlo simulations. In this work we predicted the frequency of close encounters and how often (35107) 1991 VH has been a PHA.

The description of the orbits of NEOs in long-term timescales must be done statistically. We showed how the different elements can be represented by different distributions, and how the time it takes for the elements to become statistical depends on the frequency of close encounters. Through the sampling of different NEO cases we inspect the stochastic models that represent the long-term evolution of the orbital elements.

The use of a fast semi-analytical propagation tool allows an efficient study of the dynamics of asteroids. We studied in detail the orbital histories of the Janus mission targets: (35107) 1991 VH and (175706) 1996 FG3. 
%In order to generalize the study we shall consider multiple analytical solutions of the perturbation. For example, this would allow to characterize asteroids under resonances or the Lidov-Kozai mechanism. However, this was left outside the scope of this paper and is left as future work. 
We characterized the encounters that can cause a potentially disruption of the binary orbits and computed the frequency of such encounters.%, to find that there is a small probability that the Janus targets have been

Additional modeling of the effects of close encounters to other physical properties of asteroids will allow the study of the frequency of disruptive events. These are just a few potential examples of the benefits of a fast propagation tool for Solar System studies in the fashion of the presented tool.

%\note[OFM]{In order to find the timescales of such encounters we need to account for the probability of the timing of the flyby to be right.}

% {\color{red} Notes to re-write:
% \begin{itemize}
%     \item We derive and demonstrate a new semi-analytical method for rapid propagation of NEO orbits. 
%     \item We show that NEO orbits become statistical over time, with the different elements showcasing different distributions. 
%     \item We characterize the statistical nature of NEO
% \end{itemize}}

%\newpage
%\section{Acknowledgments}
%MOID computation algorithm generously provided by Jesus Peláez from Space Dynamics Group, ETSIAE, UPM. Simulations in this paper made use of the REBOUND code which is freely available at \url{http://github.com/hannorein/rebound}. This work utilized resources from the University of Colorado Boulder Research Computing Group, which is supported by the National Science Foundation (awards ACI-1532235 and ACI-1532236), the University of Colorado Boulder, and Colorado State University. This research has made use of data and/or services provided by the International Astronomical Union's Minor Planet Center. This research is funded by the Balsells Fellowship Program at the University of Colorado Boulder and by a grant from the California Institute of Technology / Jet Propulsion Laboratory. 

\newpage
\appendix
\section{Initial Conditions and uncertainties}\label{app:uncerts}

The statistical representation of the uncertainty in the orbit solutions can be done using the covariance matrix. {This information is available for multiple asteroids in  JPL’s SSD/CNEOS Small-Body DataBase \citep{SBDB}}. The covariance matrices for (175706) 1996 FG3 and (35107) 1991 VH that are used in this work are:

\begin{deluxetable*}{ccccccc}[h]
\tablenum{4}
\tablecaption{Initial covariance of the orbit of NEO binary (35107) 1991 VH}\label{t:uncertVH}
\tablewidth{0pt}
\tablehead{
\colhead{1991 VH} & \colhead{e}  & \colhead{q (au)} & \colhead{$t_p$ (TDB)} & \colhead{$\Omega$ (deg)} & \colhead{$\omega$ (deg)} & \colhead{$i$ (deg)} }
\startdata
e                & 3.0691e-16             & -3.5095e-16                 & -8.8918e-14                      & -7.2651e-15                         & -5.2217e-14                         & 3.6255e-15                     \\ \hline
q (au)           & -3.5095e-16            & 4.0175e-16                  & 1.0484e-13                       & 8.2268e-15                          & 6.1722e-14                          & -4.1006e-15                    \\ \hline
$t_p$ (TDB)      & -8.8918e-14            & 1.0484e-13                  & 7.5479e-11                       & -7.8662e-12                         & 6.4146e-11                          & -3.8102e-12                    \\ \hline
$\Omega$ (deg)   & -7.2651e-15            & 8.2268e-15                  & -7.8662e-12                      & 3.104e-11                           & -3.4282e-11                         & -4.876e-12                     \\ \hline
$\omega$ (deg)   & -5.2217e-14            & 6.1722e-14                  & 6.4146e-11                       & -3.4282e-11                         & 8.1951e-11                          & 1.9766e-12                     \\ \hline
$i$ (deg)        & 3.6255e-15             & -4.1006e-15                 & -3.8102e-12                      & -4.876e-12                          & 1.9766e-12                          & 8.4244e-12                     \\ \hline
\enddata
\tablecomments{{As obtained from JPL’s SSD/CNEOS Small-Body DataBase \citep{SBDB}}. Using DE431 and SB431-N16, orbit solution date 2021 April 15 for {epoch JD = 2456902.5.}}
\end{deluxetable*}

\begin{deluxetable*}{ccccccc}[h]
\tablenum{5}
\tablecaption{Initial covariance of the orbit of NEO binary (175706) 1996 FG3}\label{t:uncertFG3}
\tablewidth{0pt}
\tablehead{
\colhead{1996 FG3} & \colhead{e}  & \colhead{q (au)} & \colhead{$t_p$ (TDB)} & \colhead{$\Omega$ (deg)} & \colhead{$\omega$ (deg)} & \colhead{$i$ (deg)} }
\startdata
e                & 1.2362e-16             & -1.3026e-16                 & 7.9966e-15                       & -4.1649e-14                         & 3.9787e-14                          & -2.3726e-14                    \\ \hline
q (au)           & -1.3026e-16            & 1.3752e-16                  & -9.0726e-15                      & 4.7436e-14                          & -4.5397e-14                         & 2.5036e-14                     \\ \hline
$t_p$ (TDB)      & 7.9966e-15             & -9.0726e-15                 & 2.9877e-12                       & 6.6638e-11                          & -6.6225e-11                         & -2.4698e-12                    \\ \hline
$\Omega$ (deg)   & -4.1649e-14            & 4.7436e-14                  & 6.6638e-11                       & 7.0256e-09                          & -6.9647e-09                         & -6.8055e-11                    \\ \hline
$\omega$ (deg)   & 3.9787e-14             & -4.5397e-14                 & -6.6225e-11                      & -6.9647e-09                         & 6.9046e-09                          & 6.7284e-11                     \\ \hline
$i$ (deg)        & -2.3726e-14            & 2.5036e-14                  & -2.4698e-12                      & -6.8055e-11                         & 6.7284e-11                          & 6.7136e-12                     \\ \hline  
\enddata
\tablecomments{{As obtained from JPL’s SSD/CNEOS Small-Body DataBase \citep{SBDB}}. Using DE431 and SB431-N16, orbit solution date 2021 April 26 {for epoch JD = 2454796.5.}}
\end{deluxetable*}

In the case of the artificial cases used to illustrate the long-term dynamics, we set the covariance matrix to be a diagonal matrix of 1e-6 in the Keplerian set $\{a,e,i,\Omega,\omega,\sigma \}$. While this is orders of magnitude larger than the uncertainties of (175706) 1996 FG3 and (35107) 1991 VH, the uncertainty without further observations increases exponentially after only tens of encounters. Thus, it is adequate for the studies in long-term simulations of this work.

The individual particles are sampled considering a multidimensional normal distribution centered around the nominal values shown in Table  \ref{t:cases}. Then, we use the Cholesky factorization of the covariance matrices to add the corresponding perturbation from the nominal.

\newpage
\section{Computation of Laplace coefficients} \label{App:LaplaceCoeffs}
%{\color{red} Written in previous paper. Do you think it is okay to include it again?}
The expansion of the potential requires the computation of Laplace coefficients, as introduced by Laplace (1798). \cite{Brouwer1961,Murray2000} detail both the expansion and computation of coefficients. In the case of the expansion in equation \ref{eq:pot-exp}:

\begin{equation*} 
\left \langle R_{j} \right \rangle = n_ja^2_j \left [ \frac{1}{2}A_{jj}e^2_j + \frac{1}{2}B_{jj}I^2_j + \sum^{N}_{\substack{k=1 \\ k\neq j}}  A_{jk} e_je_k\cos{ \left ( \varpi_j-\varpi_k\right )}  + B_{jk} I_jI_k\cos{ \left ( \Omega_j-\Omega_k\right )} \right ]
\end{equation*}

The coefficients $A_{jk}$,$B_{jk}$,$B_{jj}$ and $A_{jj}$ are:

\begin{equation}
A_{jk} = - n_j \frac{1}{4} \frac{m_k}{m_c+m_j} \alpha_{jk} \bar{\alpha}_{jk} b^{(2)}_{3/2}\left( \alpha_{jk} \right)
\end{equation}
\begin{equation}
B_{jk} = + n_j \frac{1}{4} \frac{m_k}{m_c+m_j} \alpha_{jk} \bar{\alpha}_{jk} b^{(1)}_{3/2}\left( \alpha_{jk} \right)
\end{equation}
\begin{equation}
A_{jj} = \sum^{N}_{k=1,k\neq j} B_{jk}
\end{equation}
\begin{equation}
B_{jj} = - \sum^{N}_{k=1,k\neq j} B_{jk}
\end{equation}

%\noindent
The definition of Laplace coefficient is:

\begin{equation}
\frac{1}{2} b_{s}^{(k)} (\alpha) = \frac{1}{2 \pi} \int_{0}^{2 \pi} \frac{\cos{ (k\psi) }d\psi}{(1-2\alpha \cos \psi + \alpha^2)^s} 
\end{equation}

This integral can be rewritten in a series expansion that simplifies the computation of the Laplace coefficients numerically as function of the rising factorial or Pochhammer symbol. However, it is found to be computationally more efficient to compute the quadrature integral above. There are many recursion and derivative rules that avoid computing the coefficients based on the definition. These expressions use the nomenclature of $D$ being the derivative operator $\frac{\mathrm{d} }{\mathrm{d} \alpha}$. 

\begin{equation}
b_{s+1}^{(j)} = \frac{s+j}{s} \frac{(1+\alpha^2)}{(1-\alpha^2)^2}b_{s}^{(j)} - \frac{2(j-s+1)}{s} \frac{\alpha}{(1-\alpha^2)^2} b_{s}^{(j+1)}
\end{equation}

\begin{equation}
b_{s+1}^{(j+1)} = \frac{j}{j-s} \left( \alpha + \frac{1}{\alpha} \right)b_{s+1}^{(j)} - \frac{j+s}{j-s} b_{s+1}^{(j-1)}
\end{equation}

\begin{equation}
D b_s^{(j)} = s \left( b_{s+1}^{(j-1)} - 2\alpha b_{s+1}^{(j)} + b_{s+1}^{(j+1)} \right)
\end{equation}

\begin{equation}
D^n b_s^{(j)} = s \left( D^{n-1} b_{s+1}^{(j-1)} - 2\alpha D^{n-1} b_{s+1}^{(j)} +  D^{n-1} b_{s+1}^{(j+1)} - 2(n-1) D^{n-2} b_{s+1}^{(j)}\right)
\end{equation}

\newpage
\bibliography{main-aastex-v2}{}

\begin{thebibliography}{}
\expandafter\ifx\csname natexlab\endcsname\relax\def\natexlab#1{#1}\fi
\providecommand{\url}[1]{\href{#1}{#1}}
\providecommand{\dodoi}[1]{doi:~\href{http://doi.org/#1}{\nolinkurl{#1}}}
\providecommand{\doeprint}[1]{\href{http://ascl.net/#1}{\nolinkurl{http://ascl.net/#1}}}
\providecommand{\doarXiv}[1]{\href{https://arxiv.org/abs/#1}{\nolinkurl{https://arxiv.org/abs/#1}}}

\bibitem[{Alessi \& S{\'{a}}nchez(2015)}]{Alessi2015}
Alessi, E.~M., \& S{\'{a}}nchez, J.~P. 2015, Journal of Guidance, Control, and
  Dynamics, 39, Number 2, \dodoi{10.2514/1.g001237}

\bibitem[{Armellin {et~al.}(2010)Armellin, {Di Lizia}, Berz, \&
  Makino}]{Armellin2010}
Armellin, R., {Di Lizia}, P., Berz, M., \& Makino, K. 2010, Celestial Mechanics
  and Dynamical Astronomy, \dodoi{10.1007/s10569-010-9281-7}

\bibitem[{Benitez \& Gallardo(2008)}]{Benitez2008}
Benitez, F., \& Gallardo, T. 2008, Celest Mech Dyn Astr, 101, 289,
  \dodoi{10.1007/s10569-008-9146-5}

\bibitem[{Brouwer \& Clemence(1961)}]{Brouwer1961}
Brouwer, D., \& Clemence, G.~M. 1961, in Methods of Celestial Mechanics
  (Academic Press), 610, \dodoi{10.1016/b978-1-4832-0075-0.50013-7}

\bibitem[{Chambers(1999)}]{Chambers1999}
Chambers, J.~E. 1999, Monthly Notices of the Royal Astronomical Society, 304,
  793, \dodoi{10.1046/j.1365-8711.1999.02379.x}

\bibitem[{Dones {et~al.}(1999)Dones, Gladman, Melosh, Tonks, Levison, \&
  Duncan}]{Dones1999}
Dones, L., Gladman, B., Melosh, H.~J., {et~al.} 1999, Icarus, 142, 509,
  \dodoi{10.1006/icar.1999.6220}

\bibitem[{Fang \& Margot(2011)}]{fang2011binary}
Fang, J., \& Margot, J.-L. 2011, The Astronomical Journal, 143, 25

\bibitem[{Fang \& Margot(2012)}]{Fang2012}
Fang, J., \& Margot, J.~L. 2012, Astronomical Journal,
  \dodoi{10.1088/0004-6256/143/1/25}

\bibitem[{Farinella {et~al.}(1994)Farinella, Froeschl{\'{e}}, Froeschl{\'{e}},
  Gonczi, Hahn, Morbidelli, \& Valsecchi}]{Farinella1994}
Farinella, P., Froeschl{\'{e}}, C., Froeschl{\'{e}}, C., {et~al.} 1994, Nature,
  371, 314, \dodoi{10.1038/371314a0}

\bibitem[{Folkner {et~al.}(2014)Folkner, Williams, Boggs, Park, \&
  Kuchynka}]{Folkner2014}
Folkner, W.~M., Williams, J.~G., Boggs, D.~H., Park, R.~S., \& Kuchynka, P.
  2014, Interplanet. Netw. Prog. Rep

\bibitem[{Fuentes-Munoz \& Scheeres(2020)}]{Fuentes-Munoz}
Fuentes-Munoz, O., \& Scheeres, D.~J. 2020, in 2020 AAS/AIAA Astrodynamics
  Specialist Conference - Lake Tahoe

\bibitem[{Giorgini \& {JPL Solar System Dynamics}(2021)}]{Giorgini}
Giorgini, J., \& {JPL Solar System Dynamics}. 2021, {NASA/JPL Horizons On-Line
  Ephemeris System}.
\newblock \url{http://ssd.jpl.nasa.gov/?horizons}

\bibitem[{Gladman {et~al.}(1997)Gladman, Migliorini, Morbidelli, Zappal{\`{a}},
  Michel, Cellino, Froeschl{\'{e}}, Levison, Bailey, \& Duncan}]{Gladman1997}
Gladman, B.~J., Migliorini, F., Morbidelli, A., {et~al.} 1997, Science, 277,
  197, \dodoi{10.1126/science.277.5323.197}

\bibitem[{Greenberg {et~al.}(1988)Greenberg, Carusi, \&
  Valsecchi}]{Greenberg1988}
Greenberg, R., Carusi, A., \& Valsecchi, G.~B. 1988, Icarus,
  \dodoi{10.1016/0019-1035(88)90125-X}

\bibitem[{Gronchi(2005)}]{Gronchi2005}
Gronchi, G.~F. 2005, Celestial Mechanics and Dynamical Astronomy,
  \dodoi{10.1007/s10569-005-1623-5}

\bibitem[{Hedo {et~al.}(2020)Hedo, Fantino, Ru{\'{i}}z, \&
  Pel{\'{a}}ez}]{Hedo2020}
Hedo, J.~M., Fantino, E., Ru{\'{i}}z, M., \& Pel{\'{a}}ez, J. 2020, Astronomy
  {\&} Astrophysics, \dodoi{10.1051/0004-6361/201936502}

\bibitem[{Hedo {et~al.}(2018)Hedo, Ru{\'{i}}z, \& Pel{\'{a}}ez}]{Hedo2018}
Hedo, J.~M., Ru{\'{i}}z, M., \& Pel{\'{a}}ez, J. 2018, Monthly Notices of the
  Royal Astronomical Society, \dodoi{10.1093/mnras/sty1598}

\bibitem[{Heggie \& Rasio(1996)}]{HeggieRasio1996}
Heggie, D.~C., \& Rasio, F.~A. 1996, Monthly Notices of the Royal Astronomical
  Society, 282, 1064, \dodoi{10.1093/mnras/282.3.1064}

\bibitem[{Jones {et~al.}(2018)Jones, Slater, Moeyens, Allen, Axelrod, Cook,
  {\v{Z}}eljko, Juri{\'{c}}, Myers, \& Petry}]{Jones2018}
Jones, R.~L., Slater, C.~T., Moeyens, J., {et~al.} 2018, {The Large Synoptic
  Survey Telescope as a Near-Earth Object discovery machine},
  \dodoi{10.1016/j.icarus.2017.11.033}

\bibitem[{{JPL Solar System Dynamics} \& {Center for NEO Studies
  (CNEOS)}(2021)}]{JPLSolarSystemDynamics}
{JPL Solar System Dynamics}, \& {Center for NEO Studies (CNEOS)}. 2021, {JPL
  Small-Body Database Search Engine}

\bibitem[{Kinoshita \& Nakai(2007)}]{kinoshita2007general}
Kinoshita, H., \& Nakai, H. 2007, Celestial Mechanics and Dynamical Astronomy,
  98, 67

\bibitem[{Meyer \& Scheeres(2021)}]{MEYER2021114554}
Meyer, A.~J., \& Scheeres, D.~J. 2021, Icarus, 367, 114554,
  \dodoi{https://doi.org/10.1016/j.icarus.2021.114554}

\bibitem[{{Meyer} {et~al.}(2021){Meyer}, {Scheeres}, {Naidu}, {Benner},
  {Pravec}, \& {Scheirich}}]{2021DDA....5240506M}
{Meyer}, A.~J., {Scheeres}, D.~J., {Naidu}, S., {et~al.} 2021, in AAS/Division
  of Dynamical Astronomy Meeting, Vol.~53, AAS/Division of Dynamical Astronomy
  Meeting, 405.06

\bibitem[{Michel \& Froeschl{\'{e}}(1997)}]{Michel1997-2}
Michel, P., \& Froeschl{\'{e}}, C. 1997, Icarus, \dodoi{10.1006/icar.1997.5727}

\bibitem[{Michel {et~al.}(1996)Michel, Froeschl{\'{e}}, \&
  Farinella}]{Michel1996}
Michel, P., Froeschl{\'{e}}, C., \& Farinella, P. 1996, in Worlds in
  Interaction: Small Bodies and Planets of the Solar System, ed. H.~Rickman \&
  M.~J. Valtonen (Dordrecht: Springer Netherlands), 151--164

\bibitem[{Michel {et~al.}(1997)Michel, Froeschl{\'{e}}, \&
  Farinella}]{Michel1997}
Michel, P., Froeschl{\'{e}}, C., \& Farinella, P. 1997, Celestial Mechanics and
  Dynamical Astronomy, 69, 133, \dodoi{10.1007/978-94-017-1321-4-11}

\bibitem[{Michel {et~al.}(2005)Michel, Morbidelli, \& Bottke}]{Michel2005}
Michel, P., Morbidelli, A., \& Bottke, W.~F. 2005, Comptes Rendus Physique, 6,
  291, \dodoi{10.1016/j.crhy.2004.12.013}

\bibitem[{Milani {et~al.}(1989)Milani, Carpino, Hahn, \& Nobili}]{Milani1989}
Milani, A., Carpino, M., Hahn, G., \& Nobili, A.~M. 1989, Icarus, 78, 212,
  \dodoi{10.1016/0019-1035(89)90174-7}

\bibitem[{Milani {et~al.}(2005)Milani, Chesley, Sansaturio, Tommei, \&
  Valsecchi}]{Milani2005}
Milani, A., Chesley, S.~R., Sansaturio, M.~E., Tommei, G., \& Valsecchi, G.~B.
  2005, Icarus, \dodoi{10.1016/j.icarus.2004.09.002}

\bibitem[{Milani \& Kne{\v{z}}evi{\'{c}}(1990)}]{Milani1990}
Milani, A., \& Kne{\v{z}}evi{\'{c}}, Z. 1990, Celestial Mechanics and Dynamical
  AStronomy, 49, 347, \dodoi{10.1007/BF00049444}

\bibitem[{Morbidelli {et~al.}(2009)Morbidelli, Brasser, Tsiganis, Gomes, \&
  Levison}]{Morbidelli2009}
Morbidelli, A., Brasser, R., Tsiganis, K., Gomes, R., \& Levison, H.~F. 2009,
  Astronomy and Astrophysics, 507, 1053, \dodoi{10.1051/0004-6361/200912876}

\bibitem[{Murray \& Dermott(2000)}]{Murray2000}
Murray, C.~D., \& Dermott, S.~F. 2000, {Solar System Dynamics} (Cambridge
  University Press), \dodoi{10.1017/cbo9781139174817}

\bibitem[{Nesvorn{\'{y}} \& Bottke(2004)}]{Nesvorny2004}
Nesvorn{\'{y}}, D., \& Bottke, W.~F. 2004, Icarus, 170, 324,
  \dodoi{10.1016/J.ICARUS.2004.04.012}

\bibitem[{{\"{O}}pik(1976)}]{opik1976interplanetary}
{\"{O}}pik, E.~J. 1976, {Interplanetary encounters: close-range gravitational
  interactions}, Developments in solar system- and space science (Elsevier
  Scientific Pub. Co.).
\newblock \url{https://books.google.com/books?id=-HXvAAAAMAAJ}

\bibitem[{Pokorn{\'{y}} \& Vokrouhlick{\'{y}}(2013)}]{Pokorny2013}
Pokorn{\'{y}}, P., \& Vokrouhlick{\'{y}}, D. 2013, Icarus,
  \dodoi{10.1016/j.icarus.2013.06.015}

\bibitem[{Pravec {et~al.}(2016)Pravec, Scheirich, Ku{\v{s}}nir{\'a}k, Hornoch,
  Gal{\'a}d, Naidu, Pray, Vil{\'a}gi, Gajdo{\v{s}}, Korno{\v{s}},
  {et~al.}}]{pravec2016binary}
Pravec, P., Scheirich, P., Ku{\v{s}}nir{\'a}k, P., {et~al.} 2016, Icarus, 267,
  267

\bibitem[{Rein \& Spiegel(2014)}]{Rein2014}
Rein, H., \& Spiegel, D.~S. 2014, Monthly Notices of the Royal Astronomical
  Society, \dodoi{10.1093/mnras/stu2164}

\bibitem[{Roy(2004)}]{Roy2004}
Roy, A. 2004, {Orbital Motion, Fourth Edition} (Routledge),
  \dodoi{10.1201/9781420056884}

\bibitem[{Scheeres {et~al.}(2020)Scheeres, McMahon, Bierhaus, Wood, Benner,
  Hartzell, Hayne, Hopkins, Jedicke, Corre, Meyer, Naidu, Pravec, Ravine, \&
  Sorli}]{Scheeres2020Janus}
Scheeres, D., McMahon, J., Bierhaus, E., {et~al.} 2020, Bulletin of the AAS,
  52.
\newblock \url{https://baas.aas.org/pub/2020n6i217p06}

\bibitem[{{SSD/CNEOS API Service}(2021)}]{SBDB}
{SSD/CNEOS API Service}. 2021, {SSD/CNEOS Small-Body DataBase API}.
\newblock \url{https://ssd.jpl.nasa.gov/tools/sbdb_lookup.html#/}

\bibitem[{Tancredi(1998)}]{Tancredi1998}
Tancredi, G. 1998, Celestial Mechanics and Dynamical Astronomy, 70, 181,
  \dodoi{10.1023/A:1008331422678}

\bibitem[{Valsecchi {et~al.}(2015)Valsecchi, Alessi, \& Rossi}]{Valsecchi2015}
Valsecchi, G.~B., Alessi, E.~M., \& Rossi, A. 2015, Celestial Mechanics and
  Dynamical Astronomy, 123, 151, \dodoi{10.1007/s10569-015-9631-6}

\bibitem[{Valsecchi {et~al.}(2003)Valsecchi, Milani, Gronchi, \&
  Chesley}]{Valsecchi2003}
Valsecchi, G.~B., Milani, A., Gronchi, G.~F., \& Chesley, S.~R. 2003, Astronomy
  {\&} Astrophysics, 408, 1179, \dodoi{10.1051/0004-6361:20031039}

\bibitem[{Vokrouhlick{\'{y}} {et~al.}(2000)Vokrouhlick{\'{y}}, Milani, \&
  Chesley}]{Vokrouhlicky2000}
Vokrouhlick{\'{y}}, D., Milani, A., \& Chesley, S.~R. 2000, Icarus, 148, 118,
  \dodoi{10.1006/ICAR.2000.6469}

\bibitem[{Vokrouhlick{\'{y}} {et~al.}(2012)Vokrouhlick{\'{y}}, Pokorn{\'{y}},
  \& Nesvorn{\'{y}}}]{Vokrouhlicky2012}
Vokrouhlick{\'{y}}, D., Pokorn{\'{y}}, P., \& Nesvorn{\'{y}}, D. 2012, Icarus,
  \dodoi{10.1016/j.icarus.2012.02.021}

\bibitem[{Wisdom \& Holman(1991)}]{Wisdom1991}
Wisdom, J., \& Holman, M. 1991, The Astronomical Journal, 102, 1528,
  \dodoi{10.1086/115978}

\bibitem[{Wi{\'{s}}niowski \& Rickman(2013)}]{Wisniowski2013}
Wi{\'{s}}niowski, T., \& Rickman, H. 2013, Acta Astronomica

\end{thebibliography}
\bibliographystyle{aasjournal}



\end{document}